# Mechanical behavior of high-entropy alloys: A review

Yuanyuan Shang [a], Jamieson Brechtl [b], Claudio Psitidda [a], and Peter K. Liaw [c*]

[a] Department of Materials Design, Institute of Hydrogen Technology, Helmholtz-Zentrum Geesthacht, 21502, Geesthacht, Germany

[b] Energy and Transportation Sciences Division, Oak Ridge National Laboratory, Oak Ridge, TN 37831, USA

[c] Department of Materials Science and Engineering, The University of Tennessee, Knoxville, Tennessee 37996, USA

* Correspondence to: pliaw@utk.edu

High-entropy alloys (HEAs) are materials that consist of equimolar or near-equimolar multiple principal components but tend to form single phases, which is a new research topic in the field of metallurgy, have attracted extensive attention in the past decade. The HEAs families contain the face-centered-cubic (fcc), body-centered-cubic (bcc), and hexagonal-close-packed (hcp)-structured HEAs. On one hand, mechanical properties, *e.g.* hardness, strength, ductility, fatigue, and elastic moduli, are essential for practical applications of HEAs. Scientists have explored in this direction since the advent of HEAs. On the other hand, the pursuit of high strength and good plasticity is the critical research issue of materials. Hence, strengthening of HEAs is a crucial issue. Recently, many articles are focusing on the strengthening strategies of HEAs[1-14].

In this chapter, we reviewed the recent work on the room-temperature elastic properties and mechanical behavior of HEAs, including the mechanisms behind the plastic deformation of HEAs at both low and high temperatures. Furthermore, the present work examined the strengthening strategies of HEAs, *e.g.* strain hardening, grain-boundary strengthening, solid-solution strengthening, and particle strengthening. The fatigue, creep, and fracture properties were briefly introduced. Lastly, the future scientific issues and challenges of HEAs were discussed.

# Table of Contents

# 1. Introduction

For thousands of years, metals have been of great significance in the human’s daily life. Generally speaking, the design of different alloys has a very large impact on their performance. Therefore, based on some main performance requirements, one or two main elements are selected for the traditional alloy design. Then small amounts of other elements are added to regulate certain properties. For example, the major component in the Fe-18Ni3Al4Mo0.8Nb0.08C0.01B [weight percent (wt.%)][15] steel is Fe, and other elements are added to achieve the desired properties.

In recent years, high-entropy alloys (HEAs)[16] have broken the traditional alloy-design rule. These special alloys consist of multiple principal elements (with a principal element number of $\geq$ 5) in equimolar or near-equimolar ratios, which tend to form single-phase solid solutions. The concept of HEAs (or multi-principal element alloys) was first proposed by Yeh *et al.*[17] and Cantor *et al.*[18] in 2004. According to the preliminary definition, there are three basic empirical criteria for the design of HEAs: the principal number of elements in HEAs is no less than 5, The atomic percentage (at.%) of each element is between 5 % and 35 %, and the mixed entropy of HEAs is not lower than 12.47 $J \cdot mol^{-1} \cdot K^{-1}$ [19]. However, in recent years, the concept of HEAs has been expanded to include alloys that have four major elements[20-22].

As novel alloy systems, HEAs have attracted extensive attention in recent years[2, 23-27]. According to the traditional alloy-design theory, the more principal components there are, the easier the alloy is to form complex phases (such as intermetallic compounds). Consequently, there is a reduction in the desirable properties of the alloy[28]. Although HEAs are composed of multiple principal components, they can form a single solid-solution phase. For example, the FeCoNiCrMn[18] HEA has a face-centered-cubic (fcc) phase, the TaNbHfZrTi[29] HEA is body-centered-cubic (bcc)-structured, and the GdHoLaTbY[30] HEA consists of an hexagonal-close-packed (hcp) structure. Due to the novel composition of HEAs, they have the following characteristics: (1) high mixing entropy[31], which makes the HEAs single phases; (2) severe lattice distortion [32-34], which results in the solid-solution strengthening effect; (3) sluggish diffusion[35, 36], which reduces the phase-transformation rate (however, it should be noted that this effect has not been proved in HEAs[36, 37]); and (4) cocktail effect[38], by changing compositions, the properties of HEAs can be adjusted. In terms of their properties, HEAs have high hardness and strength[2, 26], good wear resistance[39], strong corrosion[40-43] resistance, fatigue[44-49] resistance, and fracture[50] resistance. For example, some HEAs have shown excellent corrosion behavior in $H_2SO_4$, which is better than that of some Ni alloys, Ti alloys, and steels[51-53]. Additionally, the excellent high-temperature stability and corresponding corrosion resistance of HEAs make them potential candidates for use as catalysts[54] and in high-temperature water-vapor environments that are typical of thermal power plants[55], oil detection environments in the ocean[56], extreme irradiation conditions[57], and cryogenic conditions[58]. Moreover, HEAs exhibit great fatigue resistance[44-49], and excellent mechanical properties at both low[27] and high[59] temperatures. Hence,

HEAs have the potential to serve as structural materials in harsh conditions at low and high temperatures, as well as in corrosive environments.

It is important to note that there have been numerous studies regarding the mechanical properties of HEAs. The goal of this chapter, therefore, is to systematically summarize the findings of these previous investigations and then discuss the future scientific issues and challenges of HEAs regarding lightweight HEAs, HEA films and coatings, additive manufacturing, high-throughput (HT) methods[60], as well as machine-learning (ML) techniques[61]. It is thus anticipated that this chapter will provide guidance on the future structural applications of HEAs.

# 2. Major characteristics of HEAs

## 2.1. Single phase

The high-entropy effect and sluggish diffusion of HEAs make the solid-solution phases stable, as compared to intermetallic compounds, thus allowing them to form single phases[19]. There are three major families of HEAs, namely the fcc-, bcc-, and hcp-structured HEA families. So far, about 37 elements consisting of 3d-transition metals, refractory metals, and rare-earth metals have been used to design and fabricate HEAs[62]. The increasing application potential and scientific value of HEAs therefore require the development of new alloy families.

The fcc-structured HEAs, which are based on the Fe, Co, Ni, Cr, Mn, and Cu elements, are the most studied HEAs. Compared with the conventional alloys, the fcc-structured HEAs usually have higher strength and better plasticity than some twinning-induced plasticity (TWIP) steels, austenitic stainless steels, and transformation-induced plasticity (TRIP) steels. For instance, the product of the tensile and fracture elongation of fcc-structured HEAs can reach 15 ~ 45 GPa %[5, 63], which is higher than some TRIP steels (only about 15 ~ 30 GPa %)[64-66].

For the bcc-structured HEAs, the FeCoNiCrAl[67] alloy is a representative alloy. Moreover, the refractory HEAs, such as TaNbHfZrTi and VNbMoTaW[68] alloys, are also bcc-structured. For these kinds of HEAs, the strength is higher than that of the fcc-structured HEAs, but their plasticity is poor. Further information regarding the mechanical properties of fcc- and bcc-structured HEAs can be seen in Table 5.1.

Most of the widely-studied HEAs are fcc/bcc-structured, and their compositions are mainly transition metals. In recent years, a large number of HEAs[30, 69-71] with an hcp structure have been fabricated. These kinds of HEAs are dominated by lanthanide rare earth elements, among which YGdTbDyHo is the canonical representative.

## 2.2. Super solid solutions

There are no ‘solute’ and ‘solvent’ in HEAs, and the multi-principal components make them the so-called ‘super solid solutions’[72]. Senkov *et al.*[29] used the Labusch model[73] of solid-solution strengthening to explain the mechanical behavior of the bcc-structured TaNbHfZrTi HEA. They calculated the interaction force ($f_m$) between a mobile dislocation and a solute atom of the HEA by the following equation[74]:

$$f_m = Gb^2(\eta_i + \alpha\delta_i) \tag{2.1}$$

where $G$ is the shear modulus, $b$ is the Burgers vector, $\eta$ is the modulus-misfit parameter, $i$ is the solute species, $\alpha$ is a dimensionless parameter related to the type of dislocations (for screw dislocations, $3 < \alpha < 16$ and for edge dislocations, $\alpha > 16$[75]), and $\delta$ is the lattice-mismatch parameter.

The $\eta_i$ and $\delta_i$ can be calculated by:

$$\eta_i = \frac{1}{G}\frac{\mathrm{d}G}{\mathrm{d}x_i}, \delta_i = \frac{1}{a}\frac{\mathrm{d}a}{\mathrm{d}x_i} \tag{2.2}$$

where $a$ is the lattice constant of the metal, and $x$ is the solute concentration.

The solute-induced stress increase ($\Delta\sigma_{solute\text{-}strengthening}$) correlates with $f_m$, the solute concentration ($x$), and dislocation line tension ($E_L$) as[74, 76]:

$$\Delta\sigma_{solute-strengthening} b^2 = A f_m^{4/3} x^{2/3} E_L^{-1/3} \tag{2.3}$$

$$E_L = \frac{Gb^2}{2} \tag{2.4}$$

where $A$ is a dimensionless constant of the material.

According to the calculated results, Senkov *et al.*[29] suggested that the solid-solution strengthening model might not be applied to HEAs, but they tried to estimate the effects of the modulus misfit and lattice mismatch on the $f_m$, and then the $\Delta\sigma_{solute\text{-}strengthening}$. The estimated yield stress ($\sigma_y$) of the HEA is 1,094 MPa, which is only 18 % higher than that of the experimental value, indicating that the solid solution-strengthening theory can explain the deformation behavior well.

Wu *et al.*[77] treated the HEA as an N-element alloy and other M-elements as effective solutes. They then applied the Labusch model to explain the solid-solution strengthening effect in some fcc-structured HEAs, *e.g.* the subsystems of the quinary FeCoNiCrMn HEA system (FeCoNiCrMn, FeCoNiCr, FeNiCoMn, NiCoCrMn, FeNiCo, FeNiMn, NiCoCr, NiCoMn, FeNi, and NiCo alloys). Based on their results, it is the Labusch solid-solution strengthening effect, rather than the lattice friction that contributed to the deformation behavior in the studied HEAs.

For the Labusch model, the increase of the yield stress caused by one solute can be calculated by[73-75, 78-81]:

$$\Delta\sigma_{solute-strengthening} = fG\left[\left(\frac{\eta_i}{1+\frac{1}{2}|\eta_i|}\right)^2 + \alpha^2\delta_i^2\right]^{2/3} x_i^{2/3} \tag{2.5}$$

where $f$ is the dimensionless parameter, $G$ is the shear modulus, $\eta$ is the modulus-misfit parameter, $i$ is the solute species, $\alpha$ is a dimensionless parameter related to the type of dislocations, $\delta$ is the lattice-mismatch parameter, and $x$ is the solute concentration.

For the solid solutions with multiple solute species, $i, j, \ldots$, the strengthening effect can be calculated as follows[75]:

$$\Delta\sigma'_{solute-strengthening} = fG\left\{\left[\left(\frac{\eta_i}{1+\frac{1}{2}|\eta_i|}\right)^2 + \alpha^2\delta_i^2\right]x_i + \left[\left(\frac{\eta_j}{1+\frac{1}{2}|\eta_j|}\right)^2 + \alpha^2\delta_j^2\right]x_j + \cdots\right\}^{2/3} \tag{2.6}$$

When introducing the Labusch model for HEAs, the contribution of the M-element to the N-element alloy is given by the equation[77]:

$$L_{M\to N} = G_N\left\{\sum_{i=M+1}^{N}\left[\left(\frac{\eta_i}{1+\frac{1}{2}|\eta_i|}\right)^2 + \alpha^2\delta_i^2\right]x_i\right\}^{2/3} \quad (2.7)$$

where $G_N$ is the shear modulus of the N-element alloy. The Labusch strengthening factors in Eq. (2.7), as a function of $\alpha$, are shown in Fig. 1. Figure 1(a) indicates that from the Ni to NiCo alloy, the weak dependence on $\alpha$ is due to a small lattice mismatch. Thus, the dependence of the Labusch-strengthening factors of the FeNi to FeNiCo alloy and Ni to NiCo alloy on $\alpha$ is weaker than the other curves. While in Fig. 1(b), the curves indicate that the dependence of the Labusch-strengthening factors of the Ni to NiCoCr alloy and Ni to FeNiCoCr alloy on $\alpha$ is strong.

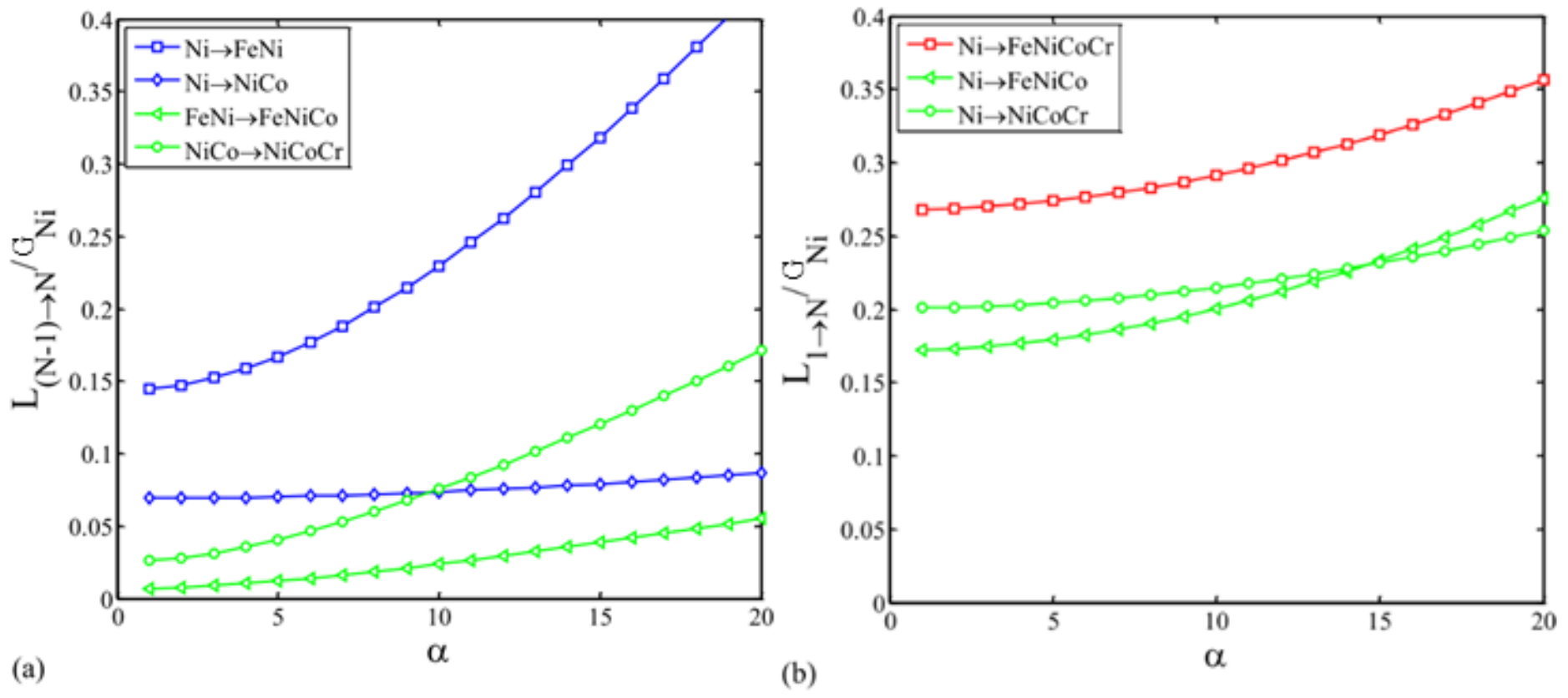

**Fig. 1.** The correlation between the Labusch-strengthening factors and $\alpha$: (a) $L_{(N-1)\to N}/G_{Ni}$ for Ni to FeNi and NiCo, FeNi to FeNiCo and NiCo to NiCoCr; (b) $L_{1\to N}/G_{Ni}$ for Ni to FeNiCoCr, FeNiCo, and NiCoCr. Figures from Wu et al. [77].

Wu *et al.*[77] compared the Labusch-strengthening factors with the athermal-strengthening effect, $\Delta\sigma_{intrinsic}$, and found that the two factors are in good agreement, suggesting that the Labusch-strengthening analysis is valid for HEAs [see Figs. 2(a) and (b)]. Figure 2(a) shows the strengthening of the intrinsic yield stress. The high increase from NiCo to NiCoCr results from the highest mismatch of the Cr element. Note that the invar effect[82], which causes the alloy to have no thermal expansion, may lead to the large deviation of the FeNi alloy. The results also show that the 1 to *N* model is more reliable than the *(N-1)* to *N* model. Figure 2(b) presents the corresponding normalized Labusch factors ($\alpha$ = 16, normalized by $G_{Ni}$). The results are in good agreement with those in Fig. 2(a). Therefore, for the alloys with simple phases but complex components, the Labusch solid-solution strengthening method gives a useful design criterion.

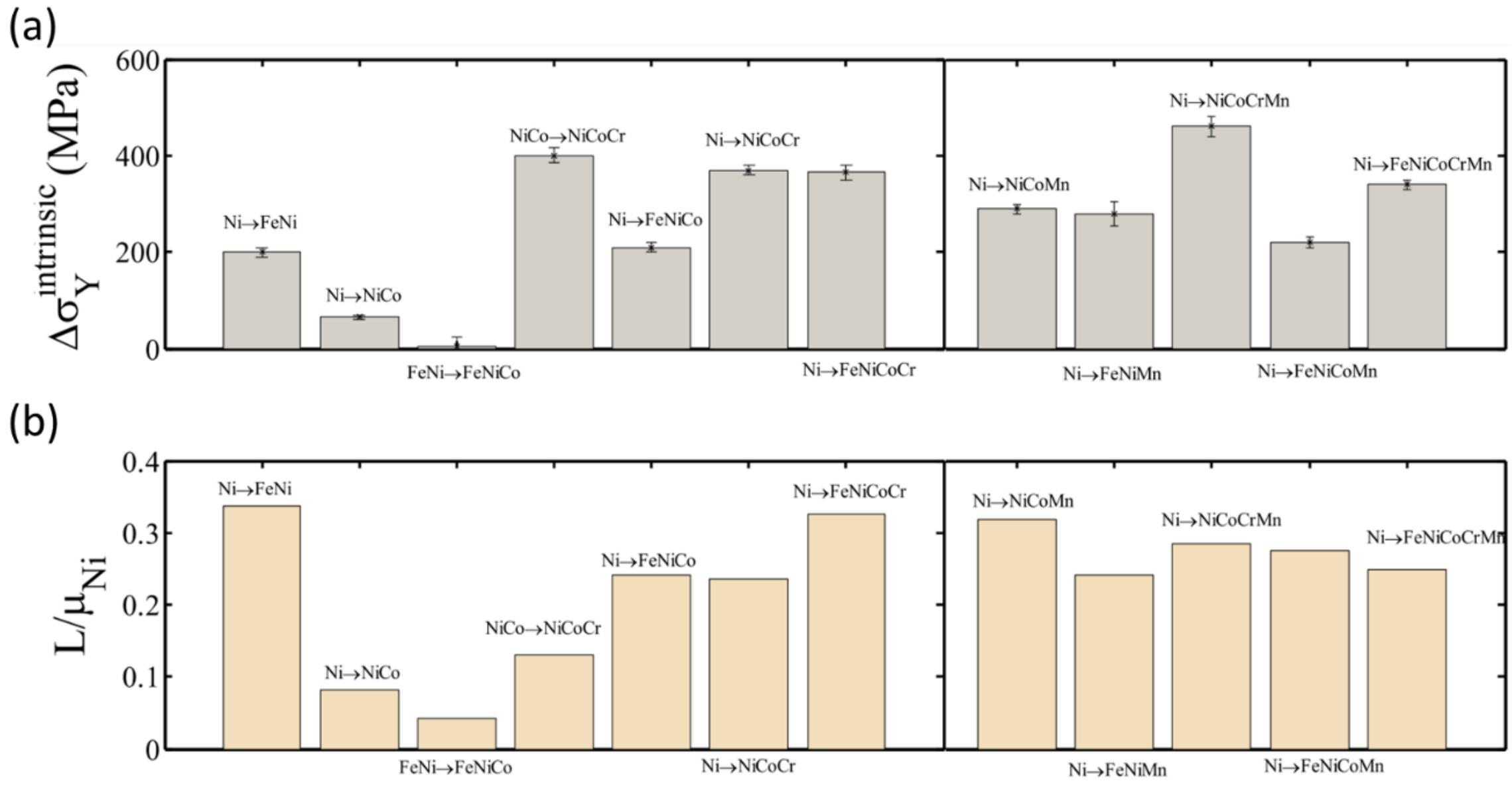

**Fig. 2.** The comparison of the (a) athermal-strengthening effect and (b) Labusch-strengthening factor of the family of Fe, Co, Ni, Cr, and Mn containing materials. Figures from Wu et al. [77].

Varvenne *et al.*[83, 84] developed a theory to better understand the solid-solution strengthening mechanism in fcc HEAs. They added a wavy configuration into a straight dislocation to minimize the potential energy of the dislocations, in which the typical feature is displayed in Fig. 3. The figure shows that when the dislocation passes the random solutes, the dislocation has a transverse length of $2\zeta$ and an amplitude of $\omega$. The energy change of the straight line segment of a length, $\zeta$, is of great importance, because it slips a distance of $\omega$ in the random solute field.

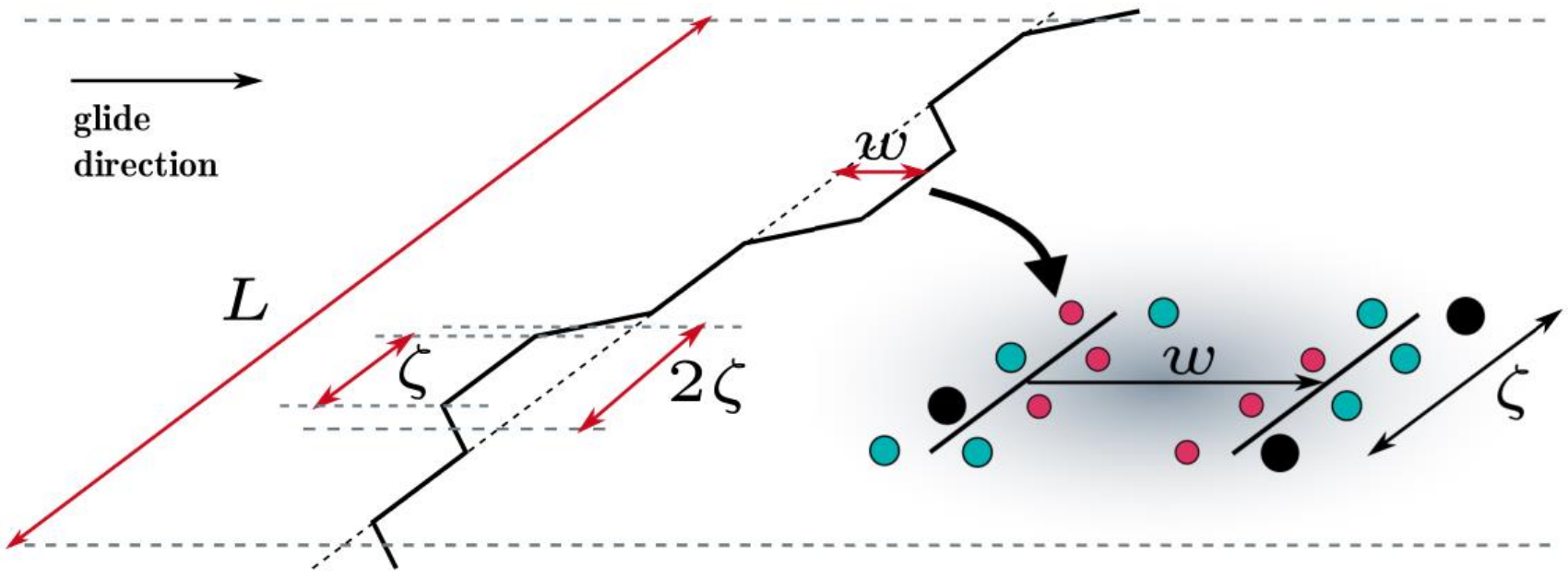

**Fig. 3.** The sketch map of a low-energy dislocation wavy configuration when the dislocation is passing through a field of solute atoms. Figure from Varvenne et al. [84].

Then they minimize the total energy for the arbitrary $\zeta$ and $\omega$ to obtain the characteristic lengths, $\zeta_c$ and $\omega_c$, by the following equations [84]:

$$\Delta E_{tot}(\zeta,\omega) = \Delta E_{LT}(\zeta,\omega) - \sigma_{\Delta U_{tot}}(\zeta,\omega)\left(\frac{L}{2\zeta}\right)$$

$$= \left[\Gamma\frac{\omega^2}{2\zeta} - \left(\frac{\zeta}{\sqrt{3}b}\right)^{\frac{1}{2}}\Delta\tilde{E}_p(\omega)\right]\left(\frac{L}{2\zeta}\right) \qquad (2.8)$$

where $L$ is the length of the dislocation, $\Delta E_{tot}$ is the total energy, $\Delta E_{LT}$ is the energy cost of the dislocation bowing, $\sigma_{\Delta U_{tot}}(\zeta,\omega)$ is the potential-energy change, $\Gamma$ is the dislocation-line tension, $b$ is the Burgers vector, and $\Delta\tilde{E}_p(\omega)$ is the potential-energy function. From the equations, both $\zeta$ and $\omega$ can be determined.

For the dislocation slip, the energy barrier, $\Delta E_b$, can be given by the following functions[84]:

$$\Delta E_b = \Delta E_b' - \Delta E_{LT} \quad (2.9a)$$

$$\Delta E_b' = \sqrt{2}\sigma_{\Delta U_{tot}} L/2\zeta_C \quad (2.9b)$$

where $\Delta E_b'$ is the potential-energy barrier.

The $\Delta E_b$ can be calculated to be $1.22(\frac{\omega_c^2 \Gamma \Delta\tilde{E}_p^2(\omega_c)}{b})^{\frac{1}{3}}$. This method then provides the guidance for developing fcc HEAs with desirable strengths as well as a better understanding regarding the strengthening behavior of HEAs.

In 2018, Varvenne *et al.*[85] used the solid-solution-strengthening method to determine the $\sigma_y$ of noble metal (Pd-Pt-Rh-Ir-Au-Ag-Cu-Ni) HEAs. Figure 4 shows the evaluated $\sigma_y$ of the Pd-Pt-Rh-Ir-Au-Ag-Cu-Ni HEAs. The results indicate that these compositions, *e.g.* $(AuAg)_{0.4}(CuNi)_{0.6}$, $(AuAg)_{0.3}(RhIr)_{0.7}$, $(CuNi)_{0.4}(RhIr)_{0.6}$, and $(AuAg)_{0.24}(CuNi)_{0.38}(RhIr)_{0.38}$, are all possible candidates for high-strength and phase-stable HEAs.

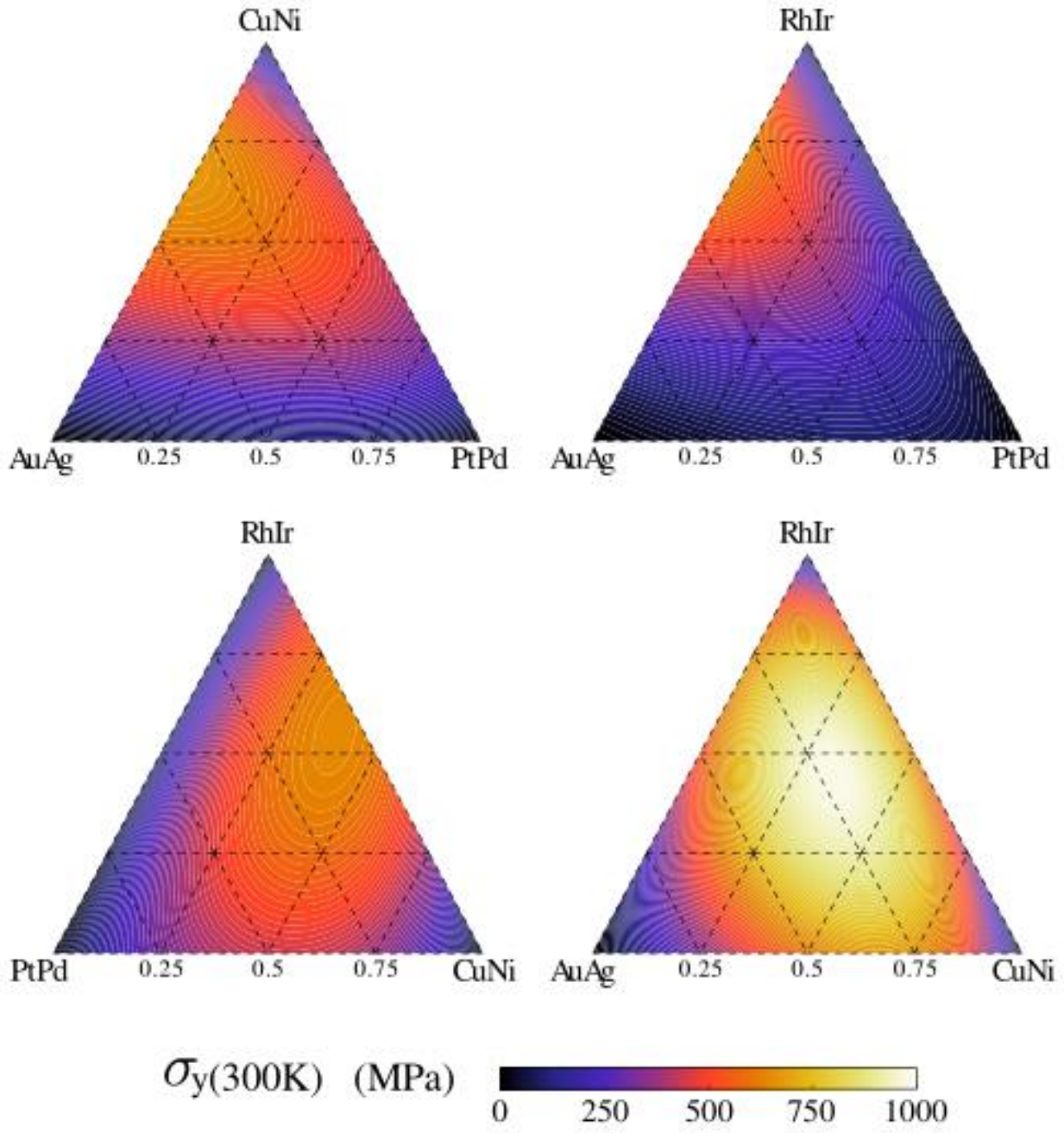

**Fig. 4.** Predicted $\sigma_y$ of different pseudo-ternary Pd-Pt-Rh-Ir-Au-Ag-Cu-Ni HEAs at 300 K. Figure from Varvenne et al. [85].

The study of Bracq *et al.*[86] has provided the information on the solid-solution strengthening effect of HEAs (the Fe-Co-Ni-Cr-Mn system). They prepared 24 different kinds of alloys and then combined the experimental and computational results to show that the strength varies with the composition. For instance, CoCrMnNi and $(CoCrFeMn)_{40}Ni_{60}$ alloys have the highest strengths, while $(CoCrFeMn)_8Ni_{92}$ and $(CoCrMnNi)_{50}Fe_{50}$ alloys have the lowest strengths. Their results also show that this model is valid to predict material strengths, which can be extended to other HEAs.

Maresca *et al.*[87] attempted to explain the theory of screw-dislocation strengthening in bcc HEAs. They presented a new model to elaborate the typical structure and motion mode of a screw dislocation under the random solute condition, as shown in Fig. 5. For the unrelaxed screw dislocation of length, $L$, its low-energy configuration is that the kink of width, $w_k$, and height, $a$, are spontaneously formed and then replaced by a characteristic length, $\zeta_c$. Due to the symmetrical compact cores of screw dislocations, it can be kinked in three different [34] planes (including their line directions). The blue region represents the dislocation bending in one plane, while the red region represents the dislocation, which bends in another plane (called the cross-slip). Therefore, the symmetrical compact core will lead to the so-called cross-slip event. When a certain stress is applied, the lateral kink glide will lead to the cross-kink formation, resulting in the pinning of the cross-kink. The Peierls mechanism or the lateral kink gliding can move the line segment between two kinks placed on the same slip plane forward. The advancement of the segment occurs over the characteristic sliding length, $2a$, where it passes through the unstable interference configuration. A molecular dynamics (MD) method[88] was used to clarify and verify the typical features of the screw dislocations, which helped determine the strength versus composition and strain rate. The results show that the theory can be applied to Ti-Nb-Zr, Nb-W, Nb-Mo, and Fe-Si-based HEAs, which paves the way for designing new bcc HEAs. Yin *et al.*[89] predicted the $\sigma_y$ of the RhIrPdPtNiCu HEA using a first-principles method. The predicted value was 583 MPa, which is in a good agreement with the measured one (527 MPa), indicating that the density functional-theory (DFT)[90] is a viable way to help in the development of HEAs with excellent mechanical properties.

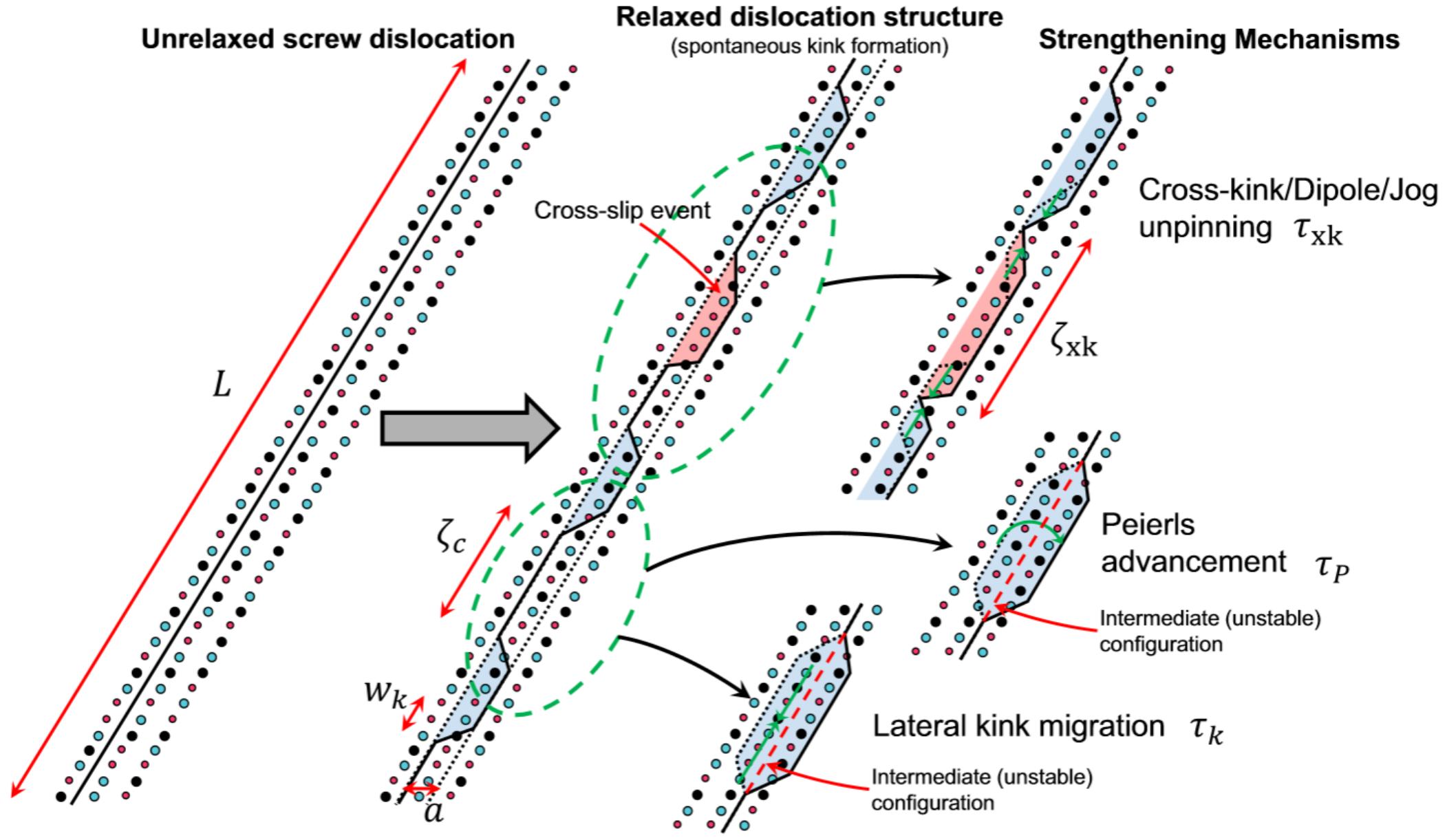

**Fig. 5.** Schematic model of a screw dislocation under the random solute condition. Figure from Maresca et al. [87].

In summary, different strategies, such as the MD method and the DFT have been used to predict the structures or the mechanical properties of HEAs. Investigating the theory of solid solutions in HEAs is of great importance for understanding the relationship among mechanical properties, microstructures, and compositions. The solid-solution model gives a practical means for the design and exploration of HEAs with desirable strengths.

Recently, scientists have observed some unusual dislocation behaviors in HEAs[91-95]. For instance, Ma[92] reported that the lattice distortion and local chemical order will lead to HEAs with inhomogeneous compositions. Thus, the dislocation motion has to pass through a rugged atomic and energy landscape, which provides new insights into dislocation energetics and is beyond the traditional solid-solution hardening method. Chen *et al.*[93] simulated the dislocation mobility in a $Co_{16.67}Fe_{36.67}Ni_{16.67}Ti_{30}$ bcc-structured HEA. Their results showed that the rate-limiting step for screw dislocations is the trapping from kinks, rather than the kink nucleation. Meanwhile, the edge dislocations have a similar activation process and comparable activation barrier as screw dislocations to detrap from the kinks, leading to the comparable contribution of edge dislocations to the strength. Moreover, the dominant presence of edge dislocations in refractory bcc HEAs of NbTaVTi and CrMoNbV[94, 95], and a refractory bcc medium-entropy alloy of MoNbTi[91] has been experimentally found during deformation. The existence of edge dislocations in controlling the strengths of refractory bcc entropic alloys is in contrast with the conventional bcc metals that are usually controlled by screw dislocations[96-100]. Moreover, the multiple slip paths could enhance the plasticity and ductility in bcc refractory HEAs[91, 101].

# 3. Elastic properties of HEAs

Haglund *et al.*[102] measured the elastic constants of the FeCoNiCrMn HEA via the resonant ultrasound spectroscopy (RUS) technique, where the HEA was assumed to be elastically isotropic. Subsequently, the Poisson's ratio ($v$), Young's modulus ($E$), and bulk modulus ($B$) were obtained by the following equations[103]:

$$v = \frac{L-2G}{2(L-G)} \quad (3.1)$$

$$E = \frac{G(3L-4G)}{L-G} \quad (3.2)$$

$$B = L - \frac{4G}{3} \quad (3.3)$$

where $L$ is the longitudinal modulus, and $G$ is the shear modulus. The results are shown in Table 3.1.

Laplanche *et al.*[104] measured the elastic moduli of medium-entropy subsystems of the FeCoNiCrMn HEAs. By analyzing the torsional deformation of plates, $G$ can be determined from the following equation[105]:

$$G = \rho F_T^2 \frac{(h_T^2+e_T^2)l_T^2}{\left(1-\frac{e_T}{\sqrt{3}h_T}\right)e_T^2} \quad (3.4)$$

where $\rho$ is the density, $F_T$ is the torsional-resonance frequency, and $e_T$, $l_T$, and $h_T$ are the thickness, gauge length, and width of the plate, respectively.

The Young's modulus, $E$, was obtained by first bending the beam samples and then calculating according to[106]:

$$E = 0.9464 \rho F_B^2 \frac{l_B^4}{e_B^2} U\left\{\frac{e_B}{l_B}, \nu\right\} \quad (3.5)$$

where $\rho$ is the density, $F_B$ is the bending-resonance frequency, $e_B$, $l_B$, and $h_B$ are the thickness, gauge length, and width of the beam, respectively. Furthermore, $U\left\{\frac{e_B}{l_B}, \nu\right\}$ is a correction factor. Equation 3.5, therefore, provides a new way to measure the moduli of HEAs.

Table 3.1 provides a summary of the elastic constants for some reported HEAs, and the elastic moduli of some pure metals and steels are also shown. The CoCrFeNi HEA has the largest $G$ and $E$ value among the fcc HEAs, and the $B$, $G$, and $E$ values of fcc HEAs are larger than those of the two steels listed here. For the refractory HEAs, the NbMoTaW HEA has the largest $B$, $G$, and $E$ values. The HEAs have greater $c_{11}$, $c_{12}$, $c_{44}$ values than those of the steels listed here.

**Table 3.1 Elastic constants of reported HEAs**

| Composition | B (GPa) | G (GPa) | E (GPa) | ν | $c_{11}$ | $c_{12}$ | $c_{44}$ |
|---|---|---|---|---|---|---|---|
| FeCoNiCrMn[102, 107] | 143 | 80 | 202 | 0.265 | 172.1 | 107.5 | 92 |
| FeCoNiCrMn[108] | 129 | 89.76 | 218.6 | 0.218 | 179.2 | 103.9 | 158.5 |
| CoFeNi[109] | 177 | 65 | 174 | 0.34 | 214 | 159 | 114 |
| CoFeNi[108] | 167.5 | 73.0 | 191.2 | 0.310 | 205.1 | 148.7 | 135.0 |
| NiCoCr[109] | 187 | 91 | 234 | 0.29 | 249 | 156 | 142 |
| CrFeNi | 165.7 | 78.02 | 202.3 | 0.296 | 203.4 | 146.8 | 149.4 |
| CoCrFeNi[108] | 182 | 98.5 | 250.4 | 0.266 | 238.3 | 150.8 | 168.2 |
| FeCoNiCrAl[110] | 178 | 78 | 204 | 0.309 | 214 | 160 | 152 |
| $FeCoNiCrAl_{0.3}$[111] | 188 | 88 | - | - | 229 | 168 | 122 |
| $FeCoNiCrAl_{0.3}$[112] | 177 | 85 | 220 | 0.29 | 234 | 148 | 134 |
| $FeCoNiCrAl_{0.3}$[112] | 172 | 81 | 211 | 0.30 | 225 | 145 | 129 |
| TiZrVNb[113] | 118.6 | 45.70 | 121.1 | 0.330 | 166.4 | 94.7 | 53.8 |
| TiZrNbMo[113] | 137.3 | 53.33 | 141.7 | 0.328 | 209.9 | 101.0 | 52.6 |
| TiZrNbMoV[113] | 138.5 | 53.17 | 141.1 | 0.330 | 213.7 | 100.7 | 50.9 |
| TiZrHfNbV[114] | 126.6 | 34.6 | 95.0 | 0.375 | 149.5 | 115.1 | 55.0 |
| TiZrHfNb[115] | 136.7 | 31.9 | 88.9 | 0.392 | 154.3 | 127.9 | 56.7 |
| TaNbHfZrTi[114] | 136.3 | 37.9 | 104.1 | 0.373 | 160.2 | 124.4 | 62.4 |
| TiZrHfNbCr[114] | 117.2 | 38.6 | 104.4 | 0.352 | 153.1 | 99.3 | 49.2 |
| NbMoTaW[114] | 261.6 | 84.4 | 228.7 | 0.354 | 413.5 | 185.6 | 69.0 |
| VNbMoTaW[114] | 245.1 | 75.1 | 204.5 | 0.361 | 380.8 | 177.3 | 61.2 |
| ZrHfNb[114] | 113.2 | 29.7 | 81.89 | 0.379 | 127.5 | 106.1 | 57.1 |
| CrMoTi[116] | 185.7 | 80.1 | 210.1 | 0.31 | 313.9 | 121.6 | 70.9 |
| MoNbTi[116] | 170.9 | 67.3 | 179.5 | 0.32 | 267.3 | 122.7 | 64.9 |
| MoNbV[116] | 200.2 | 71.7 | 192.1 | 0.34 | 326.3 | 137.2 | 59.5 |
| MoTiV[116] | 171.4 | 67.3 | 178.6 | 0.33 | 268.8 | 122.7 | 63.8 |
| AlMoNbV[116] | 174.2 | 75.0 | 196.7 | 0.31 | 251.4 | 135.6 | 89.1 |
| CrMoTiV[116] | 193.4 | 73.3 | 195.2 | 0.33 | 313.0 | 133.6 | 64.0 |
| MoNbTiV[116] | 172.6 | 65.1 | 173.5 | 0.33 | 266.3 | 125.8 | 61.9 |
| NbTaTiV[90] | 146.6 | 42.9 | 117.3 | 0.368 | 196.8 | 121.4 | 46.7 |
| NbTaTiV[90] | 137.8 | 35.7 | 98.6 | 0.381 | 190.2 | 111.6 | 33.5 |
| Steel Fe-19Cr-10Ni[117] | 168.2 | 77.4 | 199.6 | 0.290 | 261.4 | 106.6 | - |
| TWIP steel Fe-22Mn-3Al-3Si[118] | 114.7 | 71.8 | 179 | 0.24 | 175 | 83 | 46 |
| TWIP steel Fe-24Mn-3Al-2Si-1Ni-0.06C[119] | 128.9 | 75.0 | 188 | 0.25 | 167 | 110 | 140 |
| Fe[120] | 196.1 | 75.8 | 200 | 0.32 | 237.0 | 141.0 | 116.0 |
| Al[120] | 63.5 | 27.5 | 72.4 | 0.32 | 108.2 | 61.3 | 28.5 |
| Co[120] | 107.84 | 41.4 | 110 | 0.33 | 168.4 | 121.4 | 75.4 |
| W[120] | 289.9 | 157 | 400 | 0.27 | 501.0 | 198.0 | 151.4 |

To better understand the elastic-deformation behavior of the FeCoNiCrMn HEA, single-crystal elastic constants, $c_{ij}$, were quantified, via the Kroner's model[121-124]. By using the experimental elastic constant values, the regression of the experimental data could be carried out by minimizing the $\chi^2$ in Eq. (5.6)[123, 124]:

$$\chi^2 = \sum_{i=1}^{n}\left[\left(\frac{(\frac{1}{E_{hkl}})_{meas,i}-(\frac{1}{E_{hkl}})_{calc,i}}{e_{1,i}}\right)^2+\left(\frac{(\frac{\nu_{hkl}}{E_{hkl}})_{meas,i}-(\frac{\nu_{hkl}}{E_{hkl}})_{calc,i}}{e_{2,i}}\right)^2\right] \quad (3.6)$$

where $n$ is the number of ($hkl$) diffraction planes used in the model. As for the measurement values, the $hkl$-specific reciprocal diffraction-elastic constants, $(\frac{1}{E_{hkl}})_{meas}$ and $(\frac{\nu_{hkl}}{E_{hkl}})_{meas}$, can be obtained from the neutron diffraction, $e_{1,i}$ and $e_{2,i}$ are the corresponding measurement errors. The $(\frac{1}{E_{hkl}})_{calc}$ and $(\frac{\nu_{hkl}}{E_{hkl}})_{calc}$ can be calculated by the Kroner's model [123].

Reciprocal-diffraction elastic constants, $\frac{1}{E_{hkl}}$ and $\frac{\nu_{hkl}}{E_{hkl}}$, of the FeCoNiCrMn HEA calculated by the Kroner's model are plotted as a function of $\Gamma_{hkl}$ in Fig. 6 together with the experimental data, where $\Gamma_{hkl} = \frac{h^2k^2+k^2l^2+l^2h^2}{(h^2+k^2+l^2)^2}$. In this self-consistent model, the reciprocal-diffraction elastic moduli are correlated with the single-crystal elastic constants by $\Gamma_{hkl}$. As presented in Fig. 6, the Kroner's model fitting of the dependence of $\frac{1}{E_{hkl}}$ and $\frac{\nu_{hkl}}{E_{hkl}}$ on $\Gamma_{hkl}$, is in good agreement with the experimental values, confirming the applicability of the model for evaluating the lattice constants. Using the model, the single-crystal elastic constants, $c_{11}$, $c_{12}$, and $c_{44}$, for the FeCoNiCrMn HEA were determined to be 172.1 GPa, 107.5 GPa, and 92.0 GPa (see Table 3.1), respectively. From these results, the shear anisotropy, $A = 2\frac{c_{44}}{c_{11}-c_{12}}$, was calculated to be 2.84.

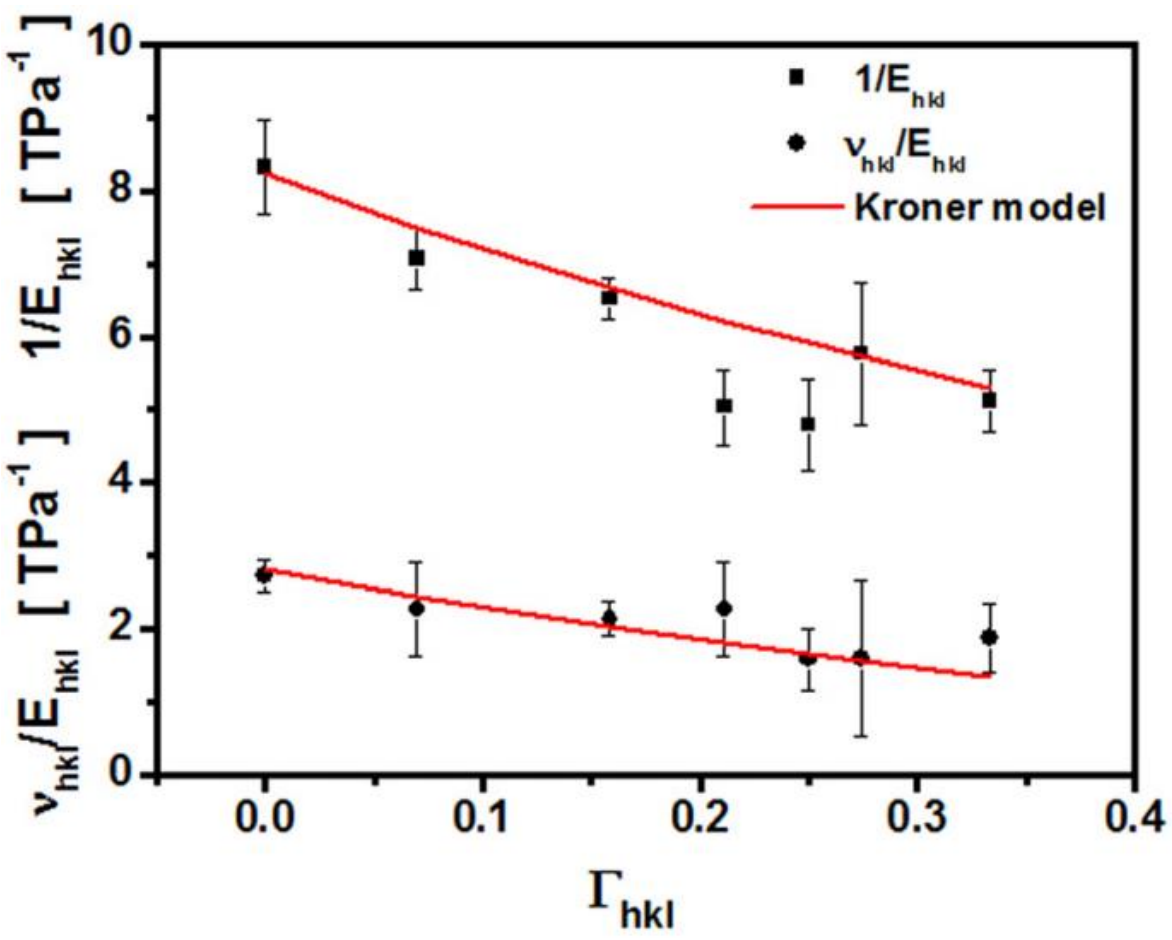

**Fig. 6.** Relationship between the elastic moduli and elastic-anisotropy factors of differently-orientated crystals of the FeCoNiCrMn HEA and its fitting to the Kroner model. Figure from Wu et al. [107].

Lee *et al.*[95] studied the elastic properties of an NbTaTiV HEA by both neutron diffraction and first-principles calculation, for which the results can be seen in Table 5.1. Figures 7(a)-(h) give three-dimensional (3D) and two-dimensional (2D) schematic illustrations for the directional dependence of calculated elastic properties. The maximum-to-minimum Young’s modulus [Figs. 7(a) and (b)] ratio is 1.16, indicating that the HEA is slightly anisotropic, while the linear compressibility [Figs. 7(c) and (d)] shows negligible anisotropy characteristic. For the shear modulus, as presented in Figs. 7(e) and (f), the maximum-to-minimum ratio is 1.17, where the minimum value is along the <110> direction, and the maximum value is along the <111> direction. Moreover, for the Poisson’s ratio, as can be seen from Figs. 7(g) and (h), the maximum-to-minimum ratio is 1.32, and the maximum and minimum values both occur along the <110> direction, which further suggests the slightly anisotropic elastic behavior of the NbTaTiV HEA.

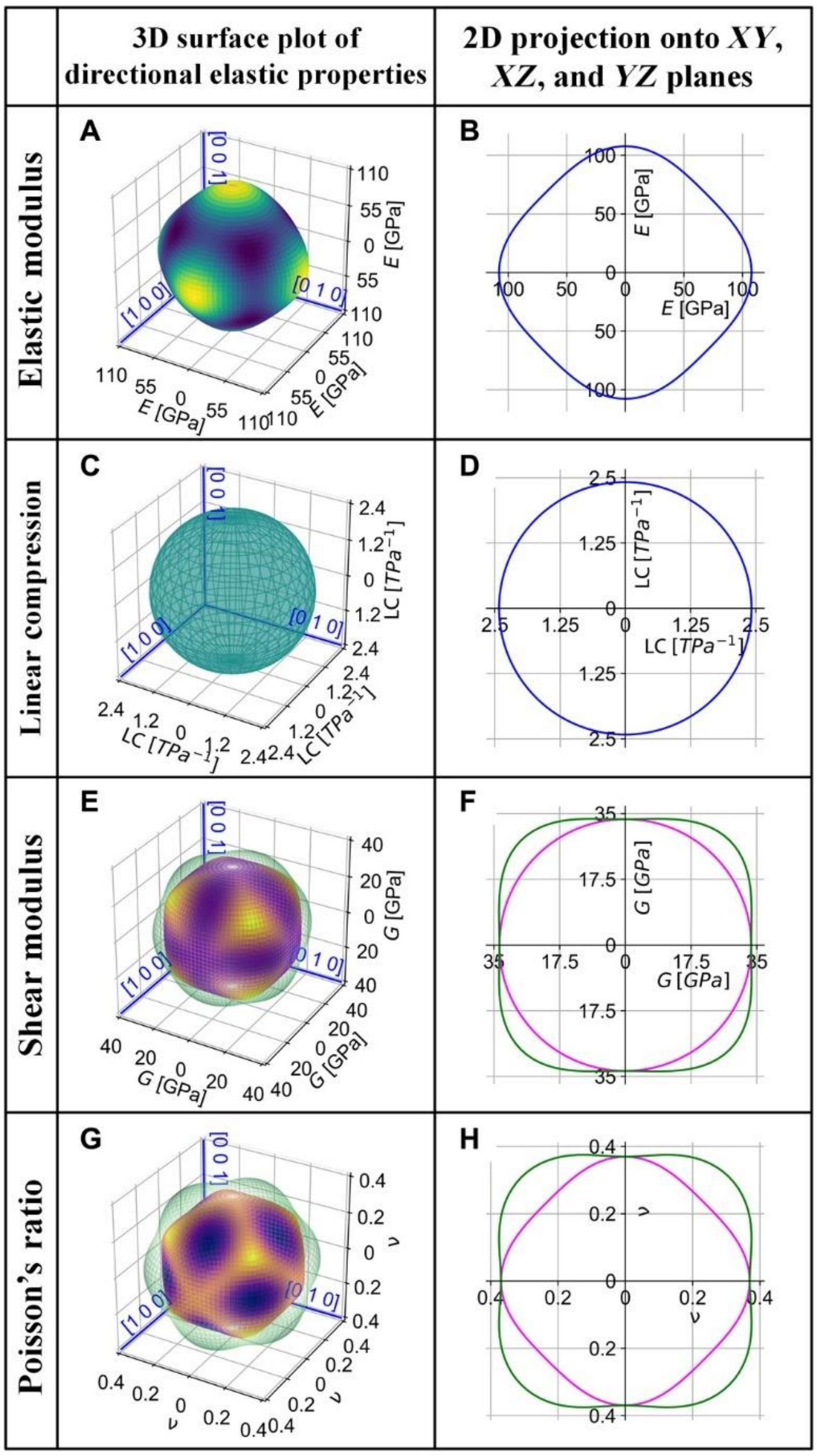

**Fig. 7.** The directional dependence of calculated elastic properties of the NbTaTiV HEA. (a) 3D-surface plot and (b) 2D projection onto *XY*, *XZ*, and *YZ* planes of directional Young's modulus; (c) 3D-surface plot and (d) 2D projection onto *XY*, *XZ*, and *YZ* planes of directional linear compression; (e) 3D-surface plot and (f) 2D projection onto *XY*, *XZ*, and *YZ* planes of directional shear modulus; (g) 3D-surface plot and (h) 2D projection onto *XY*, *XZ*, and *YZ* planes of directional Poisson's ratio. Figures from Lee et al. [95].

By measuring the longitudinal sound wave velocity ($V_l$) and transverse sound wave velocity ($V_t$) along the <100> and <110> directions of the different materials. Jin *et al.*[109] obtained the $c_{11}$, $c_{12}$, and $c_{44}$ of monocrystalline Ni-containing ternary HEAs using the following formulas:

For the <100> direction,

$$V_l = \sqrt{\frac{c_{11}}{\rho}} \tag{3.7}$$

$$V_t = \sqrt{\frac{c_{44}}{\rho}} \tag{3.8}$$

For the <100> direction,

$$V_l = \sqrt{\frac{c_{11}+c_{12}+2c_{44}}{2\rho}} \tag{3.9}$$

where $\rho$ is the density. The results can be seen in Table 3.1.

## 4. Hardness and compression behavior of HEAs

To some extent, due to the permanent deformation, Vickers hardness (HV) correlates with the $\sigma_y$ according to the following empirical equation[125]:

$$\text{HV} \sim (2.5 \text{ to } 3.0) \times \sigma_y \tag{4.1}$$

The Vickers hardness of HEAs is strongly dependent on the crystal structure, since it can range from 100 HV[126] for the fcc HEA, to 1,200 HV[127] for the bcc HEA. For example, He *et al.*[4] found that the hardness values of the $(FeCoNiCrMn)_{100-x}Al_x$ HEA increase slightly from 176 HV for the FeCoNiCrMn HEA to 182 HV for the $(FeCoNiCrMn)_{93}Al_7$ HEA (all alloys are fcc-structured). With the increase of the Al content, the structure of HEAs turns to be a mixture of fcc + bcc phases, and the hardness value significantly increases. For example, the hardness value of $(FeCoNiCrMn)_{86}Al_{14}$ reaches about 538 HV. However, when the structure of $(FeCoNiCrMn)_{100-x}Al_x$ HEAs completely transforms to a bcc phase, the hardness value almost retains the same value. From the results, they surmised that the bcc-structured B2 phase contributes significantly to the hardness.

Zhang *et al.*[127] reported that the laser-induced rapid solidification led to a relatively- high hardness of 1,152 HV in the $FeCoNiCrCuTiMoAlSiB_{0.5}$ HEA. The martensitic phase, whose nucleation resulted from the joint effect of the laser-induced rapid solidification and presence of boron interstitials, leads to the high hardness of the HEA. Based on the results, this strengthening method could be extended to other rapidly-solidified HEAs.

For the refractory bcc-structured HEAs, Senkov *et al.*[68, 128] found that the Vickers hardness results of the single bcc-phased TaNbWMo and TaNbWMoV HEAs were approximately 459 HV and 540 HV, respectively. The hardness values of these two HEAs are greater than those of their constituents, which indicates that there is an apparent solid-solution strengthening effect in HEAs. In 2014, Senkov *et al.* showed that the hardness of the $Al_{0.4}Hf_{0.6}NbTaTiZr$[129] HEA (500 HV) increases by replacing 0.4 Hf in TaNbHfZrTi[29] (406 HV) with Al without changing its phase composition. It was determined that the increased hardness was a consequence of the nano-sized precipitates located at the grain boundaries in the $Al_{0.4}Hf_{0.6}NbTaTiZr$ HEA.

The bcc-structured HEAs typically exhibit high strengths although the plasticity is poor. For instance, Fig. 8 shows the room-temperature compressive test of the VNbMoTaW[68] HEA, where the $\sigma_y$ exceeded ~1,200 MPa while its elongation was 1.7%. After compression deformation, scanning electron microscopy (SEM) imaging revealed that the fracture morphologies of the alloys contained a brittle quasi-cleavage fracture, suggesting that in this alloy, the primary failure mode is tensile rather than shear.

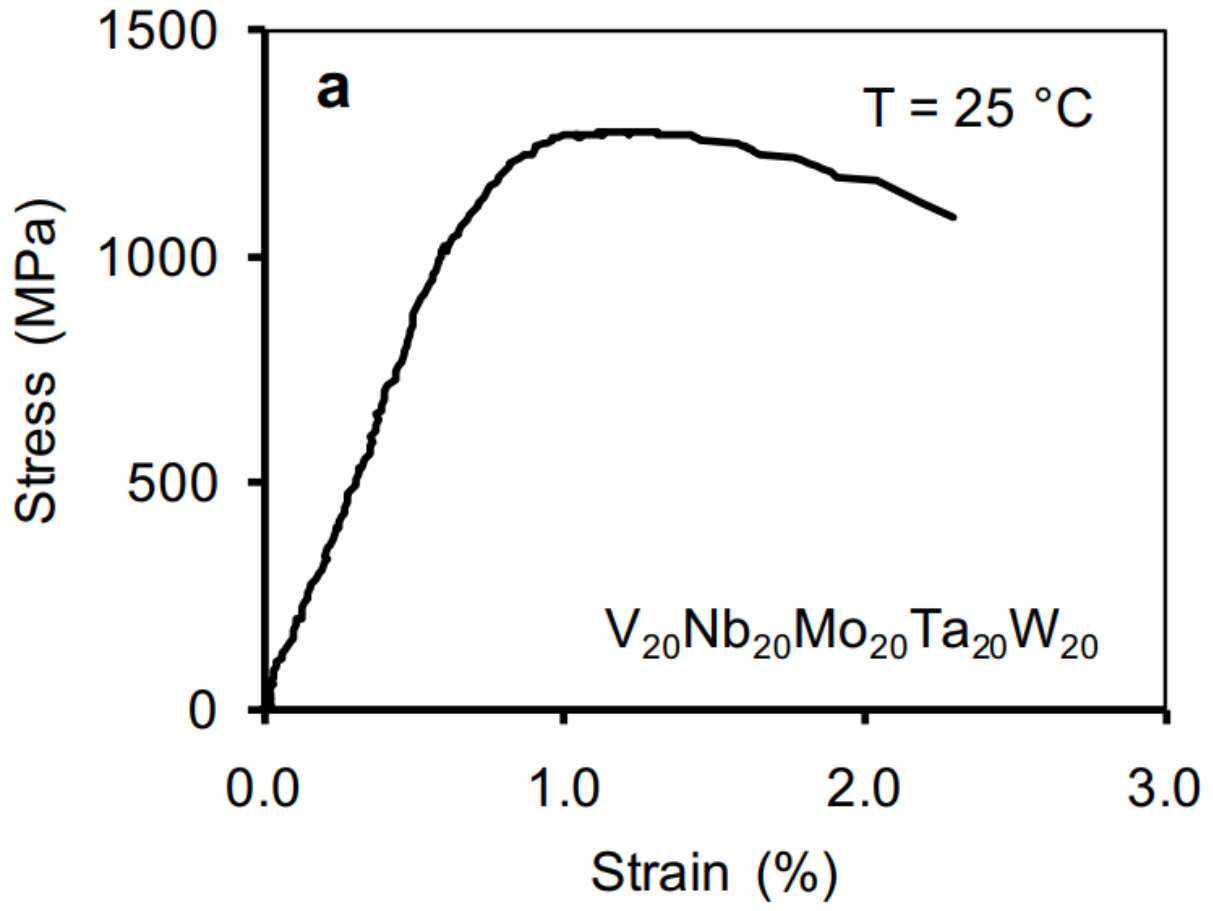

**Fig. 8.** The compressive engineering stress-strain curve of the VNbMoTaW HEA. Figure from Senkov et al. [68].

Khaled *et al.*[130] observed that the as-milled $Al_{20}Li_{20}Mg_{10}Sc_{20}Ti_{30}$ HEA has an fcc structure, with an average grain size of 12 nm, and hardness of ~591 HV. However, after annealing at 773 K, the structure of the HEA changes to an hcp phase, with a grain size of 26 nm. At the same time, its hardness is ~500 HV, which is a relatively-high value. It was reported that the nanocrystalline grain size is what leads to the high hardness in the HEA. Zhou *et al.*[131] reported that the compressive $\sigma_y$ of $AlCoCrFeNiTi_x$ (x = 0, 0.5, 1, and 1.5) HEA rods with a diameter of 5 mm ranges from 1.50 to 2.26 GPa, which is superior to most of the high-strength alloys reported, *e.g.* bulk metallic glasses[132-134]. The different $\sigma_y$ values result from various phase compositions and dendritic structures, which gives a new way to design HEAs with high strengths.

The refractory TaNbHfZrTi HEA has been observed to exhibit relatively-higher strength and good plasticity[29]. For instance, it was found that this alloy exhibited a $\sigma_y$ approaching 929 MPa, and a compression plasticity surpassing 50 %. They suggested that the relatively high strength arises from the solid-solution-strengthening effect. Here, the large distortions in the HEA can act as big obstacles for dislocation movements, and the simultaneous dislocation slip and twinning can reduce the stress concentration at grain boundaries, resulting in the increased ductility of the TaNbHfZrTi HEA.

For the other kind of refractory HEAs, Senkov *et al.*[68] reported that the $\sigma_y$ of the as-cast NbMoTaW HEA was ~ 1,058 MPa, but its relatively-low compression plasticity (only 1.5 %) hinders its further application. The microstructure of the NbMoTaW HEA was found to be bcc-structured, and there was a small number of segregations (< 5 %) at the dendritic grain boundary. The fracture morphology shows that the failure mode was with longitudinal rather than shear cracks, and corresponds to the brittle quasi-cleavage fracture.

Takeuchi *et al.*[71] found that the YGdTbDyLu and GdTbDyTmLu HEAs are nearly single hcp-phased. The value of the valence electron concentration (VEC) of the HEAs is typically 3, and these two kinds of hcp-structured HEAs were first reported, indicating that there could exist other possible hcp-structured HEAs. Moreover, according to the results of X-ray diffraction (XRD) characterization (Fig. 9), the YGdTbDyHo HEA[135] is hcp-structured ($P6_3$), in which the lattice constants are $a$ = 0.7229 nm, $c$ = 0.5695 nm, and the ratio of $c$ to $a$ is 0.7879 (where $a$ and $c$ are the edge lengths of the axes of the hcp unit cell). Note that the red-indexed peaks in Fig. 9 are not the $P6_{3/}$mmc hcp crystal structure. However, the calculated $a$ value is two times that calculated by Feuerbacher *et al.*[69], which is due to the presence of the Gd-enriched oxide particles with small sizes and low volume fractions.

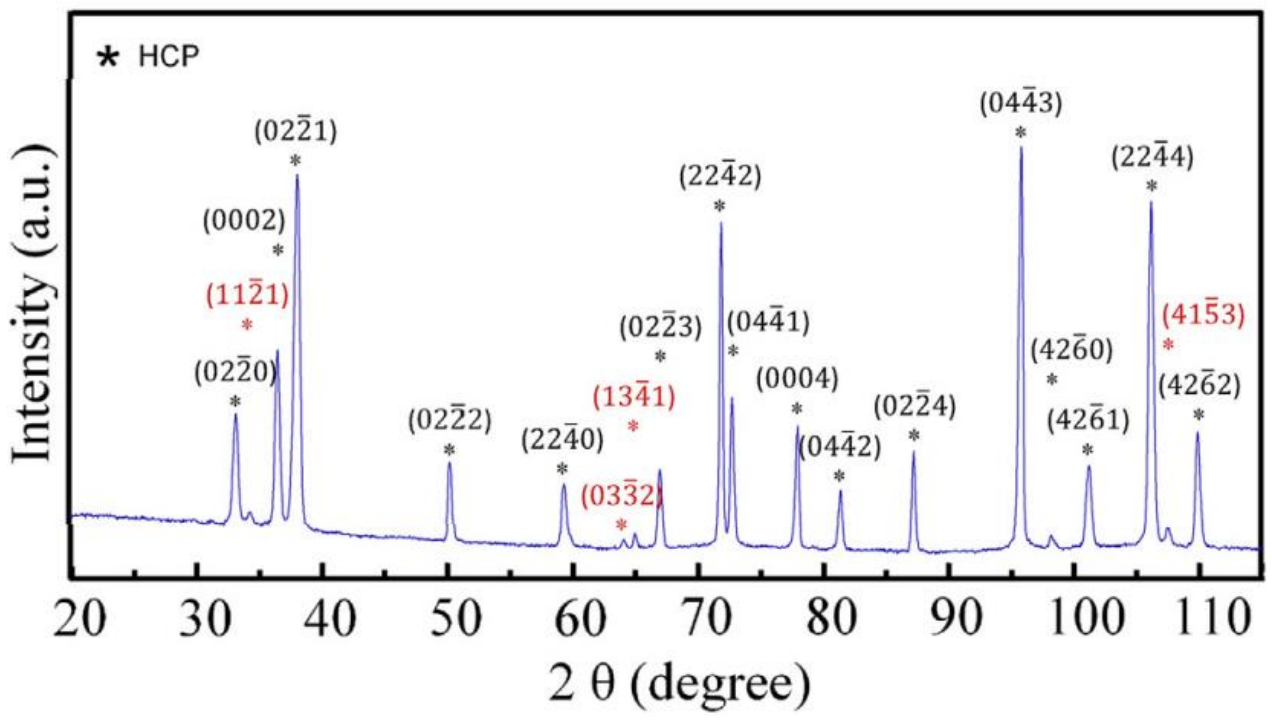

**Fig. 9.** The XRD pattern of the YGdTbDyHo HEA. Figure from Soler et al. [135].

For the hcp-structured YGdTbDyHo HEA, a small amount of $\gamma$ phases and oxides can be found in the alloy. To eliminate the effect of the $\gamma$ phase on the mechanical behavior, the compression experiments of micropillar samples with different diameters were carried out by the focused ion beam (FIB). Figure 10 displays the stress-strain curve of the YGdTbDyHo micropillars, with diameters ranging from 1.25 to 10 μm. Note that in the four samples, there is no obvious hardening behavior observed. However, the decrease of the intermittent stress was found in the 1-μm micropillar. While in the 5-μm micropillar, the decrease of intermittent was subtle. In the largest pillars (10 μm), the decrease of intermittent was completely lacking. Furthermore, it was observed that the flow stress decreases gradually, relative to the maximum flow stress reached in the previous strain. When the internal deformation of the segment occurs, the flow stress increases at a higher rate, which is more apparent when the strain reaches 5 %. The results show that with the increase in the sample diameter, the volume fraction of the oxide-precipitation phase also increases, and the strength of oxide precipitation is also enhanced.

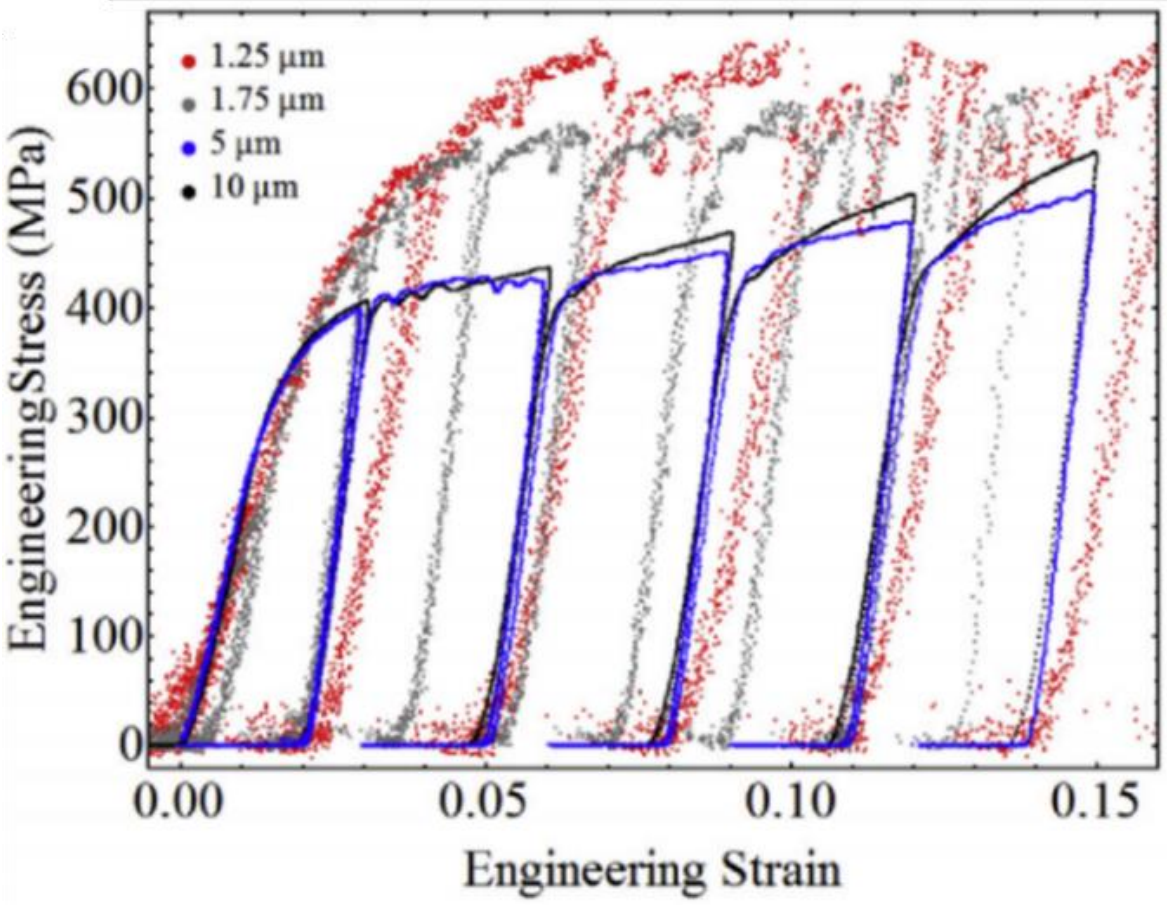

**Fig. 10.** Micropillar compression stress-strain curves of the [1212]-oriented YGdTbDyHo HEA single crystals with diameters ranging from 1.25 to 10 μm. Figure from Soler et al. [135].

The TiZrHfSc alloy is one type of the hcp-structured HEA, which displays a good balance between strength and plasticity [136]. The compressive $\sigma_y$ of the as-cast HEA is 700 MPa, and the fracture elongation is close to 20 %. Both the sliding systems with the base surface of {0001} and the cylindrical surface of {1010} as the sliding surfaces can be started, and the main slip direction is <2110>. Furthermore, this HEA exhibits good thermal stability. When aging at 1,273 K for 5 h, the Ti-rich solid solution is dissolved into the matrix. Then a plate-like phase is formed, but the fraction is very low, which does not affect the compressive properties much.

## 5. Tensile behavior of HEAs

For the fcc-structured HEAs, FeCoNiCrMn is the most representative alloy, and other ternary, quaternary, or even six-and higher component HEAs can be obtained from this composition[137]. The FeCoNiCrMn HEA was first proposed by Cantor[18], and thus, it is also called the 'Cantor alloy'. This alloy has been found to exhibit good thermodynamic stability[138], excellent fracture toughness[23], and ductility[139]. Otto *et al.*[139] studied the tensile properties of the FeCoNiCrMn HEA, as presented in Fig. 11. For the fine-grained (grain size of 4.4 μm) HEA, the $\sigma_y$ and ultimate tensile strength (UTS) are 460 MPa and 630 MPa, respectively. It was also observed that the fracture elongation can reach 60 % at room temperature. For the coarse-grained (grain size of 155 μm) HEA, the $\sigma_y$ is 125 MPa, and the UTS is 450 MPa, with a fracture elongation of 80 %.

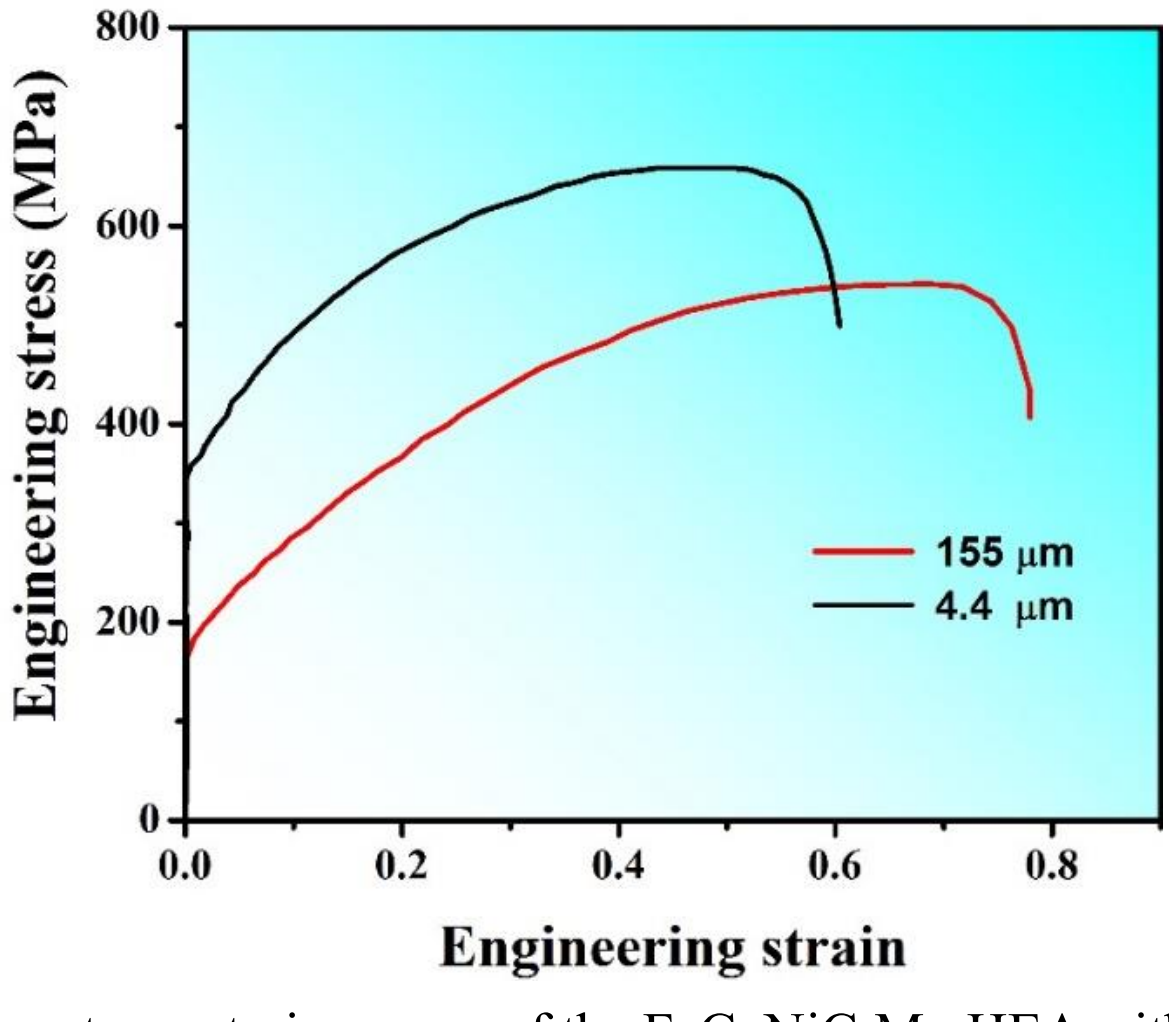

**Fig. 11.** Engineering stress-strain curves of the FeCoNiCrMn HEA with a grain size of 4.4 μm and 155 μm, respectively. Figure from Otto et al. [139].

In the initial stage of plastic deformation (a strain of 2.1 %), the dominant dislocation-slip mode is planar, as presented in Fig. 12(a). Slip is initiated along the {111}-type planes, and the slip direction is 1/2 <110>. Then the full dislocation decomposes into 1/6 <112> Shockley partial dislocations and a large number of stacking faults, similar to the conventional fcc-structured alloys. Otto *et al.* argued that it is the short-range ordering (SRO) or the short-ranged clustering (SRC) that leads to the formation of the localized dislocation slip on {111}-type planes. This phenomenon was also seen in the binary concentrated solid solutions[140].

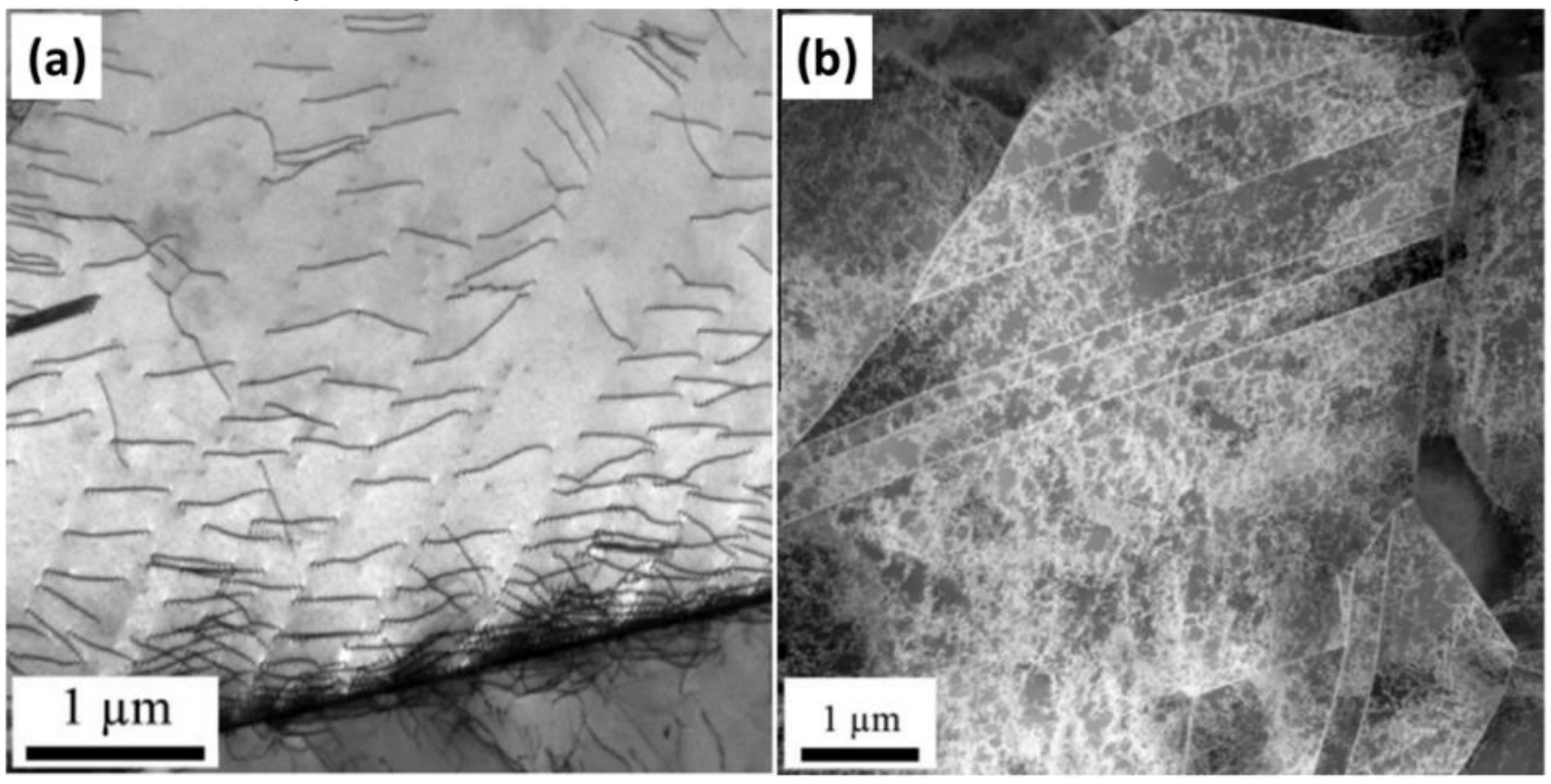

**Fig. 12.** The dislocation structure of the FeCoNiCrMn HEA pre-strained to (a) a strain of 2.1 % and (b) a strain of 20.6 %, respectively. Figures from Otto et al. [139].

Figure 12(b) shows that when straining to 20.6%, the dislocation structure has evolved into dislocation cells, which consist of the 'walls' formed by three-dimensional tangles of dislocations and the internal regions with a lower dislocation density. This phenomenon is similar to those fcc-structured alloys with the low or medium stacking fault energy (SFE)[141]. At the same time, the observed twin consists of an annealing twin, whereas no deformation twin was observed.

In 2014, Wu *et al.*[142] studied the mechanical properties of the family of equiatomic quaternary, ternary, and binary alloys with compositions of Fe, Co, Ni, Cr, and Mn. Figure 13 summarizes the engineering stress-strain behaviors of these alloys, using the data in Ref.[142]. This result indicates that there is no direct relationship between the strength and the number of principal elements, and usually the Cr-containing alloys exhibit superior mechanical properties. The findings show that at room temperature, the NiCoCr medium-entropy alloy (MEA) demonstrates superior mechanical properties, *e.g.* the UTS surpasses 800 MPa, and the fracture elongation reaches 60 %.

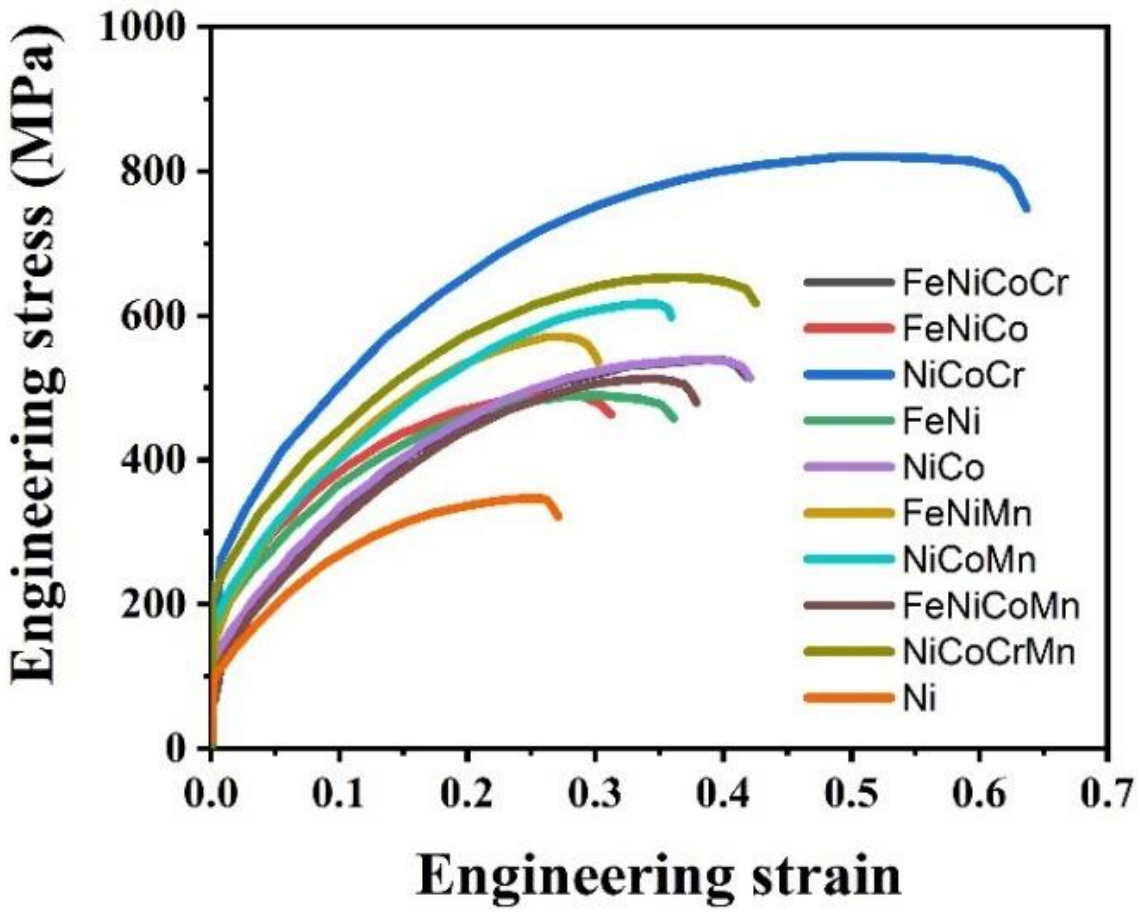

**Fig. 13.** The engineering stress-strain curves for the family of equiatomic quaternary, ternary, binary alloys, as well as pure Ni, with compositions of Fe, Co, Ni, Cr, and Mn. Figure from Wu et al. [142].

In order to explore the reasons for the superior mechanical properties of the NiCoCr ternary alloy, as compared to the FeCoNiCrMn HEA, Laplanche *et al.*[143] performed tensile tests on these two alloys. The results in Figs. 14 show that when the plastic strain is small, the higher $\sigma_y$ and work-hardening rate cause the applied stress to approach the critical twinning stress at a faster rate such that the nano-twinning can occur in a larger strain range. The calculated onset true tensile stress and resolved shear stress value of twinning in the NiCoCr alloy are 790 ± 100 MPa and 260 ± 30 MPa, respectively, which are higher than the reported values of 720 ± 30 MPa and 235 ± 10 MPa in the FeCoNiCrMn HEA, respectively. Therefore, the delayed necking of NiCoCr is the factor, which contributes to its better combination of strength and ductility.

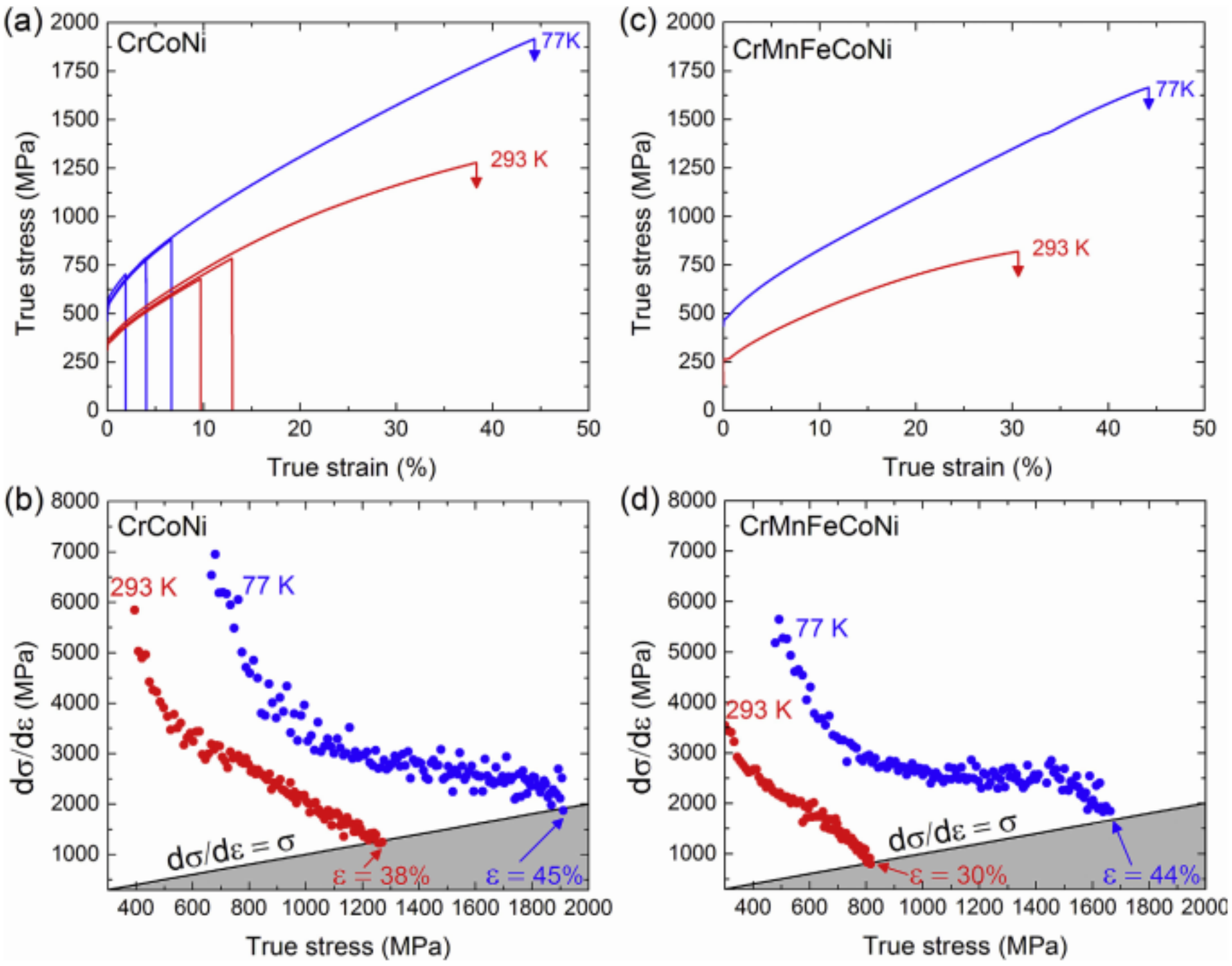

**Fig. 14.** (a) True stress-strain curves and (b) work-hardening rate ($d\sigma \cdot d\varepsilon^{-1}$) versus true stress of the CrCoNi MEA; (c) true stress-strain curves; and (d) work-hardening rate ($d\sigma \cdot d\varepsilon^{-1}$) versus true stress of the CrMnFeCoNi HEA. Figure from Laplanche et al. [143].

They also estimated the SFE for these two kinds of alloys by measuring the disassociation width of the partial dislocation under transmission electron microscopy (TEM)[5, 143]. According to the dissociation width, the value of the SFE ($\gamma$) can be calculated by the following formula[144]:

$$\gamma = \frac{Gb_p^2}{8\pi d_{act}}\left(\frac{2-v}{1-v}\right)\left(1 - \frac{2v\cos(2\beta)}{2-v}\right) \tag{5.1}$$

where $v$ is the Poisson’s ratio, $G$ is the shear modulus, $b_p$ is the magnitude of the Burgers vector of the partials, $d_{act}$ is the actual spacing between partials, and $\beta$ is the angle between the Burgers vector and the full dislocation.

Figure 15 shows the dependence of the Shockley partial separation on the angle between the full dislocation and the Burgers vector for the NiCoCr alloy. The result shows that the NiCoCr MEA has a lower $\gamma$ (22 ± 4 mJ·m$^{-2}$)[143] than that of the FeCoNiCrMn HEA (30 ± 5 mJ·m$^{-2}$)[5], which may affect the subsequent deformation behavior, *e.g.* the critical twinning stress[145, 146]. According to Venables[147, 148] and Remy[149], the SFE often has some correlation with twinning, *e.g.* as the SFE increases, the critical twinning stress increases.

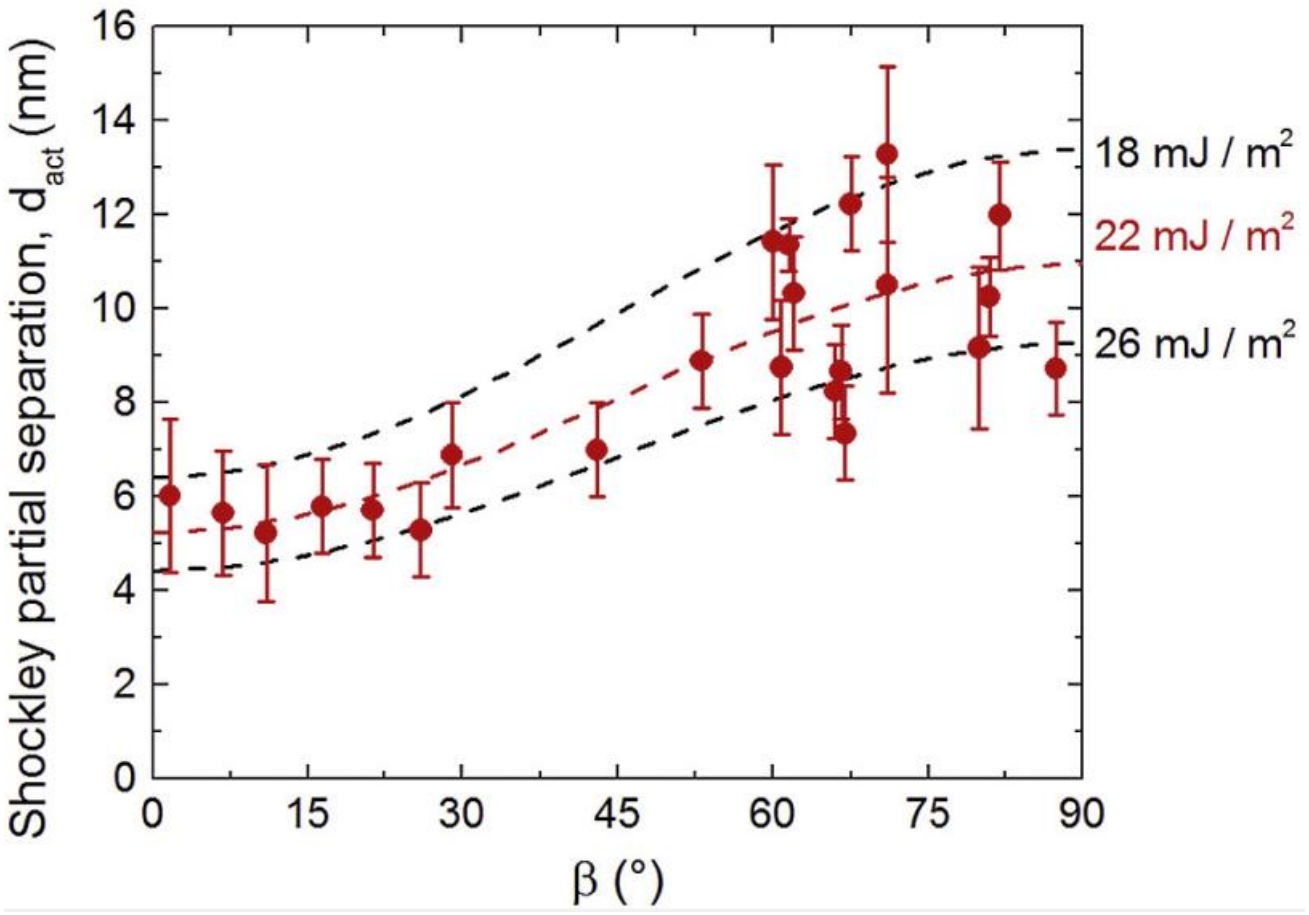

**Fig. 15.** The corresponding spacing between partials, as a function of the angle between the full dislocation and the Burgers vector, of the NiCoCr ternary alloy. Figure from Laplanche et al. [143].

Along this direction, Zhao *et al.*[150] calculated the SFE for different equiatomic fcc-structured alloys by the ab initio calculation method. The data featured in Fig. 16 were calculated by averaging the results from all the special-quasirandom structures (SQS) supercell planes, thus representing the total fault energies. Note that the calculation did not take the short range order into consideration. Otherwise, the inhomogeneity should be included[151]. The results (see Fig. 16) show that the SFE value of the NiCoCr MEA is lower than that of the FeCoNiCrMn HEA, which is in agreement with the experimental data[5, 143].

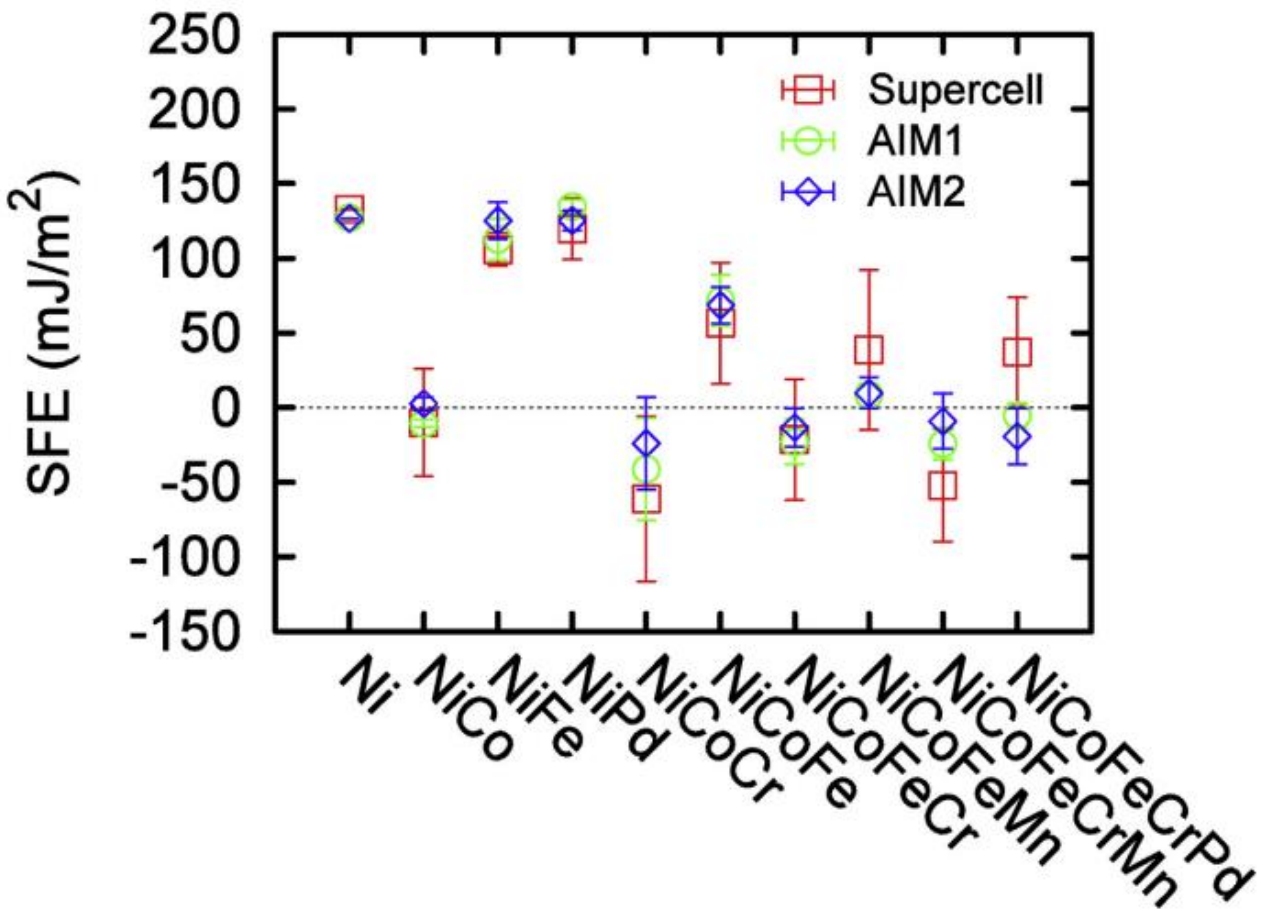

**Fig. 16.** The calculated SFE of fcc-structured concentrated solid-solution alloys, as determined by an ab initio method. Figure from Zhao et al. [150].

However, some HEAs with a bcc structure also display certain tensile plasticity behavior, such as the TaNbHfZrTi HEA. Senkov *et al.*[152] examined the tensile properties of the TaNbHfZrTi HEA after being cold rolled and then annealed at 1,073 K and 1,273 K, respectively. Figure 17 shows that the $\sigma_y$ of the annealed TaNbHfZrTi

HEA (cold rolled plus annealed at 1,073 K) can exceed ~ 1,100 MPa, and the fracture elongation was able to exceed ~ 9.7 % at room temperature.

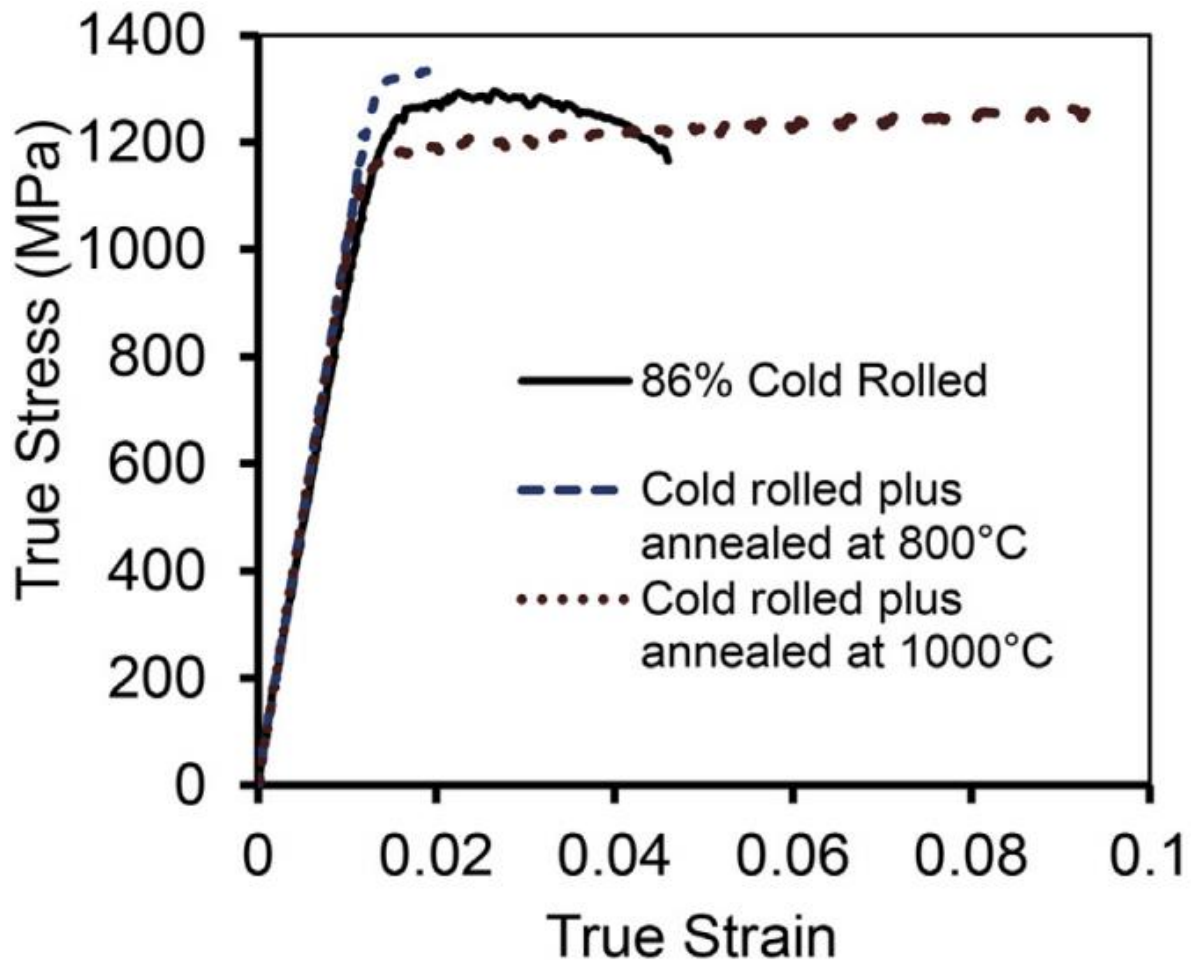

**Fig. 17.** The true stress-strain curve of the TaNbHfZrTi HEA treated at different cold rolling conditions. Figure from Senkov et al. [152].

Lilensten *et al.*[153] studied the underlying deformation mechanism of the TaNbHfZrTi HEA by SEM, electron backscattering diffraction (EBSD), and TEM. When being strained to 2.65 %, as can be seen from the SEM and EBSD images in Fig. 18, active dislocation slips with a planar slip mode were present, and no deformation twinning is observed. Moreover, the activities of dislocations were evident at grain boundaries.

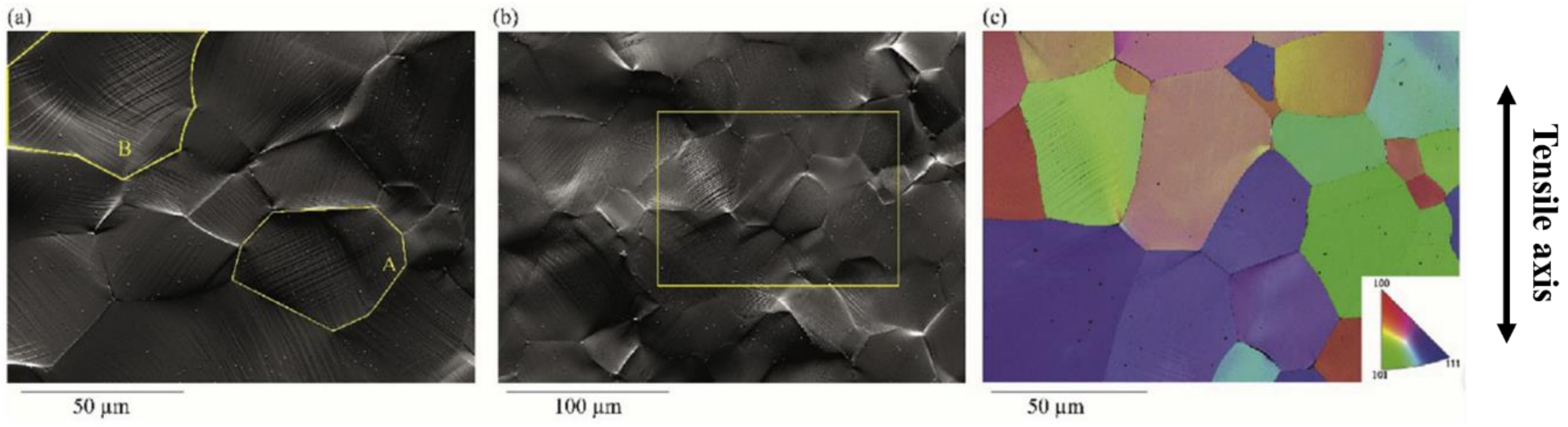

**Fig. 18.** (a) and (b): SEM images for the surface of the TaNbHfZrTi HEA with a true strain of 2.65 %; (c) the corresponding EBSD map of the selected region in (b), whereas the inserted figure is an inverse pole plot along the tensile axis direction. Figures from Lilensten et al. [153].

Figures 19 (a)-(d) present the bright-field TEM images of the TaNbHfZrTi HEA pre-strained to different strains, *e.g.* 0.22 %, 2.89 %, and 10.12 %. When the plastic deformation initiates, as presented in Fig. 19(a), there are two slip systems that correspond to <111> screw dislocations. Besides straight dislocations, there are some dislocation dipoles and loops [see Fig. 19(b)]. When the strain is higher, the two slip systems can form dislocation bands and dislocation-free zones. With an increase in the strain, the distance between the bands decreases, as exhibited in Figs. 19(c) and (d). These results indicate that screw dislocations are present during the entire deformation process.

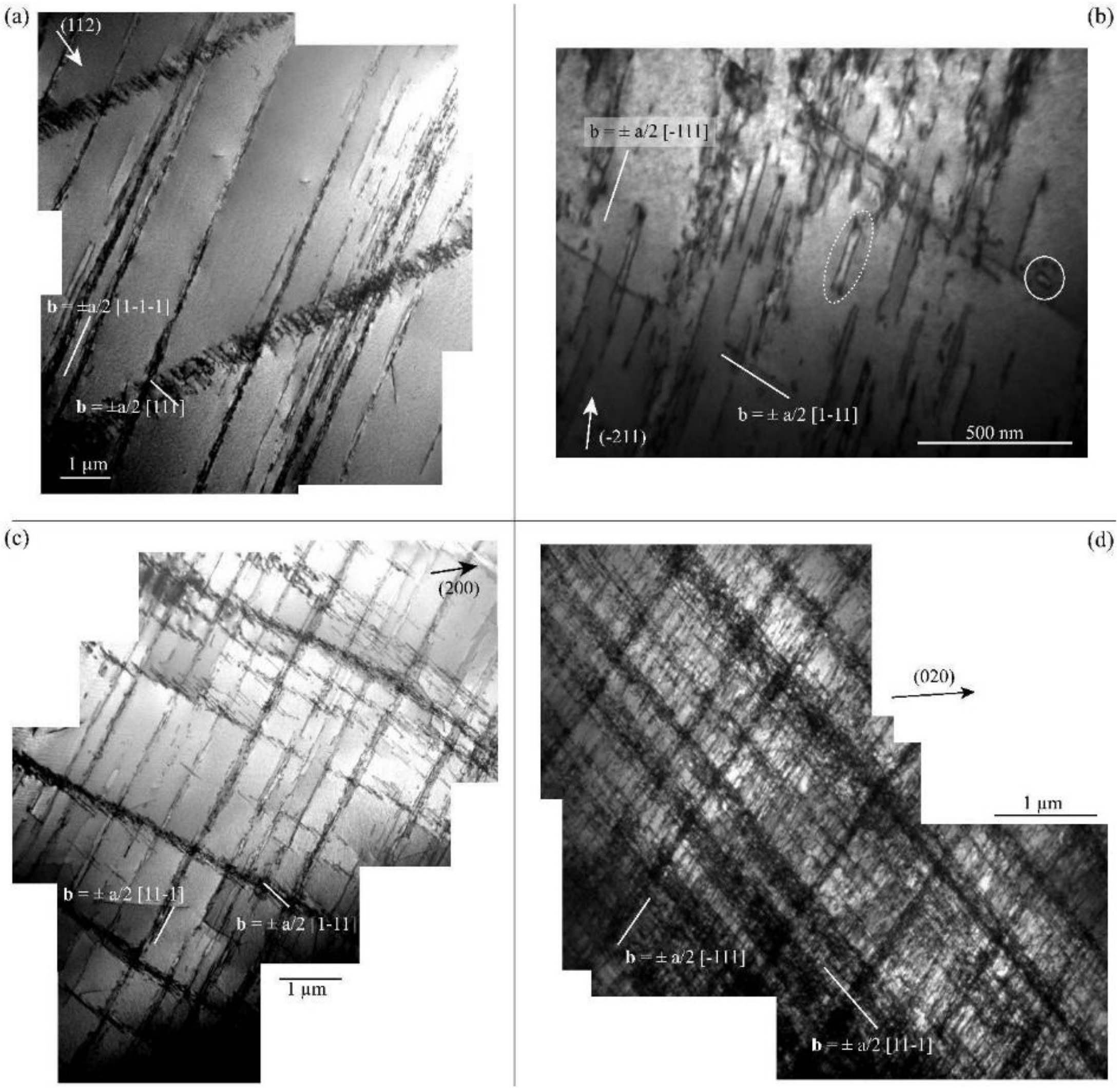

**Fig. 19.** (a) and (b): Bright-field TEM images of the dislocation structure of the TaNbHfZrTi HEA pre-strained to 0.22 %, (c) 2.89 %, and (d) 10.12 %. Figures from Lilensten et al. [153].

Table 5.1 summarizes the hardness, $\sigma_y$, UTS, and fracture elongation of some HEAs with different compositions, structures, grain sizes, and processing methods. The fcc HEAs exhibits higher fracture elongation than the bcc HEAs, but with lower $\sigma_y$ and UTS values. Moreover, the different compositions, processing methods, grain sizes, and microstructures can influence the mechanical properties of the listed HEAs.

**Table 5.1 Summary of mechanical properties of some HEAs**

| Composition | Structure | Average grain size (μm) | Processing | Hardness (HV) | $\sigma_y$ (MPa) | US (MPa) | Fracture elongation (%) |
|---|---|---|---|---|---|---|---|
| TCoCrMnNi[142] | fcc | 36 | AC+1,373K 24h+CR 90%+1,273K 1h | | 280 | 699 | 43 |
| TCoCrFeNi[154] | fcc | 11 | AC+1,273K 24h+HR92%, | | 300 | 671 | 42 |
| | | | 1,273K+1,173K 1h | | | | |
| $^{T}$CoFeMnNi[142] | fcc | 48 | AC+1,373K 24h+CR 90%+1,273K 1h | | 175 | 551 | 41 |
| $^{T}$FeCoNiCrMn[155] | fcc | 4.4 | AC+1,473K 48h+CR87%+1,073K 1h | | 362 | 651 | 51 |
| | | 50 | AC+1,473K 48h+CR 87%+1,273K 1h | | 197 | 568 | 60 |
| | | 155 | AC+1,473K 48h+CR87%+1,423K 1h | | 171 | 530 | 57 |
| $^{C}$AlCoCrFeNi[156] | bcc | | AC | 501 | | | |
| $^{C}$AlCoCrFeNi[131] | bcc | | AC | | 1,500 | 2,830 | 26.9 |
| $^{C}$AlCoCrFeNiTi[131] | bcc | | AC | | 1,860 | 2,580 | 8.8 |
| $^{C}$AlCrFeNiV[157] | bcc+unknown phase | | AC | 640 | | | |
| $^{C}$CoCrCuFeNi[39] | bcc | | AC | | 230 | 888 | 50.2 |
| $^{C}$CoCrCuFeNiTi[39] | bcc | | AC | | 1,227 | 1,272 | 0 |
| $^{C}$AlNbTiV[158] | bcc | | AC+1473K 24h | | 1,000 | 1,280 | 6.0 |
| $^{C}$AlNbTiVZr[158] | bcc+fcc | | AC+1473K 24h | | 1,500 | 1,675 | 3.0 |
| $^{C}$CrNbTiVZr[159] | bcc+fcc | | AC+1473K 24h | | 1,260 | 1,270 | 0.2 |
| $^{C}$AlCrNbTiVZr[159] | bcc+hcp1+hcp2 | | AC+1473K 24h | | 850 | 850 | 0 |
| $^{C}$AlCrFeCoNiCu[160] | fcc+bcc+unknown phase | | AC | | 1,303 | 2,081 | 24 |
| $^{C}$AlCrFeCoNiCuMn[160] | fcc+bcc+unknown phase | | AC | | 1,005 | 1,480 | 15 |
| $^{C}$AlCrFeCoNiCuTi[160] | fcc+bcc+unknown phase | | AC | | 1,234 | 1,356 | 9 |

| | | | | | |
|---|---|---|---|---|---|
| $^{C}$AlCrFeCoNiCuV[160] | fcc+bcc+ unknown phase | AC | 1,469 | 1,970 | 16 |
| $^{C}$CoCrCuFeMnNiTiV[161] | fcc+bcc+ σ | AC | 1,312 | 1,312 | 0 |
| $^{C}$AlNbTiV[162] | bcc | AC+1,473K 24h | 1,020 | 1,318 | 5 |
| $^{C}$TiZrHfNbV[114] | bcc+ unknown phase | AC | 1,170 | 1,463 | 29.6 |
| $^{C}$CrHfNbTiZr[114] | bcc+Laves1+Laves2 | AC | 1,375 | 2,130 | 2.8 |
| $^{C}$CrNbTiVZr[163] | bcc+Laves | AC+HIP, 1,473K 207MPa 2h+1,473K 24h | 1,298 | - | 3 |
| $^{C}$CrNbTiZr[163] | bcc+Laves | AC+HIP, 1,473K 207MPa 2h+1,473K 24h | 1,260 | - | 6 |
| $^{C}$TaNbHfZrTi[164] | bcc | AC+HIP, 1,473K 207MPa 2h+1,473K 24h | 929 | - | >50 |
| $^{C}$NbTiVZr[163] | bcc | AC+HIP, 1473K 207MPa 2h+1473K 24h | 1,105 | - | >50 |
| $^{C}$HfNbTiZr[165] | bcc | AC, 1573K/6 h/SC | 879 | 969 | 14.9 |
| $^{C}$HfMoNbTiZr[166] | bcc | AC, 1373K/10 h/SC | 1,575 | - | 9 |
| $^{C}$MoNbTaVW[167] | bcc | AC | 1,246 | 1,270 | 1.7 |
| $^{C}$MoNbTaW[167] | bcc | AC | 1,058 | 1,211 | 1.5 |

*Strain rate: $10^{-3}$ $s^{-1}$.

AC (as-cast); HR (hot-rolled); CR (cold-rolled); HIP (hot isostatic pressed); SC (slow cooled).

T: tensile properties; C: compressive properties. US: ultimate strength

# 6. Strengthening of HEAs

As previously mentioned, in HEAs, the concept of solutes and solvents as in traditional alloys can no longer be applied. On this basis, the strengthening theory also needs to be verified. Therefore, the research in this field is a hot topic of HEAs. In recent years, scientists have done a great amount of work on how to strengthen HEAs[1, 2, 7, 14, 26, 168-174]. It is worth noting that the deformation behavior of metals is closely related to their defects contents, such as dislocations and twins[175, 176]. Therefore, it is of great importance to manipulate the defects to strengthen the HEAs, which will be described in the following sections. Here, the traditional strengthening methods are discussed, such as strain hardening, grain-boundary strengthening, solid-solution strengthening, and particle strengthening.

## 6.1. Strain hardening

In strain hardening, dislocations can act as obstacles for the movement of other dislocations. Generally, the $\sigma_y$ of the metal is positively related to the dislocation density ($\rho_d$). In traditional alloys, plastic deformation can lead to an increase of $\rho_d$, and the contribution of the dislocations to the strength of the material can be estimated by[125]:

$$\Delta\sigma_d = k_d M G b \sqrt{\rho_d} \qquad (6.\ 1)$$

where $k_d$ is a constant, $M$ is the Taylor factor, $G$ is the shear modulus, and $b$ is the Burgers vector.

The effect of strain strengthening on the mechanical properties of the FeCoNiCrMn HEA was studied by Lu *et al.*[72]. Their results show that the $\sigma_y$ of the HEA increases from 195 MPa to 1,100 MPa, with an increase in the pre-rolling to 60 %. However, the plasticity significantly decreased from the initial 60 % to less than 3 %. Yao *et al.*[177] observed similar behavior in the $Fe_{40}Mn_{27}Ni_{26}Co_5Cr_2$ HEA, and both the FeCoNiCrMn and $Fe_{40}Mn_{27}Ni_{26}Co_5Cr_2$ HEAs are very similar to the metals with an fcc structure.

## 6.2. Grain-boundary strengthening

Grain-boundary strengthening is a very important strategy that has been widely used in HEAs. If a material possesses a relatively finer grain size, the plastic deformation under an external force can be dispersed in more grains, leading to a uniform force distribution and smaller stress concentration. Moreover, the material with a smaller grain size has a relatively-larger grain-boundary surface area to volume ratio, leading to the unfavorable crack propagation, so that it is called grain-boundary strengthening.

The amount of grain-boundary strengthening ($\Delta\sigma_{gbs}$) can be estimated by the Hall-Petch equation[125]:

$$\Delta\sigma_{gbs} = \frac{k_{HP}}{\sqrt{d}} \qquad (6.\ 2)$$

where $k_{HP}$ is the Hall-Petch coefficient, and $d$ is the average grain size.

Generally speaking, the Hall-Petch coefficient of traditional fcc metals[178] will not exceed 600 $MPa\cdot\mu m^{-1/2}$. Liu *et al.*[179] calculated that the Hall-Petch coefficient of the FeCoNiCrMn HEA reaches 677 $MPa\cdot\mu m^{-1/2}$. It can be found that the grain-boundary- strengthening effect in HEAs is more significant than in traditional materials[179]. They thought that the large lattice distortion of the HEA can result in greater lattice resistance, leading to a higher energy barrier for the dislocation to overcome before it can become mobile.

The tensile properties of the FeCoNiCrMn HEA with average grain sizes of 155 μm and 4.4 μm were measured by Otto *et al.*[139]. The corresponding UTS values were 520 MPa and 670 MPa, respectively. They found that both the $\sigma_y$ and the UTS increase with decreasing the grain size. But for the alloy with a finer grain, the fracture elongation is 10 % lower than that with a coarser grain. Subsequently, Seol *et al.*[180] purified the grain boundary of the alloy by adding a small amount of B to FeCoNiCrMn and $Fe_{40}Mn_{40}Cr_{10}Co_{10}$ HEAs, thus reducing the grain size and greatly improving the strength of the material without deteriorating its plasticity. The results also showed that a small amount of B can enhance the cohesion of the grain boundary, thus resulting in an incremental strengthening effect of the grain boundary.

Juan *et al.*[181] refined the grain size of the bcc-structured TaNbHfZrTi HEA by controlling the annealing temperature and time, which effectively improved its strength and plasticity. For example, the TaNbHfZrTi HEA with an average grain size of 38 μm has a $\sigma_y$ of 958 MPa and a fracture elongation of 20%, while the $\sigma_y$ and fracture elongation for the same HEA with an average size of 128 μm are 940 MPa and 15%, respectively. It was reported that the Nb and Ta elements can affect the grain-growth rate, leading to the slower migration of grain boundaries. Consequently, the solute-drag effect yields the strengthening of the HEA.

### 6.3. Solid-solution strengthening

The principle of solid-solution strengthening is that the alloying elements are dissolved into the matrix material to cause lattice distortion to a certain extent, thus increasing the strength of the alloy. Solid-solution strengthening can be divided into two types: substitutional solid-solution strengthening and interstitial solid-solution strengthening. He *et al.*[4] found that when less than 8 at.% Al was added to the FeCoNiCrMn HEA, the alloy still consisted of an fcc-structured single phase. Moreover, the addition of Al slightly increased the UTS of the alloy while the plasticity decreased by about 10 %.

Stepanov *et al.*[182] found that adding a small amount of V to the FeCoNiCrMn HEA has little influence on its strength and plasticity. Liu *et al.*[145] showed that with the change of the Mn content in the $(FeCoNiCr)_{100-x}Mn_x$ (where x = 6, 14, and 20) HEAs, the UTS and plasticity of the HEAs did not noticeably change. Their results indicate that the effect of substitutional solid-solution strengthening on these types of HEAs are very limited.

Interstitial solid-solution strengthening refers to the addition of elements that can be dissolved in the interlattice positions of the matrix, generally with a radius difference of more than 41 % of the matrix atoms, such as H, C, B, N, O, etc[183-187]. Compared with substitutional atoms, interstitial atoms can interact more with the defects in HEAs, so that the strengthening effect is more significant[188, 189]. In recent years, due to the combination of this interstitial strengthening effect and relatively low fabrication costs, interstitial element doped HEAs have attracted the attention of many more scientists[171, 190-201].

Wang *et al.*[188] added different amounts of C (0, 0.07, 0.16, 0.30, 0.55, and 1.1 at.%) to the $Fe_{40.4}Ni_{11.3}Mn_{34.8}Al_{7.5}Cr_6$ HEA, and it was found that the strengths of the HEAs increased significantly with the increase of the C content. It was also observed that adding carbon improved the plasticity of the alloy [see Fig. 20(a)]. For instance, adding 1.1 at.% C increases the $\sigma_y$ from 159 to 355 MPa, UTS from 535 to 1,174 MPa, and fracture elongation from 40.8% to 49.5%. Finally, it was reported that the $\sigma_y$ increases linearly with the carbon content [Fig. 20(b)], which is a consequence of increasing the lattice friction.

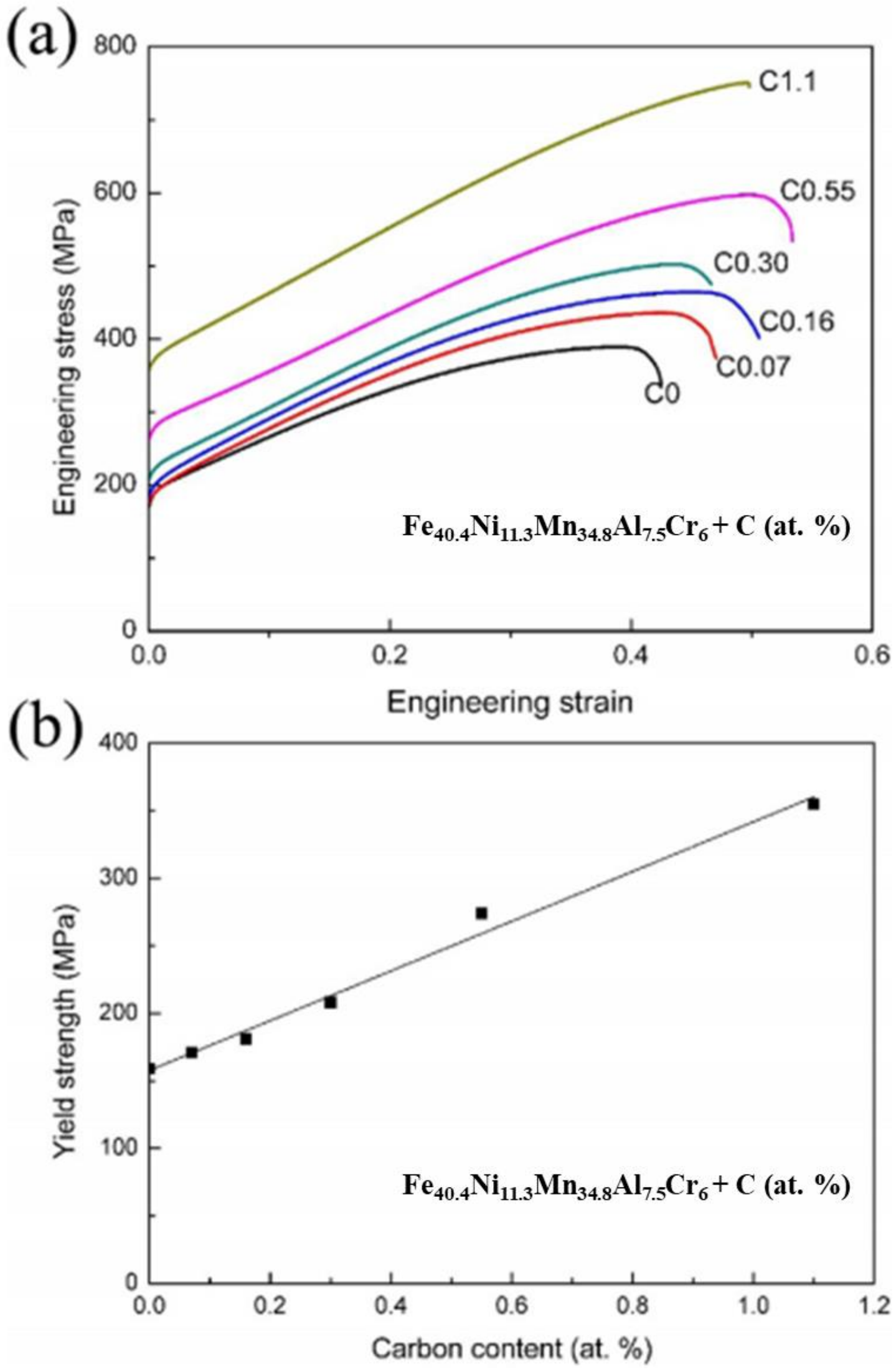

**Fig. 20.** (a) Engineering stress-strain curves and (b) $\sigma_y$ versus carbon concentration of 0, 0.07, 0.16, 0.30, 0.55, and 1.1 at.% carbon-doped $Fe_{40.4}Ni_{11.3}Mn_{34.8}Al_{7.5}Cr_6$ HEAs. Figures from Wang et al. [188].

They also found that the addition of carbon changes the slip mode of dislocations from the wavy slip (Figs. 21) to planar slip (Figs. 22). When the strain reaches to 40%, a dislocation-cell structure is typically formed in the undoped alloy[202]. However, no such structure was observed in the carbon-doped alloy. They hypothesized that the non-cell structure in the carbon-containing alloy leads to improved ductility.

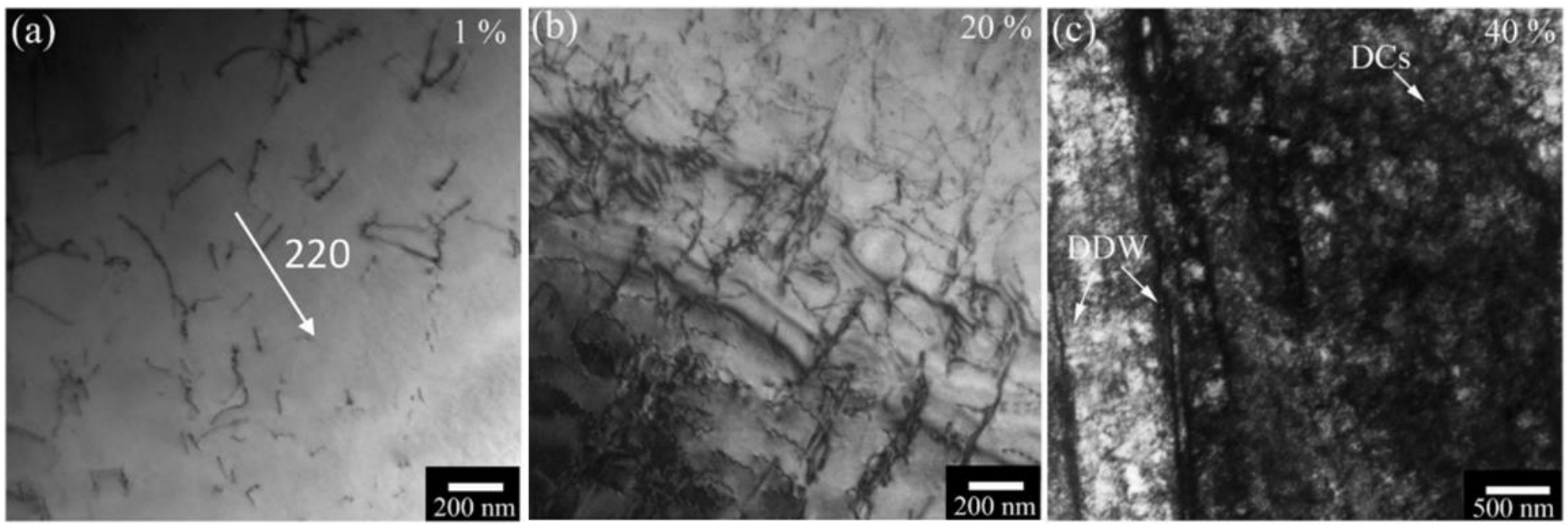

**Fig. 21.** Bright-field TEM images of the $Fe_{40.4}Ni_{11.3}Mn_{34.8}Al_{7.5}Cr_6$ HEA pre-strained to (a) 1 % strain; (b) 20 % strain; and (c) 40 % strain. Figures from Wang et al. [188].

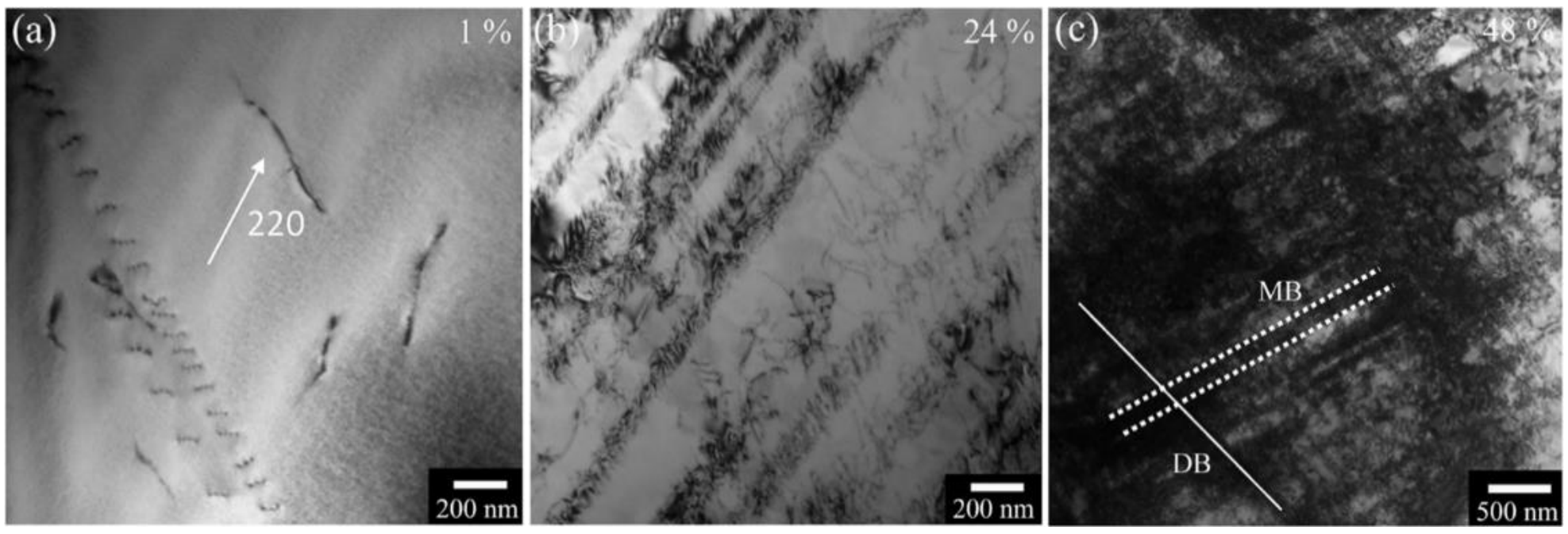

**Fig. 22.** Bright-field TEM images of the 1.1 at.% carbon-containing $Fe_{40.4}Ni_{11.3}Mn_{34.8}Al_{7.5}Cr_6$ HEA pre-strained to (a) 1 % strain; (b) 24 % strain; and (c) 48 % strain. Figures from Wang et al. [188].

When Stepanov *et al.*[203] added the 0.1 at.% C element to the FeCoNiCrMn HEA, the strength of the alloy increased significantly whereas the plasticity decreased slightly. In another study, Xie *et al.*[204] added the 0.1 at.% N element to the same alloy. The $\sigma_y$ was reported to increase by 200 MPa although the plasticity was reduced very little. The similar behavior was also found in the carbon-doped FeCoNiCr[205] HEA and NiCoCr[189] MEA. It was determined that when the carbon content is high, the $M_{23}C_6$ type carbides may form and deteriorate the ductility of the HEAs. However, if all carbon atoms are solutionlized into the matrix, the strength and ductility will increase simultaneously.

For the refractory HEAs, Chen *et al.*[206] added 0.05, 0.1, 0.2 at.% of oxygen to the $ZrTiHfNb_{0.5}Ta_{0.5}$ refractory HEA to improve the strength of the alloy. In another study, Lei *et al.*[26] added 2.0 at.% O to the TiZrHfNb HEA. It was found that the UTS is greatly enhanced by 48.5 %, and ductility is improved by 95.2 %, which solves the strength-ductility trade-off. The strengthening effect of O on HEAs is unexpected, as presented in Figs. 23.

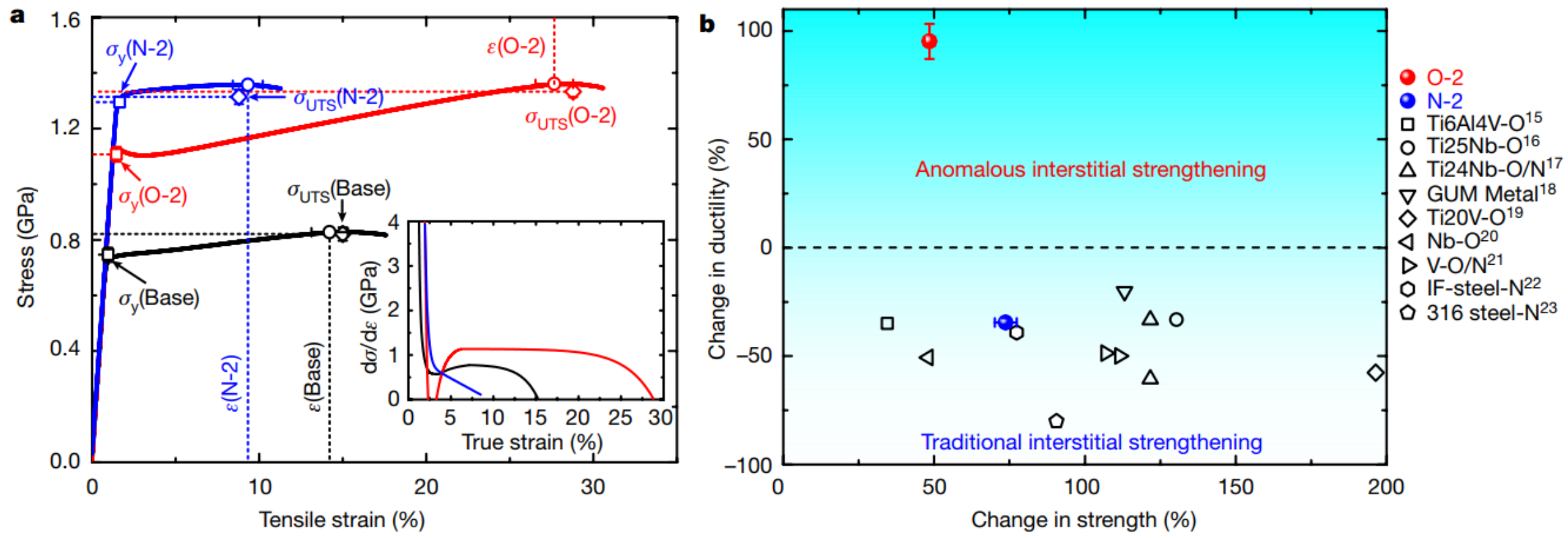

**Fig. 23.** (a) Tensile engineering stress-strain curve of O-doped and N-doped TiZrHfNb HEAs, the insert picture is the corresponding work-hardening rate curves. (b) A comparison of the strengthening effects of interstitial elements on HEAs and other conventional alloys. Figures from Lei et al. [26, 207-215].

The addition of O changes the dislocation-slip mode from the planar to wavy slip, as shown in Figs. 24. The typical dislocation structure in the base HEA is coplanar dislocation arrays. In the O-doped HEA, the dipolar walls that mainly contain primary dislocation dipoles with a high density are formed, indicating that the dislocation has a cell-forming structure. In the N-doped HEA, the typical planar-slip bands are formed, which is similar to the undoped HEA. The results suggest that oxygen significantly change the deformation mode while the nitrogen does not. They explained that the anomalous interstitial-strengthening effect comes from the ordered interstitial complexes in the HEA. The ordered complex can pin the dislocations, thus enhancing the strength and ductility. The TEM and high-angle-annular-dark-field scanning transmission electron microscopy (HAADF-STEM) were used to characterize the ordered oxygen complexes, as shown in Figs. 25(a)-(e). The ordered oxygen complex can promote the double cross-slip, which leads to dislocation multiplication during deformation. Consequently, the work-hardening rate is increased and the necking is delayed, which results in excellent mechanical properties of the HEA.

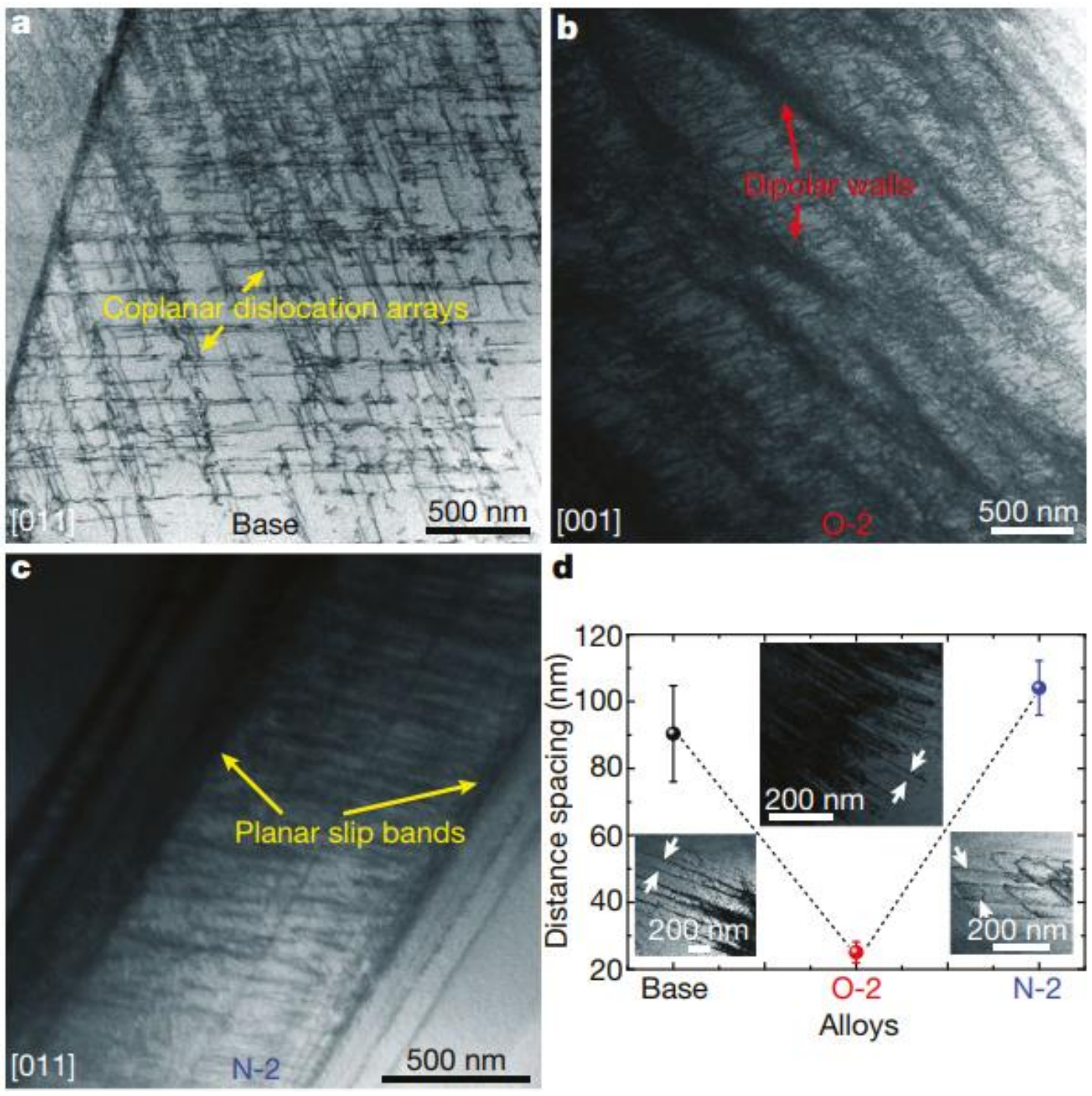

**Fig. 24.** Typical dislocation structure in (a) base HEA; (b) O-doped HEA; (c) N-doped HEA, respectively; and (d) the distance spacing between dislocation bands in the HEAs. Figures from Lei et al. [26].

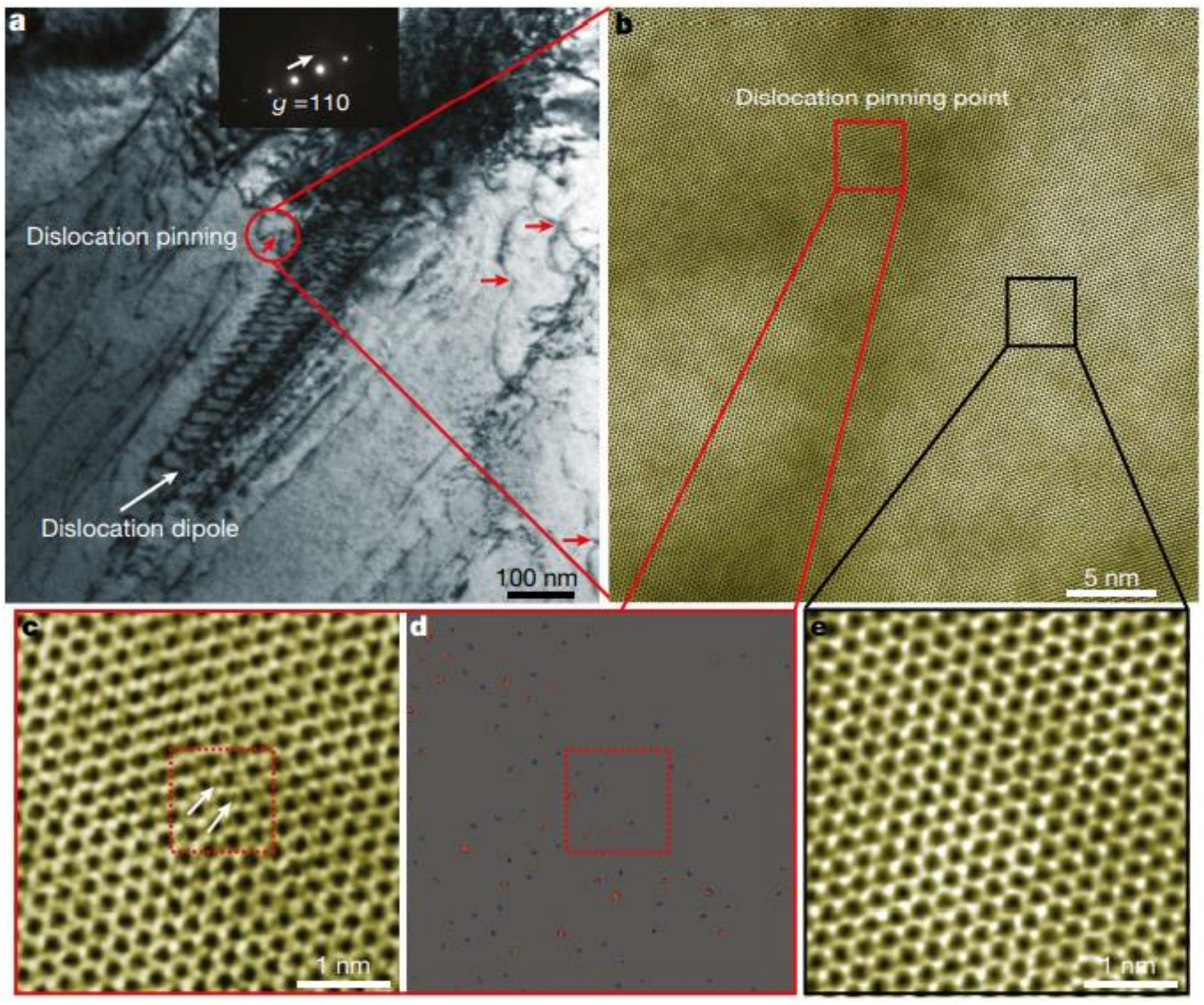

**Fig. 25.** (a) Dislocation structure in the O-2 HEA strained to 8 %. The aberration-corrected STEM annular bright-field (STEM-ABF) image of the local atomic structure (b) near the dislocation-pinning point and (c) at the pinning point; (d) STEM- HAADF image of the region near the dislocation-pinning point; and (e) STEM-ABF image of the region away from the pinning point. Figures from Lei et al. [26].

A summary plot of the fracture elongation versus UTS for the solid-solution-strengthened HEAs or MEAs[13, 26, 188, 189], as compared with single-phase HEAs[2, 155], and other kinds of metals[216] is presented in Fig. 26. The results show that there was an apparent interstitial solid-solution strengthening effects on HEAs or MEAs, such as the $(TiZrHfNb)_{98}O_2$, $Fe_{40.3}Ni_{11.2}Mn_{34.7}Al_{7.5}Cr_6C_{0.3}$, $(NiCoCr)_{99.25}C_{0.75}$, and $Fe_{49.5}Mn_{30}Co_{10}Cr_{10}C_{0.5}$ alloys.

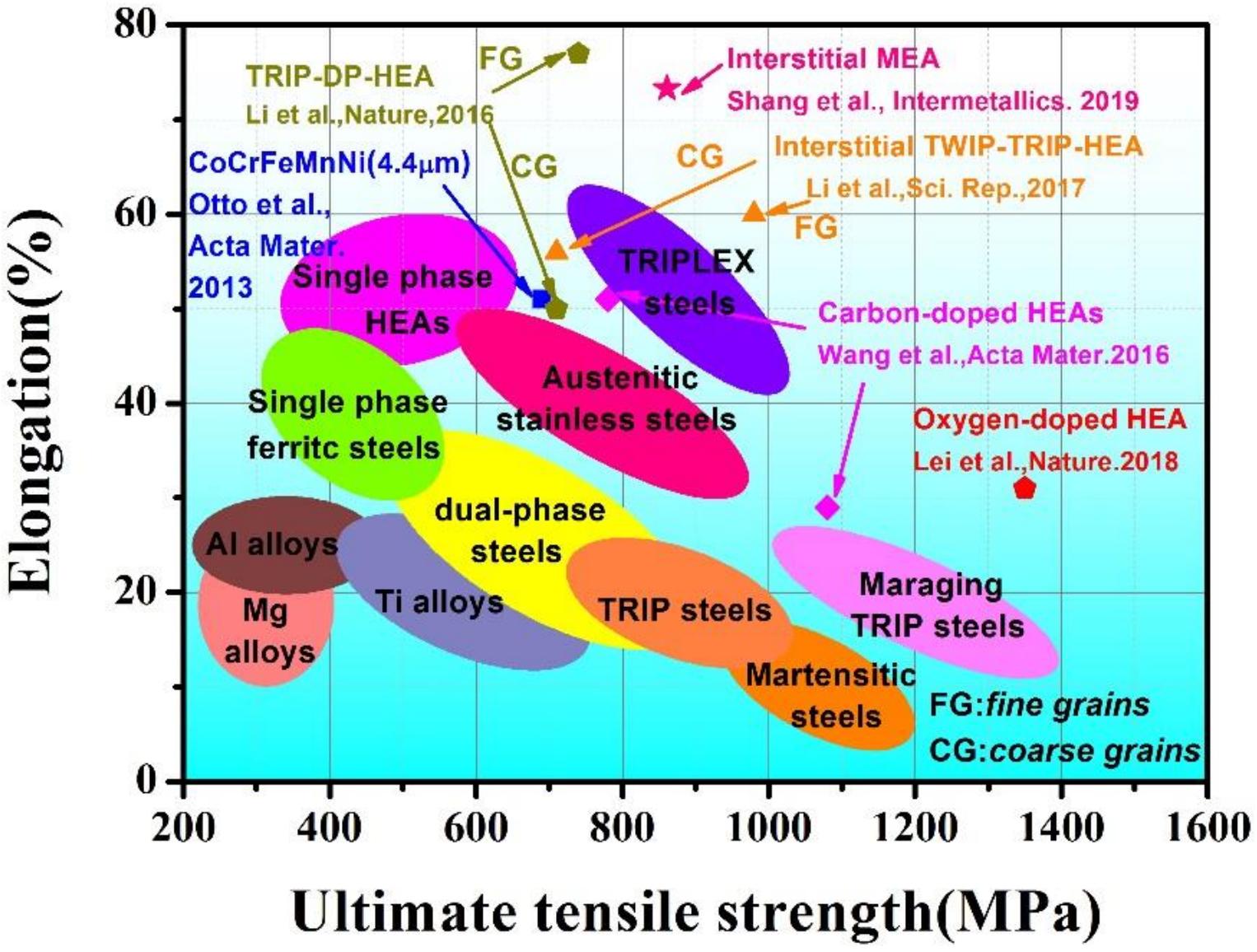

**Fig. 26.** Plots of fracture elongation versus UTS of different interstitial HEAs and MEAs, as compared with other kinds of alloys. Figure from Li et al. [2, 13, 26, 155, 188, 189, 216].

### 6.4. Particle strengthening

In this section, the coarse two-phase HEAs and the fine particles-strengthened HEAs will both discussed. The coarse two-phase HEAs always have large spaces between the particles, causing the small Orowan stress that leads to inconspicuous strengthening[217]. Note that significant strengthening can also occur in the coarse two-phase alloys. For the fine-particle strengthening, *e.g.* precipitation hardening and dispersion strengthening, the size of particles is often below a micrometer[125]. In this section, we summarize recent works on these two kinds of particle-strengthening strategies.

#### 6.4.1. Coarse two-phase HEAs

A coarse secondary phase can significantly influence the mechanical properties of HEAs. For example, if the Young's modulus of the second phase was larger than that of the matrix (when being strained), the load will spread to the particles, thus enhancing the stiffness. The transferred load also leads to the reduction of the stress in the matrix, resulting in the increase of $\sigma_y$. Note that if the strain was restricted in some local region, the second phase would result in hardening of the alloy. Furthermore, the coarse particles can also interact with dislocations, causing the formation of dislocation loops, thus affecting the mechanical properties, *e.g.* high dislocation density and more work hardening[125, 218].

(1) TRIP effect

In traditional materials, the TRIP effect has been proven to significantly improve the toughness and plasticity of irons and steels[219-221]. For the fcc-structured alloys, the SFE can markedly affect the deformation mode of the material. When the SFE is high, the deformation mode is the dislocation slip. For intermediate values of the SFE, twinning is the primary deformation mode. When the SFE is low, the deformation mode changes to phase transformation. For the alloys with the lower SFE, the TRIP effect can be very apparent. The TRIP effect has also been used to design HEAs with excellent comprehensive mechanical properties. In the fcc HEA system, with the decrease of the SFE, the plastic-deformation mechanism of the alloy changes from the dislocation slip to twin deformation. However, if the SFE continues to decrease, it changes to the martensitic-transformation deformation.

From one point of view, Li *et al.*[2] designed the $Fe_{80-x}Mn_xCo_{10}Cr_{10}$ (x = 45, 40, 35, and 30 at.%) HEAs to produce the TRIP effect in HEAs. As can be seen in the EBSD maps in Fig. 27, with the decrease of x, the stability of the fcc phase decreased gradually. The deformation mechanism of the alloy changes from the dislocation-slip-dominated plasticity to twin-induced plasticity and then to phase-transformation-induced plasticity. When x = 30, the TRIP dual-phase (DP) $Fe_{50}Mn_{30}Co_{10}Cr_{10}$ HEA was produced.

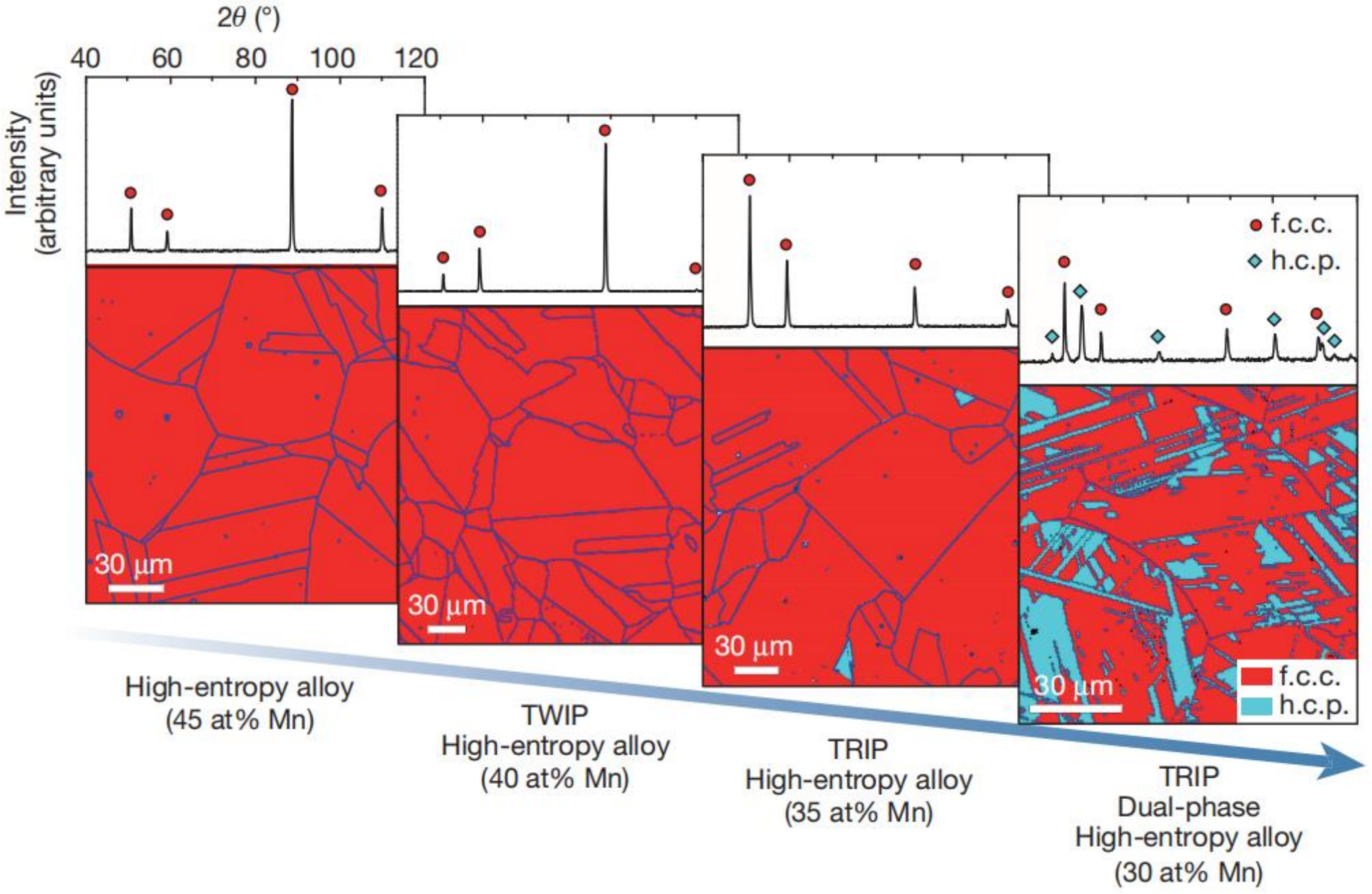

**Fig. 27.** EBSD phase maps and XRD patterns of $Fe_{80-x}Mn_xCo_{10}Cr_{10}$ (x = 45, 40, 35, and 30 at.%). Figure from Li et al. [2].

Figures 28 shows the structure evolution of the TRIP-DP-HEA. As can be seen from the EBSD maps in Fig. 28(a), the fraction of the hcp phase increases with the applied strain. The electron channeling contrast imaging (ECCI) images in Fig. 28(b) show that there were a large number of observed stacking faults caused by the slip of 1/6 <112> Shockley partial dislocations. These stacking faults become the nucleation sites of the martensite phase in the subsequent deformation process. During the deformation process, the high-density phase boundary produced during phase transformation hinders the dislocation slip, which promotes the working-hardening behavior of the alloy and delays the necking, as shown in Fig. 28(c). Furthermore, they surmised that by increasing the fraction of the hcp phase, the strength and ductility increase simultaneously. During the deformation process, the simultaneous deformation of the two phases will result in the obvious dynamic strain-stress partitioning effect. The elastic compliance of the two phases leads to the inhabitation of crack nucleation, which is different from those dual-phase alloys with a large difference in strengths between two constituents. Based on the TRIP-DP concept, Li *et al.*[222] simulated the phase stability of hcp and fcc phases in $Fe_{40-x}Co_{20}Ni_xCr_{20}Mn_{20}$ ($0 \leq x \leq 20$ at.%) HEAs by the DFT ab initio method and then established the criterion for designing TRIP dual-phase HEAs. According to their result, the selected $Fe_{34}Co_{20}Ni_6Cr_{20}Mn_{20}$ HEA possesses the superior UTS and working-hardening ability among this type of HEAs.

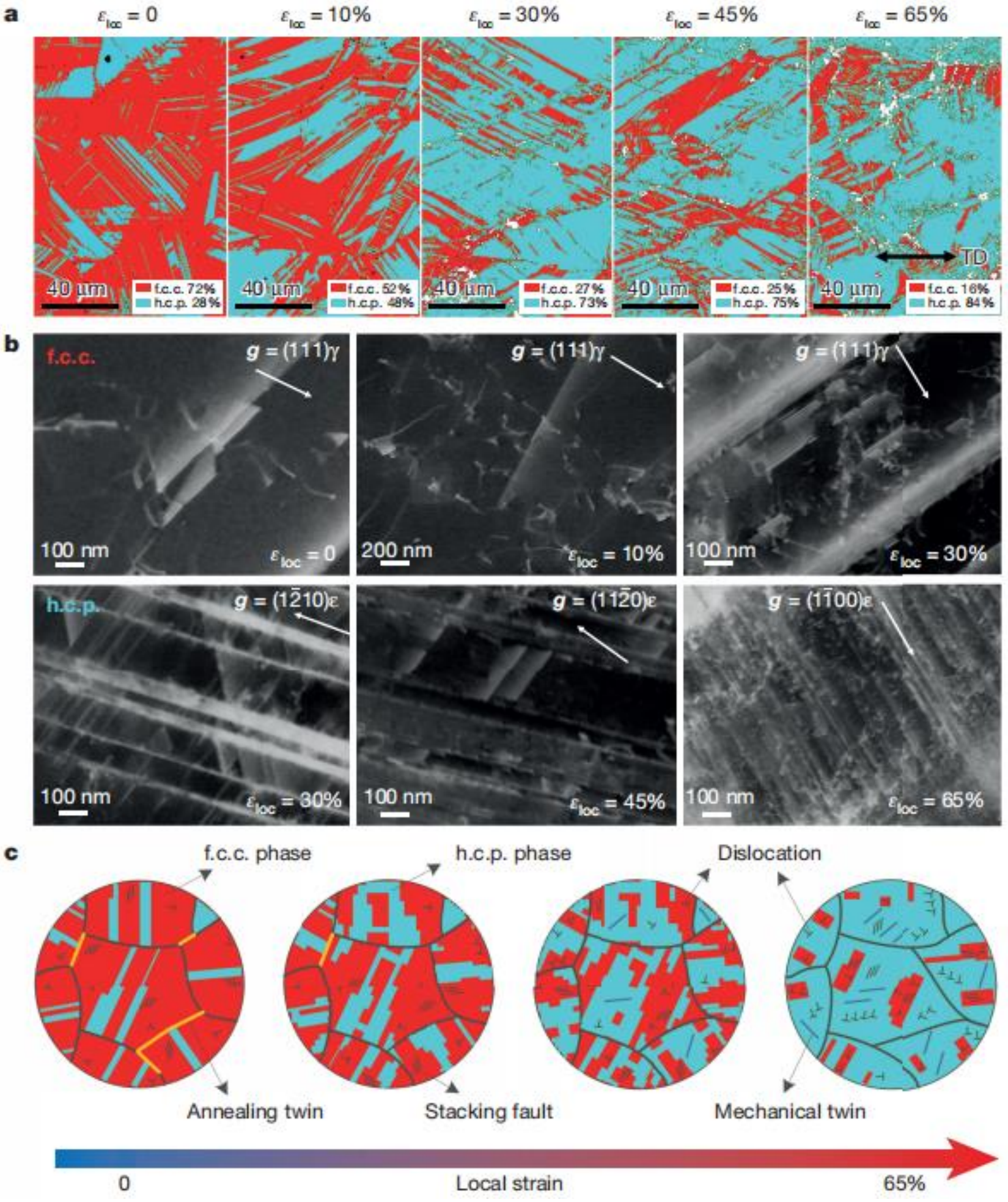

**Fig. 28.** (a) EBSD phase maps and (b) ECCI photographs of the TRIP-DP-HEA pre-strained to different strains. (c) Schematic sketches of the structure evolution during the applied strain. Figures from Li et al. [2].

Huang *et al.*[7] applied the TRIP effect to the TaNbHfZrTi refractory HEA. By reducing the content of Ta in the HEA, the stability of the bcc phase was reduced, and the two-phase microstructure of bcc and hcp phases was obtained. Consequently, when an external force was applied, the bcc phase transformed into an hcp phase. Although the strength of the TaNbHfZrTi HEA decreased with the decrease of the Ta content, the plasticity was significantly increased by the introduction of the TRIP effect, and a significant work-hardening effect was produced simultaneously.

(2) Eutectic HEAs

The concept of the eutectic HEA, which recently has been the subject of great in the HEA community[223, 224], may provide a way to solve the strength-ductility trade-off, thus simplifying the industrial production of HEAs. Since the eutectic reaction is an isothermal-transition process, and there is no solidification temperature range that reduces segregations and shrinkage-pore content at the same time. The eutectic $FeCoNi_{2.1}CrAl$ HEA designed by Lu[223] has a high-strength bcc phase and relatively high toughness fcc phase. Importantly, ingots of this HEA can be successfully fabricated to the industrial scale. Different from other bulk HEAs, there are fewer casting defects in the eutectic HEA ingots. Moreover, the eutectic HEA has a microstructure consisting of the fine fcc/B2 sheet, and its good mechanical properties can be maintained up to 973 K [223]. This new design idea is expected to improve the poor casting properties of HEAs, thereby solving some important problems such as macro-segregation.

Shi *et al.*[225] tailored the thermo-mechanical process of an $FeCoNi_{2.1}CrAl$ HEA to obtain the ultrafine-grained eutectic form of the alloy. Figures 29(a)-(h) present the microstructure and composition distribution of this HEA, as obtained using different characterization techniques, *e.g.* EBSD, energy dispersive spectroscopy (EDS) and TEM. Figure 29(a) shows the typical lamella morphology of the as-cast eutectic $FeCoNi_{2.1}CrAl$ HEA. After cold-rolling and annealing, the dual-phase heterogeneous lamellar (DPHL) bcc structure was also found in the fcc lamellae, as exhibited in Figs. 29(b) and (c). Figure 29(d) indicates that the lamellae contain recrystallized grains and annealing twins, which is a typical twin structure that forms during annealing. According to Figs. 29(e) and (f), the enriched Ni and Al lamellae consist of B2 grains, and the FeCr-rich lamellae contain fcc grains. Figure 29(g) shows the typical microstructure of annealing twins. Figure 29(h) is a diagrammatic sketch of the DPHL structure, where a two-hierarchical heterogeneity composed of fcc/B2 grains within the lamellae of fcc phase is presented.

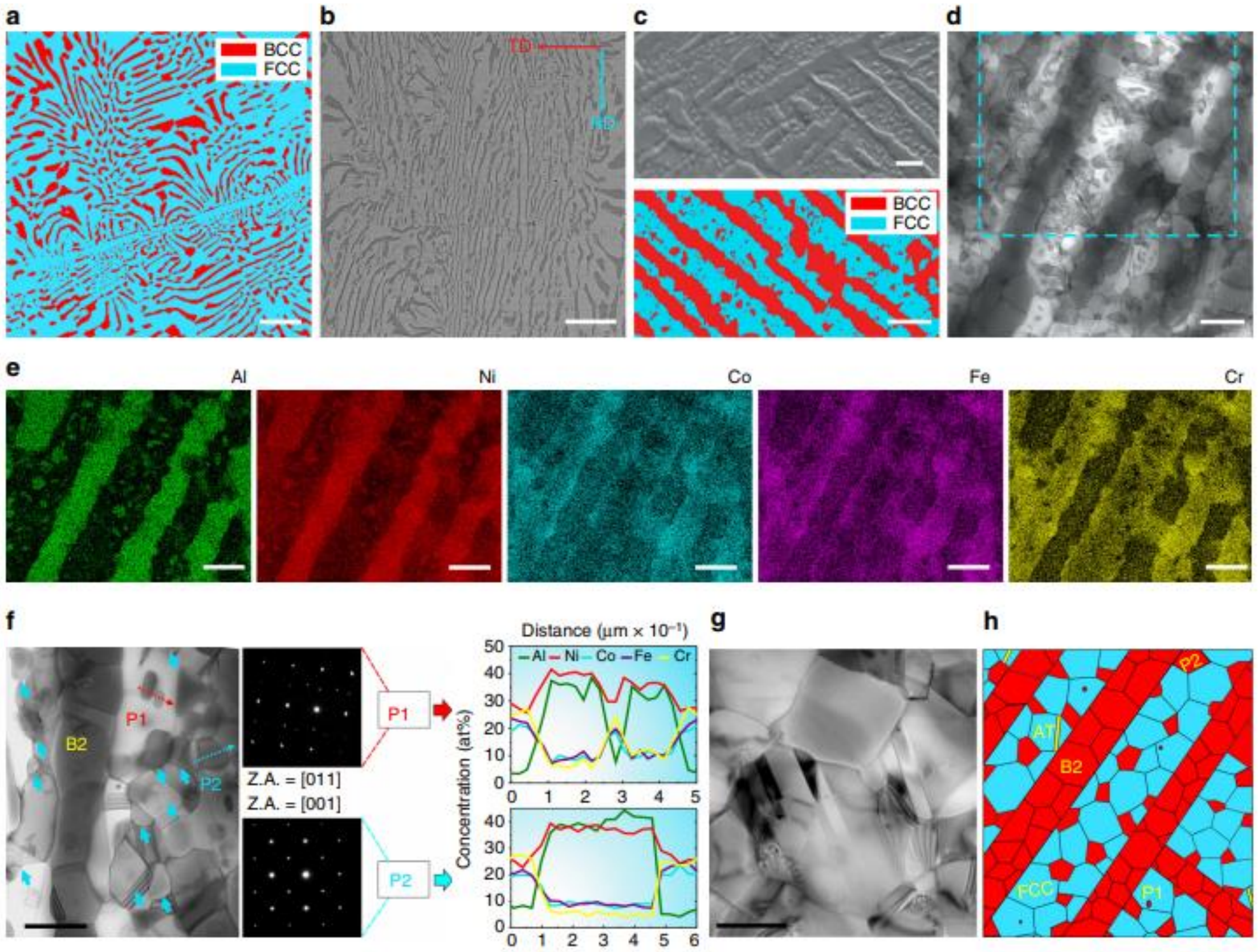

**Fig. 29.** (a) EBSD map of the as-cast $FeCoNi_{2.1}CrAl$ HEA. SEM images at (b) Low magnification, (c) High magnification, and (d) EBSD map of the cold-rolled and annealed $FeCoNi_{2.1}CrAl$ HEA, respectively. (e) EDS maps of the cold-rolled and annealed $FeCoNi_{2.1}CrAl$ HEA. (f) TEM image of the cold-rolled and annealed $FeCoNi_{2.1}CrAl$ HEA and corresponding electron diffraction spots as well as the EDS results of the composition distribution. (g) Annealing twin under TEM. (h) Schematic sketch of the microstructure. Figures from Shi et al. [225].

Room-temperature tensile tests were performed using a strain rate of $2.5 \times 10^{-4}$ $s^{-1}$ on the $FeCoNi_{2.1}CrAl$ HEAs treated in different conditions, *e.g.* as-cast HEA, and HEAs cold-rolled to 84 ~ 86 % and then annealed at 933 K, 973 K, and 1,013 K, respectively. They also compared the properties with reported ultrafine-grained HEAs[226], as well as the complex and hierarchical HEAs[227]. As shown in Fig. 30(a), the DPHL HEAs exhibit the superior strength and ductility as compared to the as-cast HEA. Figure 30(b) displays the strain-hardening rate vs. the true strain, where the graph is divided into three distinct regions that are labelled I, II, and III. The strain-hardening rate in Region I decreases rapidly, which refers to the lack of mobile dislocations. The increasing of strain-hardening rate in Region II corresponds to the dislocation multiplication and tangle. Finally, the strain-hardening rate decreases with true strain in Region III, corresponding to the further movement of dislocations.

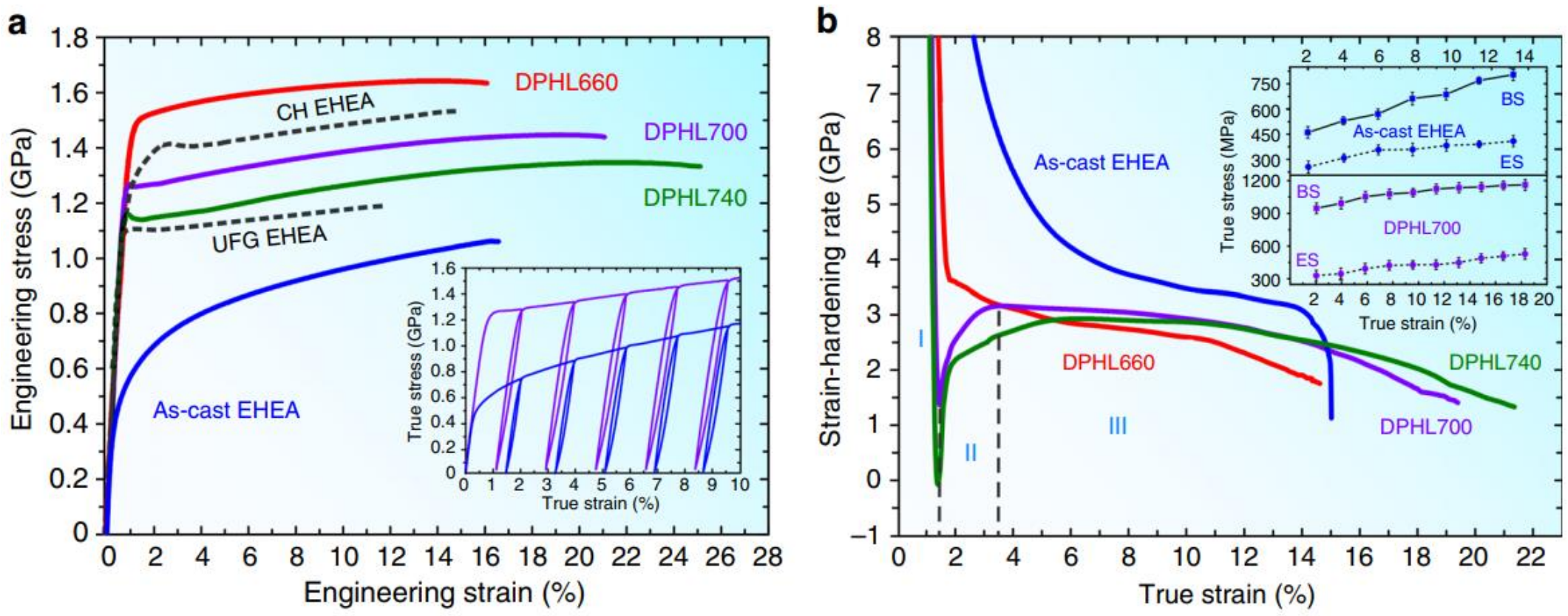

**Fig. 30.** (a) Engineering stress-strain curves of the $FeCoNi_{2.1}CrAl$ HEA treated in various conditions, *e.g.* as-cast HEA, and HEAs cold-rolled to 84 ~ 86 % and then annealed at 933 K, 973 K, and 1,013 K, respectively. (b) Corresponding strain-hardening curves, the inset figure exhibits the back stress (BS) and effective stress (ES) versus true strain for the as-cast HEA and DPHL700 HEA. Figures from Shi et al. [225].

To understand the deformation mechanisms, they pre-strained the DPHL HEA to different strain levels, *e.g.* 4%, 13%, and 21%. The dislocation-evolution process is shown in Figs. 31(a)-(c), and Fig. 31(d) provides a schematic sketch. The results indicate that at the early deformation stage, the softer fcc grains can bear more strain and thus deform more easily, leading to the formation of strain gradients. With increasing the strain, the gradients increase, resulting in high back-stress hardening. Similar behavior was also observed in the B2 phase embedded in the fcc grains. They attributed the good combination of strength and plasticity to the inherited lamellar structure of the as-cast HEA, which provides a hierarchical constraint effect that results from the eutectic microstructure and the self-generated microcrack-arresting mechanisms.

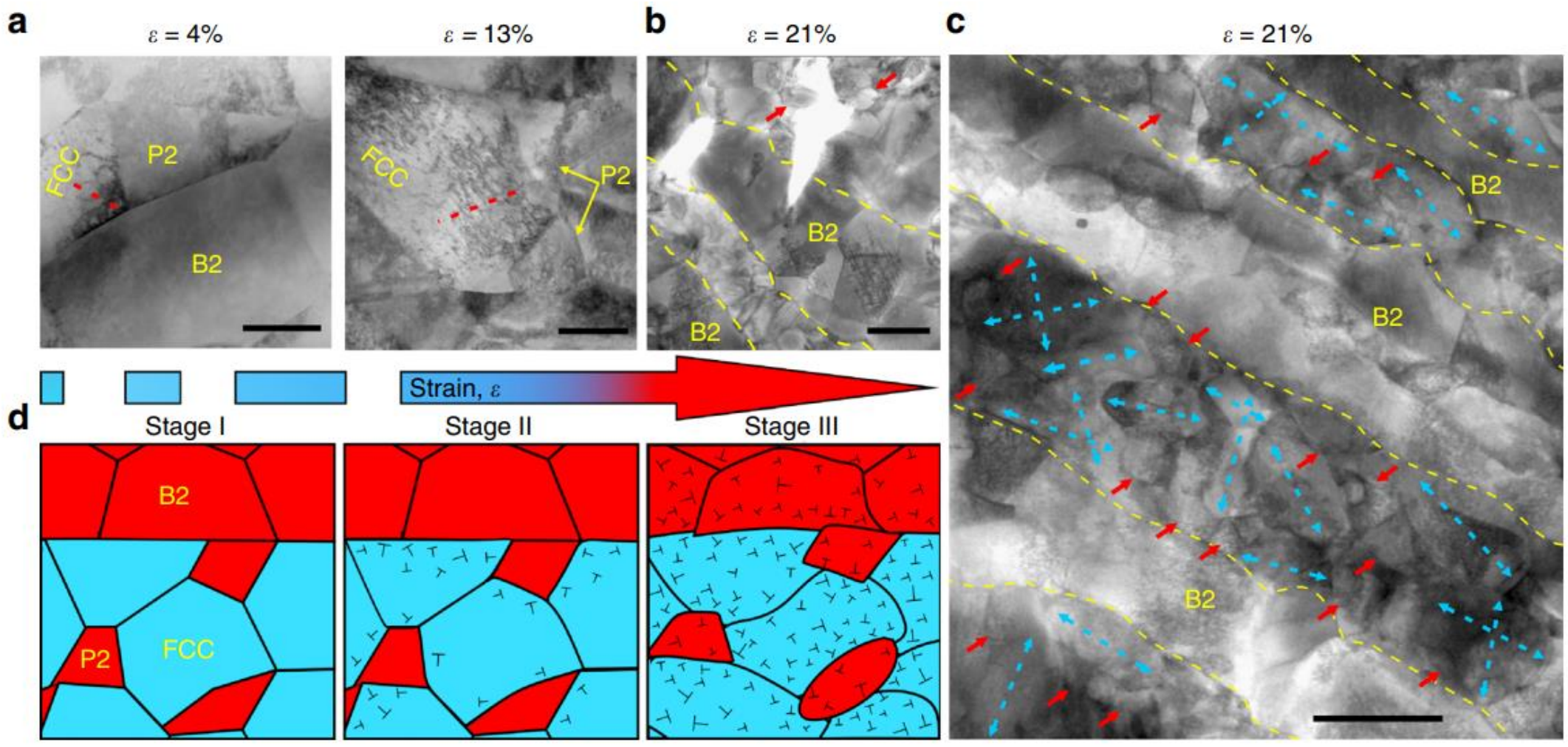

**Fig. 31.** TEM images of the $FeCoNi_{2.1}CrAl$ eutectic HEA strained to (a) 4 %, (b) 13 %, and (c) 21 % strains. (d) Schematic sketch of the dislocation evolution during deformation. Figures from Shi et al. [225].

Gao *et al.*[228] studied the microstructure of the $FeCoNi_{2.1}CrAl$ eutectic HEA in detail using TEM. It was found [see Figs. 32(a)-(d)] that the excellent mechanical properties were attributed to the synergistic deformation of the fcc and bcc phases. More specifically, the fcc (L12) phase acts as a soft phase through the planar slip of dislocations and deformation of stacking faults, while the bcc (B2) phase serves as the strengthening phase via nanometer-precipitation strengthening. In the $FeCoNi_{2.1}CrAl$ eutectic HEA, the fcc phase has a lower SFE value, and the phase interface between the L12 and B2 phases is semi-coherent, which can bear the high stress. Also, there are Cr-rich nano-precipitated phases with a diameter of about 20 nm in the B2 phase. Based on the Orowan-bypassing mechanism[229], the precipitated phases prevent the dislocation slip, thus increasing the strength of the B2 phase.

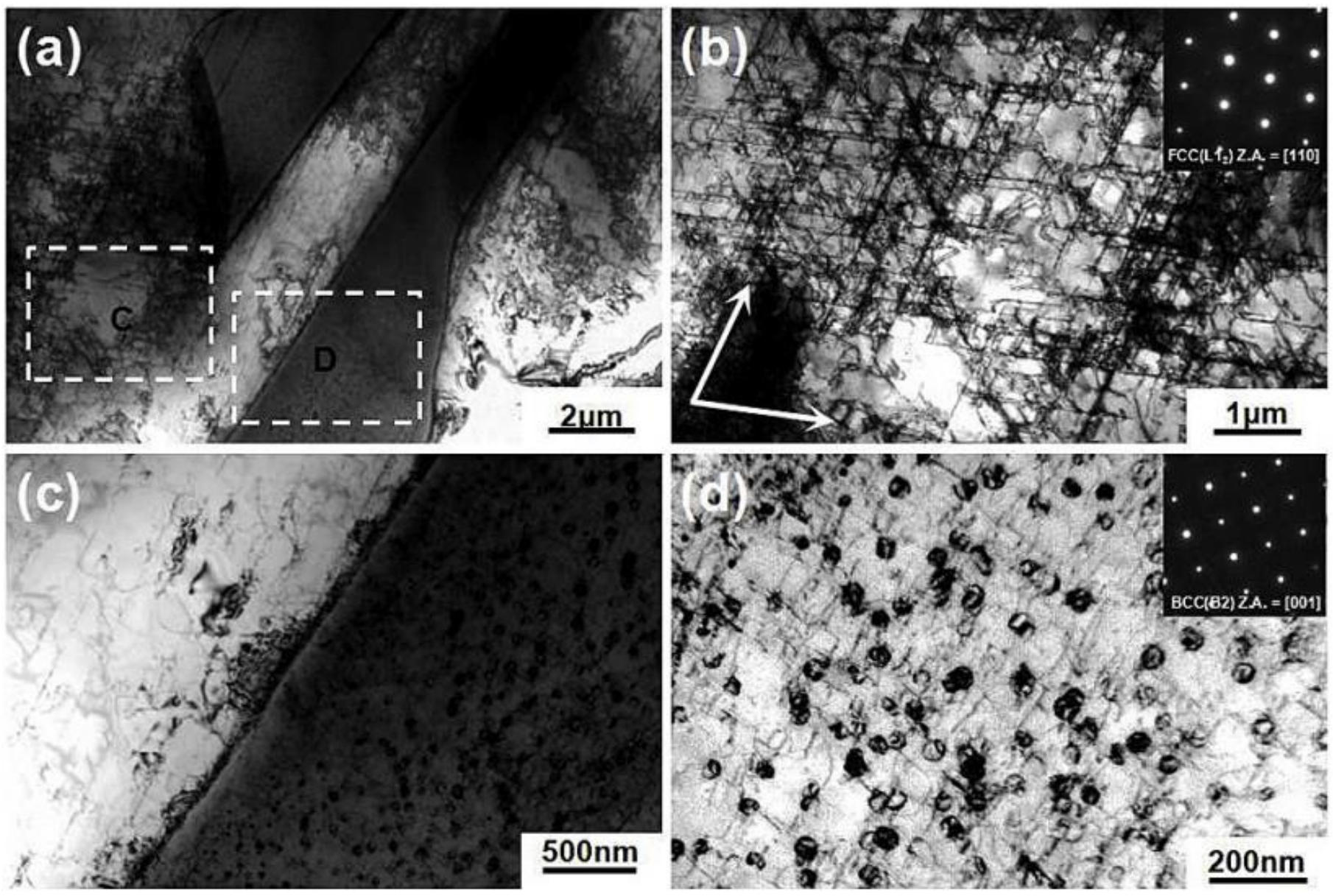

**Fig. 32.** (a) Bright-field TEM images of the deformed $FeCoNi_{2.1}CrAl$ eutectic HEA at low magnification; (b) Enlarged image in the region, C, in (a), the inserted SAED pattern is the fcc (L12) phase; the enlarged image of the region, D, at low magnification, and (d) High magnification, the inserted SAED pattern in (d) is the bcc (B2) phase. Figures from Gao et al. [228].

Jiang *et al.* [230] studied the structures and properties of the $CoFeNi_2V_{0.5}Nb_{0.75}$ eutectic HEA after heating to 773, 873, 973, 1,073, and 1,273 K. The eutectic HEA was composed of the fcc solid-solution phase and the $Fe_2Nb$ type Laves phase. It was reported that the $NbNi_4$ intermetallic compound formed in the alloy when the annealing temperature exceeded 873 K. With the increase of the annealing temperature, the volume fraction of the $NbNi_4$ intermetallic compound increased, while the volume fraction of the eutectic region decreased. When the annealing temperature reached 1,073 K, the fiber-like structure was produced in the $Fe_2Nb$ Laves phase, especially at the eutectic cell boundary. In this scenario, the two phases connect more closely in the eutectic structure, thus leading to excellent compressive properties.

He *et al.*[231] designed the eutectic $CoCrFeNiNb_x$ (x = 0.1, 0.25, 0.5, and 0.8) HEAs. The addition of the Nb element causes severe lattice distortion, which further enhances the strength of the fcc matrix. With the increase of the Nb content, the hardness of the alloy also increased, where the compression strength of the hypoeutectic $CoCrFeNinNb_{0.25}$ reached as high as 2 GPa. Furthermore, the compression plasticity was nearly 40 %, which indicates that the alloy exhibits a desirable balance between strength and toughness.

### 6.4.2. Strengthening by fine particles

Fine particles refer to those whose average sizes are below 1 micrometer and can distribute uniformly in the matrix, resulting in a strong inhibition of the dislocation slip[125]. The strengthening effect can be determined by the radii of the particles ($r$), the volume fraction of particles in the matrix ($f_V$), and the mean spacing between the particles ($2\lambda$) on the chosen slip plane. Note that the three factors are not independent. For example, when the fraction of particles is kept constant, an increase in the radius leads to an increase in the mean spacing between the particles[125].

There are two mechanisms to explicate the fine-particle strengthening effect. One mechanism is the cutting process, which means that when passing the particles, the dislocation can cut the particles to continue its slip. For the cutting mechanism, the contribution of the particles to the strength ($\Delta\tau_{ps}$) is[125]:

$$\Delta\tau_{ps} = \frac{F}{b}\frac{1}{2\lambda} = k_{ps}\sqrt{f_V r} \tag{6.3a}$$

$$f_V = N\frac{4\pi r^3}{3(2a)^3} \tag{6.3b}$$

where $F$ is the force caused by obstacles, $2\lambda$ is the mean spacing between the particles, $k_{ps}$ is a material constant for the cutting mechanism, $f_V$ is the volume fraction of particles in the matrix, $N$ is the number of spheres, $b$ is the Burgers vector, $r$ is the radius of the particle, and $2a$ is the edge length of the cube.

Another mechanism is the Orowan process, in which the dislocation overcomes the obstacles by passing rather than cutting. The contribution of the particles with the Orowan mechanism ($\Delta\tau_{ps'}$) is described as follows[125]:

$$\Delta\tau_{ps'} = \frac{Gb}{2\lambda} = k_{ps'}\frac{\sqrt{f_V}}{r} \tag{6.4}$$

where $k_{ps'}$ is the material constant for the Orowan mechanism, $G$ is the shear modulus, $2\lambda$ is the mean spacing between the particles, $r$ is the radius of the particle, $f_V = N\frac{4\pi r^3}{3(2a)^3}$ is the volume fraction of particles in the matrix, and $2a$ is the edge length of the cube.

In addition to the above classification method, fine-particles strengthening can also be divided into two types of mechanisms. One is precipitation strengthening, where the other second-phase particles will dissolve in the matrix at high temperatures, which may affect the elevated-temperature properties of the alloys[125]. For example, when heating locally, some precipitates may coarsen or even dissolve into the matrix. The other mechanism consists of dispersion strengthening, where the second-phase particles will not dissolve at high temperatures. Hence, the dispersion particles have the elevated-temperature resistance. Furthermore, the dislocations cannot overcome the obstructions by cutting these particles. Hence, with small particle radii, the strengthening effect can be large.

(1) Precipitation hardening

Tsai *et al.*[232] introduced two types of precipitates into the $Al_{0.3}CoCrCu_{0.5}FeNi$ HEA, which increased the $\sigma_y$ of the HEA from 150 MPa to 300 MPa. From Figs. 33(a)-(f), we can see these two types of precipitates. The first type consists of a plate-like geometry, whereas the other type has a spherical geometry. The plate-like precipitates contain an fcc-structured Cu-rich phase and an L12-structured $(Ni, Cu)_3Al$ phase. The spherical precipitates are also L12-structured. Using thermodynamics, they discussed the formation sequence of the precipitates. During the cooling process, the first precipitation was the Cu-Ni-Al-rich plate. When the temperature was lowered from 1,373 to 1,203 K, the plates split into a Cu-rich phase and an L12-structured $(Ni, Cu)_3Al$ phase. When the temperature fell below 1,173 K, the reduction in the solubility resulted in the formation of the spherical precipitates.

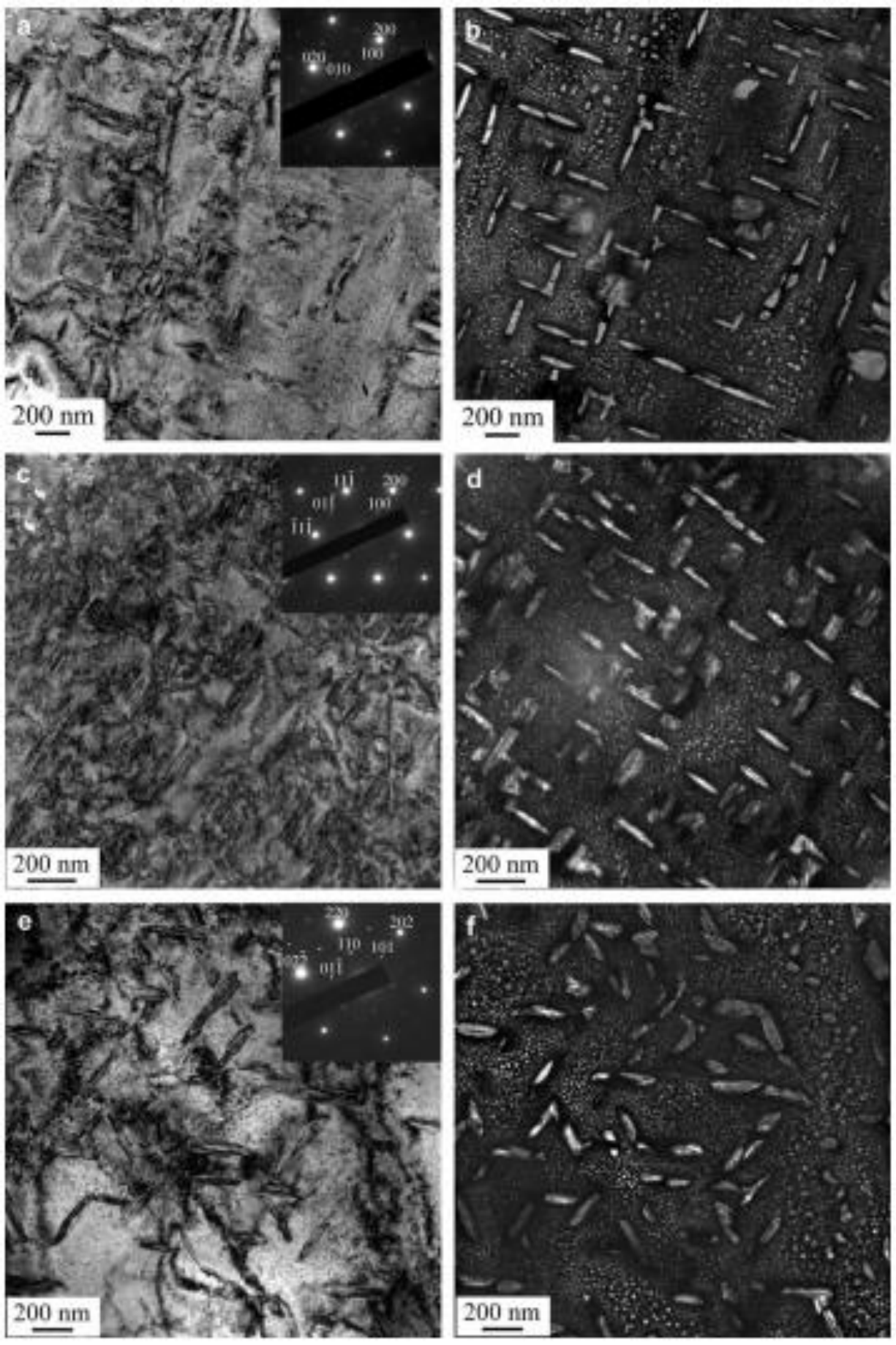

**Fig. 33.** (a) Bright-field TEM image and (b) Corresponding dark-field image of the $Al_{0.3}CoCrCu_{0.5}FeNi$ HEA observed from the zone axes of <100>; (c) Bright-field TEM image; and (d) Corresponding dark-field image of the $Al_{0.3}CoCrCu_{0.5}FeNi$ HEA observed from the zone axis of <110>; (e) Bright-field TEM image; and (f) Corresponding dark-field image of the $Al_{0.3}CoCrCu_{0.5}FeNi$ HEA observed from the zone axis of <111>. Figures from Tsai et al. [232].

He *et al.*[6] added different amounts of Ti and Al to the FeCoNiCr HEA to control the types, sizes, and distributions of precipitated phases. For the $(FeCoNiCr)_{94}Ti_2Al_4$ HEA, a large number of dispersed nano-γ' phases were observed in its matrix, suggesting that the amount of the Heusler phase in the alloy could be controlled. The $\sigma_y$ of the $(FeCoNiCr)_{94}Ti_2Al_4$ samples treated with the 30 % cold-rolling deformation and 1,073 K aging for 18 h (P1 in Fig. 34) increased from 603 MPa [for the $(FeCoNiCr)_{94}Ti_2Al_4$ HEA homogenized at 1,473 K for 4 h] to 645 MPa at room temperature. Furthermore, this process resulted in fracture elongation that was approximately 40 % and a UTS which exceeded 1 GPa.

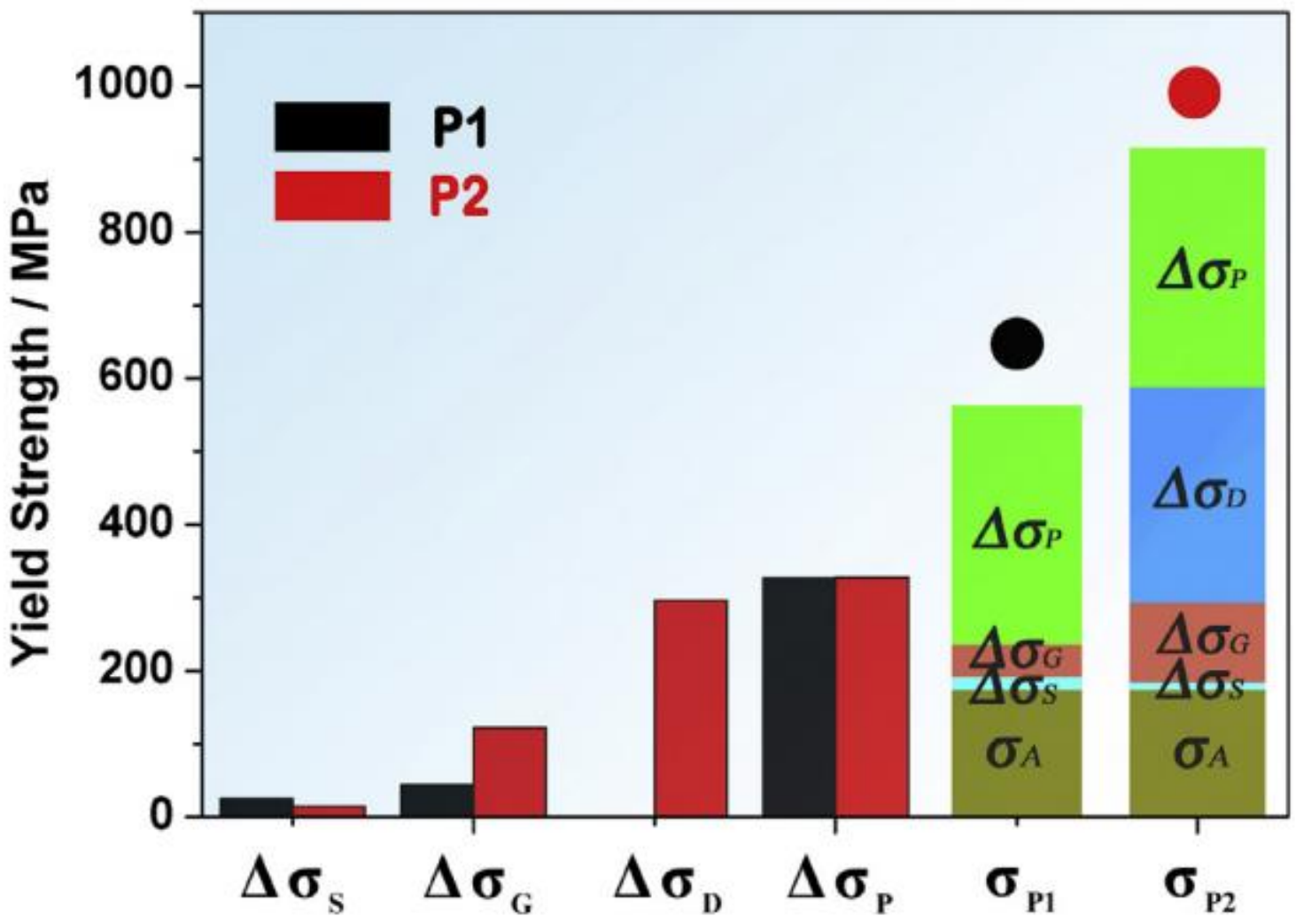

**Fig. 34.** The contributions of different strengthening strategies to the $\sigma_y$ of the $(FeCoNiCr)_{94}Ti_2Al_4$ HEA, where P1 refers to the HEA that treated with 30 % cold-rolling deformation and 1,073 K aging for 18 h, and P2 refers to the HEA that treated with 70 % cold-rolling deformation and aging at 923 K for 4 h. Figure from He et al. [6].

Additionally, the precipitated phase in the $(FeCoNiCr)_{94}Ti_2Al_4$ HEA was observed to be of the $Ni_3$(Ti, Al) type γ' phase with an L12 structure. The size of the precipitated phase ranged from 15 nm to 30 nm, while the volume fraction was about 21 %. However, the tensile $\sigma_y$ of $(FeCoNiCr)_{94}Ti_2Al_4$ samples treated with the 70 % cold-rolling deformation and aging at 923 K for 4 h (P2 in Fig. 34) was greater than 1 GPa, while still maintaining a fracture elongation of 17 %.

The contributions of different strengthening effects were also calculated. The $\sigma_y$ can be expressed as a sum of the dislocation strengthening, grain-boundary strengthening, solid-solution strengthening, and precipitation strengthening, according to the following equation[233]:

$$\sigma_y = \sigma_A + \Delta\sigma_S + \Delta\sigma_G + \Delta\sigma_D + \Delta\sigma_P \quad (6.5)$$

where $\sigma_A$ is the $\sigma_y$ of the base alloy (FeCoNiCr HEA), $\Delta\sigma_S$, $\Delta\sigma_G$, $\Delta\sigma_D$, and $\Delta\sigma_S$ are the strengthening effects of solid-solution, grain-boundary, dislocation, and precipitation strengthening, respectively. The results of the analysis are displayed in Fig. 34, which indicate that as compared to other strengthening strategies, the precipitation-strengthening effect is relatively high. Thus, the improvement of the strength is mainly due to the contribution of precipitation strengthening.

(2) Dispersion strengthening

Rogal *et al.*[234, 235] added two different types of nanoparticles to the FeCoNiCrMn HEA, namely the 5 wt.% spherical β-SiC with a diameter of 20 ~ 50 nm and $\alpha$-$Al_2O_3$ with a diameter of 5 ~ 35 nm. For the alloys that contained the $\beta$-SiC nanoparticles, the compressive $\sigma_y$ increased from 1,180 MPa for the as-cast condition to 1,480 MPa (with 5 wt.% SiC mixing at 100 rpm) and 1,940 MPa (with 5 wt.% SiC mixing at 200 rpm), which was accompanied by a decrease in the plasticity. After deformation, the formation of high-density dislocations and twins in the matrix suggested that the deformation mechanism of the nanoparticles occurs similar to that of the matrix. For the $\alpha$-$Al_2O_3$ nanoparticle-containing alloys, the $\sigma_y$ of the material increased from 1,180 MPa to 1,600 MPa, while its plasticity decreased. For the compression deformation, the fracture morphology was characteristically brittle by SEM. From the results, the authors concluded that the increase of strength comes from the nanoparticles, and the decreased plasticity derives from the presence of deformation twins.

Hadraba *et al.*[236] applied the theory of oxide-dispersion-strengthening (ODS) to HEAs and successfully prepared the yttria nanooxides-strengthened ODS FeCoNiCrMn HEA by mechanical alloying. The average grain size of the ODS HEA was found to be 0.4 μm via STEM, which is 50 % smaller than the HEA prepared in the same condition. Furthermore, the UTS at room temperature of the ODS HEA increased by 30 %, but the fracture elongation of the ODS HEA is less than 2 %. It was surmised that the presence of oxides suppresses the grain growth and dislocation movement, which results in the high strength of the alloy. However, the oxides also inhibit the twinning deformation, so that the plastic deformation is limited, resulting in low plasticity.

The above studies indicate that the strengths of HEAs can be enhanced by dispersion strengthening, although the resistance to plastic deformation is poor due to the interface discontinuity between the second-phase particles and the matrix. Moreover, the distribution of the particles can influence the mechanical properties, *e.g.* deteriorate the plasticity[1, 14]. Finally, the results suggested that the methods of production are of great importance since they can lead to a more uniform distribution of particles in the matrix, and consequently improved properties of the alloy.

## 6.5. Other strengthening mechanisms

More recently, some other strengthening strategies for other HEAs have been proposed, such as the implementation of the TWIP effect[237]. The TWIP effect refers to the deformation twins induced by material deformation under an external force, which significantly enhances the plasticity while maintaining high strength in the alloy. It should be noted that the TWIP effect is affected by the SFE of the material[238-240]. For instance, when the SFE is relatively low, the width of the extended dislocation in the deformation process is larger. Consequently, the extended dislocation is difficult to be bundled into the perfect dislocation in the slip process, which hinders the dislocation cross slip. Therefore, reducing the SFE of the alloy can change the deformation mode of the material from the dislocation-slip deformation to twin deformation, which improves the mechanical properties of the alloy[241-246].

Deng *et al.*[247] fabricated $Fe_{40}Mn_{40}Co_{10}Cr_{10}$ HEA by removing Ni and reducing the Co and Cr content in the FeCoNiCrMn HEA. The removal of Ni lowered the SFE of the alloy, while reducing the amount of Co and Cr in the matrix avoided the generation of the Cr-rich compound during fabrication. The resulting lower SFE during the deformation process led to a large number of nano-twins that could be found in the grains neighboring the <112>//TD and <111>//TD orientations (where TD is the tensile direction). Figures 35(a)-(c), as obtained by EBSD, shows that when the HEA was strained to 10 % at room temperature (strain rate of $1.0 \times 10^{-3}$ $s^{-1}$), deformation twins could be observed in the matrix and with the increase of strain, the fraction of deformation twins increased.

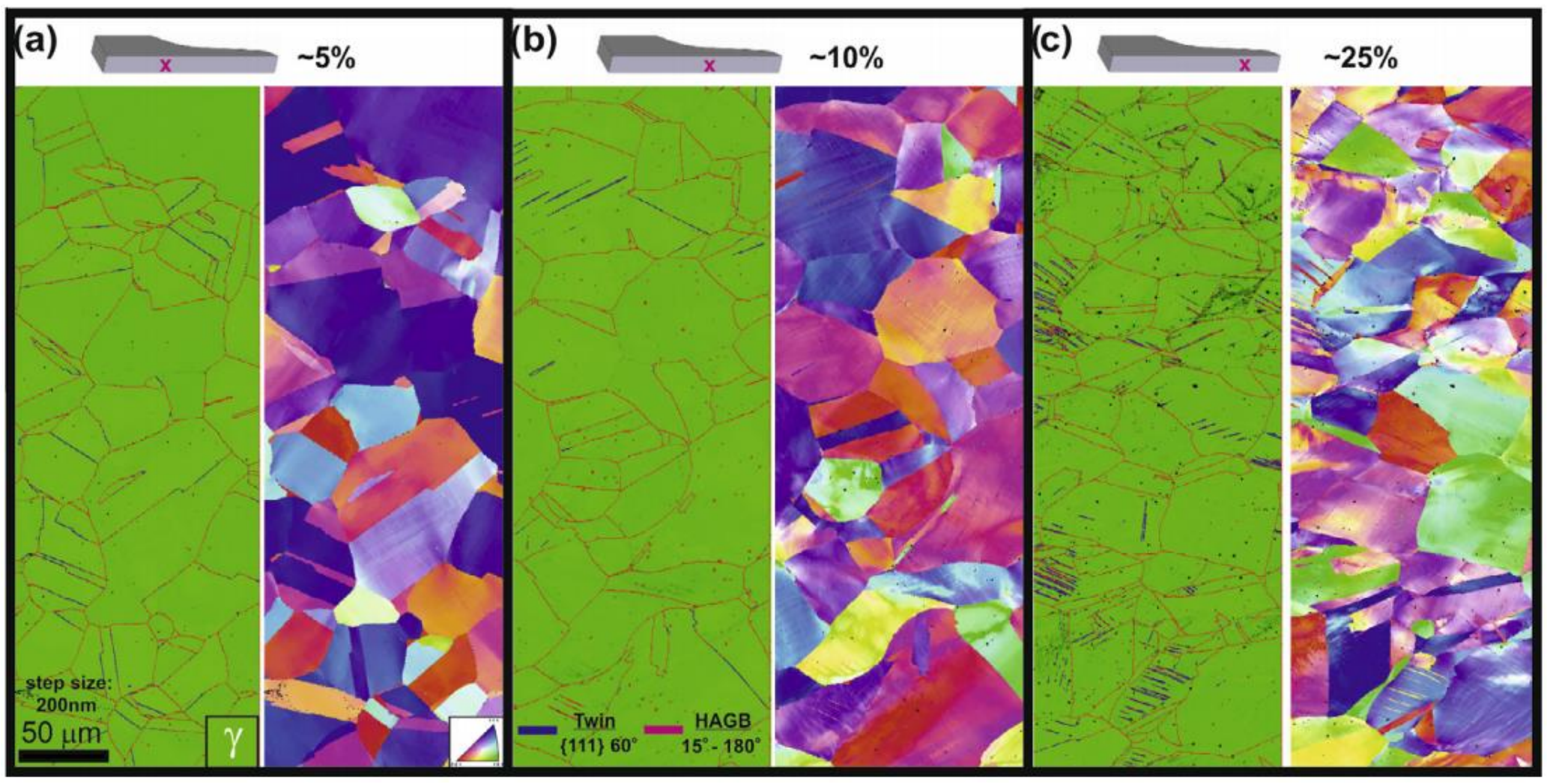

**Fig. 35.** EBSD and corresponding inverse pole figure (IPF) maps of the deformed $Fe_{40}Mn_{40}Co_{10}Cr_{10}$ HEA sample cross section tensioned to: (a) 5 % strain; (b) 10 % strain; and (c) 25 % strain at room temperature. Figures from Deng et al. [247].

The morphology of typical deformation twins in the $Fe_{40}Mn_{40}Co_{10}Cr_{10}$ HEA can be seen in Figs. 36(a)-(c), where the size of the deformation twin is on the nanometer scale. The observed twinning system is [112](111), which is very typical of the fcc alloys[248]. For the FeCoNiCrMn HEA, a large number of nano-twins in cryogenic-temperature deformation leads to the dynamic Hall-Petch effect that hinders the dislocation slip, resulting in excellent fracture toughness[23]. The results show that the number of deformation nano-twins decreases with decreasing the deformation temperature, leading to higher strengths and better ductility at sub room temperature.

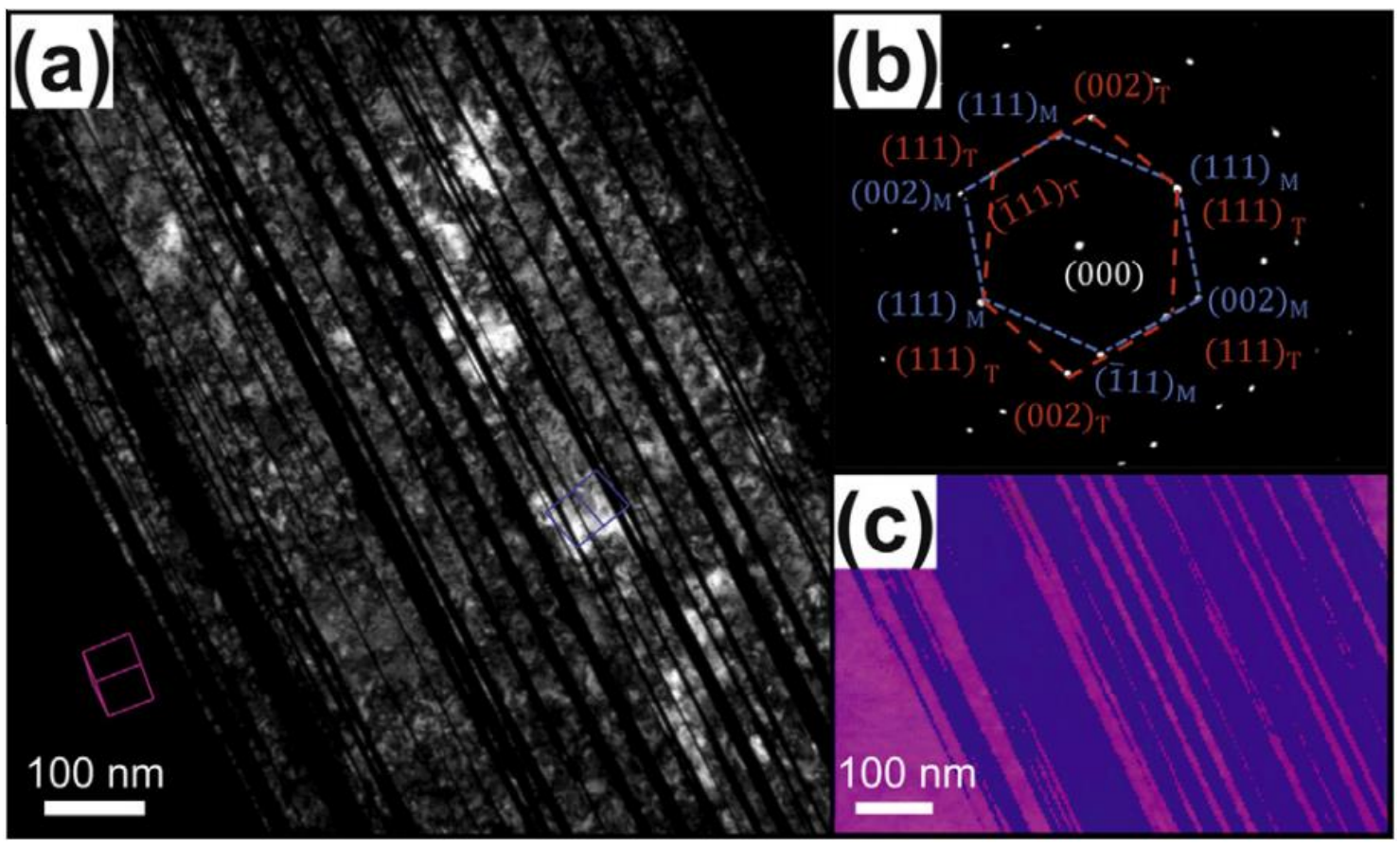

**Fig. 36.** The morphology of typical deformation twins in the $Fe_{40}Mn_{40}Co_{10}Cr_{10}$ HEA after testing at room temperature, as determined by (a) TEM dark-field image; (b) Selected-area diffraction (SAD) pattern; and (c) Corresponding map of the orientation and phase. Figures from Deng et al. [247].

Gludovatz *et al.* [27] reported that the NiCoCr MEA has excellent mechanical properties. For example, at room temperature, the UTS of the alloy can reach nearly 1 GPa, the fracture elongation is about 70 %, and the fracture toughness is as high as 275 $MPa\cdot m^{-1/2}$, which are, respectively, ≈ 15%, ≈ 30%, and ≈ 50% higher than the FeCoNiCrMn HEA at this condition. Based on this comparison, it can be said that the properties of the MEA are superior to that of the FeCoNiCrMn HEA. Furthermore, Laplanche *et al.*[143] found that the SFE of the NiCoCr MEA (22 ± 4 $mJ\cdot m^{-2}$) is lower than that of the FeCoNiCrMn HEA (30 ± 5 $mJ\cdot m^{-2}$)[143]. The results of TEM characterization revealed that during the deformation process, the nano-twinning occurred in the NiCoCr MEA, and was determined to be an additional deformation mechanism. The relatively-wide extended dislocations in the NiCoCr MEA can block the dislocation and promote the planar slip at the initial stage of the plastic deformation. Furthermore, the critical shear stress for the twinning of the NiCoCr MEA is close to that of the FeCoNiCrMn HEA, but the higher $\sigma_y$ and the work-hardening rate enable the NiCoCr MEA to reach the stress state at the lower strain level. During the deformation process, twins can be obstacles for the dislocation movement, and the interaction between twins and dislocations can induce a large strain-hardening rate, while the plastic deformation in the nanotwin-containing NiCoCr MEA can be transferred and homogenized more efficiently[175, 189, 194], resulting in the superior mechanical properties of the NiCoCr MEA.

## 7. Mechanical behaviors of HEAs at low and high temperatures

Lyu *et al.*[249] systematically summarized the mechanical behavior of HEAs at low temperatures. Figure 37 presents a comparison of the tensile $\sigma_y$ versus elongation for HEAs and other materials at temperatures of 77 and 110 K. Compared with the conventional Ti-6Al-4V titanium alloy and the high-Mn steels, HEAs exhibit comparatively-greater ductility, for example, the FeCoCrNi and CrMnFeCoNi HEAs shows comparable combination of strength and plasticity, suggesting that they could be potential candidates for cryogenic applications.

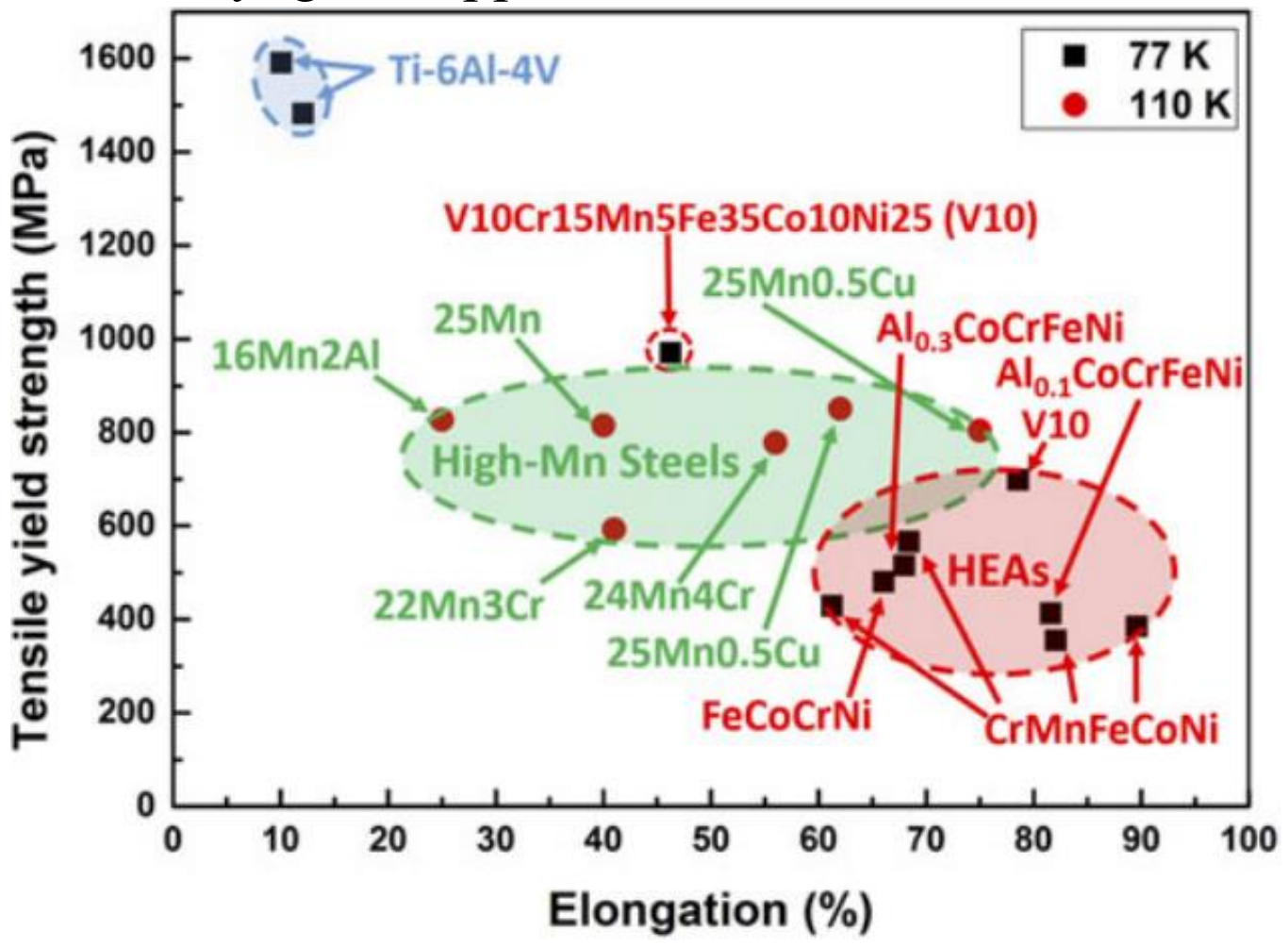

**Fig. 37.** Comparison of the tensile $\sigma_y$ versus elongation for HEAs and other materials at 77 and 110 K. Figure from Lyu et al. [154, 155, 249-253].

There are a large number of nano-twins involved in the deformation process of the FeCoNiCrMn HEA at low temperatures[23]. For instance, at the liquid-nitrogen temperature (77 K), the $\sigma_y$ and UTS increase by about 85% and 70%, thus reaching 759 MPa, and 1,280 MPa, respectively, when compared with the properties at 293K. Furthermore, the fracture elongation is more than 80%. The twin boundaries and the dislocations that reside in the twins can inhibit the dislocation slip, while the hardening rate does not decrease with the onset of deformation. Hence, the strength and plasticity increase with the decrease of temperature. Recently, Wang *et al.*[254] found that the TaNbHfZrTi HEA has an excellent combination of strength and ductility at the cryogenic temperature (77 K), in which the $\sigma_y$, UTS, and fracture elongation are 1,547 ± 11 MPa, 1,762 ± 22 MPa, and 15.2 ± 0.8%, respectively. Furthermore, the values for the same HEA tested at 277K are 875 ± 18 MPa, 994 ± 16 MPa, and 14.1 ± 1.3%, respectively, suggesting that the refractory TaNbHfZrTi HEA has great potential to be used in low-temperature extreme conditions.

Otto *et al.*[155] studied the influence of temperature on the tensile properties of the FeCoNiCrMn HEA. It was found that for a tensile strain rate of $1 \times 10^{-3}$ $s^{-1}$, the $\sigma_y$ and the UTS decrease with the temperature. He *et al.*[255] studied the high-temperature (*e.g.* 1,023 K, 1,048 K, 1,073 K, and 1,123 K) steady-state flow behavior of the FeCoNiCrMn HEA, as shown in Figs. 38. It is obvious that the steady-state flow stress depends positively on strain rates. Under a constant strain rate, the steady-state stress of the FeCoNiCrMn HEA decreases with the increase of temperatures. At a temperature of 1,073 K, the FeCoNiCrMn HEA significantly softened. For strain rates greater than $2 \times 10^{-5}$ $s^{-1}$, the deformation mechanism is dominated by the dislocation climb where the Cr and Mn elements diffuse directionally and form the second phase, resulting in the nucleation and growth of cracks in the region of the stress concentration and a corresponding decrease in the strength.

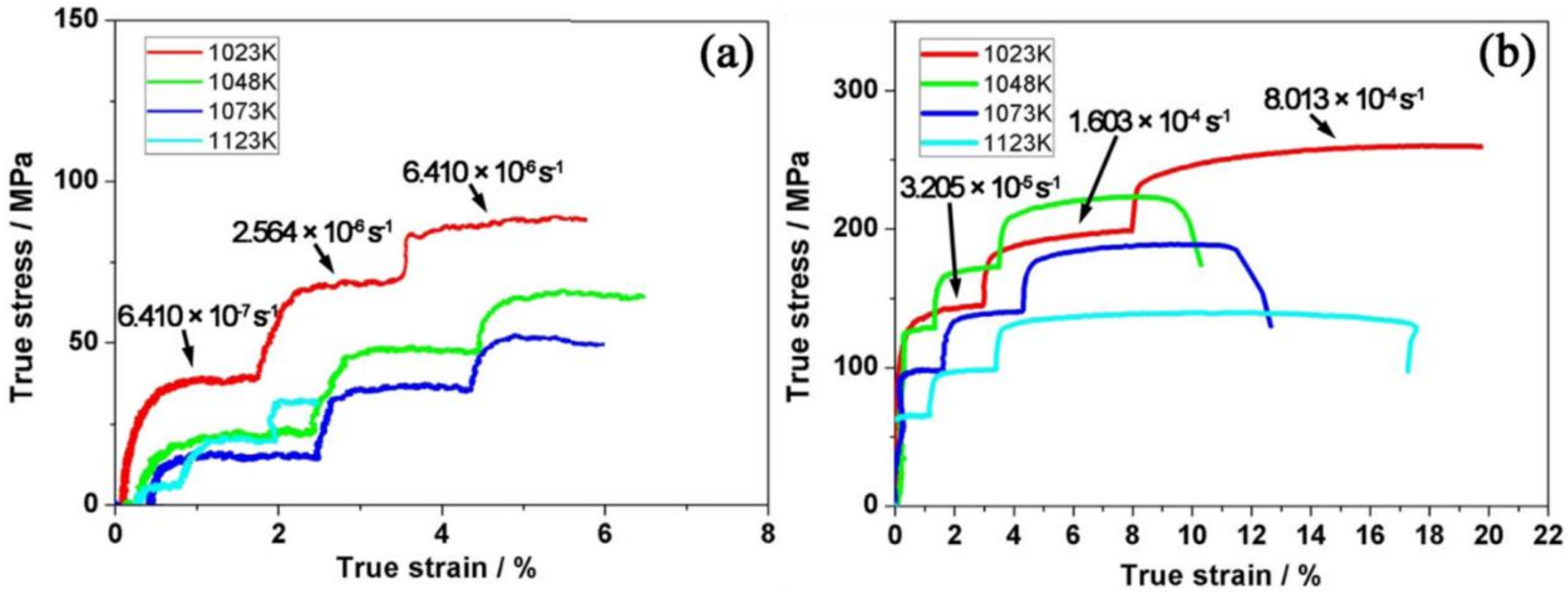

**Fig. 38.** True stress-strain curves at elevated temperatures of 1,023~1.123 K for (a) low strain rates of $6.410 \times 10^{-7}$ $s^{-1}$, $2.564 \times 10^{-6}$ $s^{-1}$, and $6.410 \times 10^{-7}$ $s^{-1}$ and (b) high strain rates of $3.205 \times 10^{-5}$ $s^{-1}$, $1.603 \times 10^{-4}$ $s^{-1}$, and $8.013 \times 10^{-4}$ $s^{-1}$ for the FeCoNiCrMn HEA. Figures from He et al. [255].

Compared with its room temperature strength, the strength of the TaNbHfZrTi HEA at elevated temperatures is markedly lower[164], suggesting that with an increase in the temperature, the strength decreases. When the temperature is lower than 873 K, the main deformation mode of the TaNbHfZrTi HEA is the dislocation slip that is supplemented by a small amount of deformation twinning. Since such a deformation process is uniform and continuous, its strength decreases slightly from 790 MPa at 673 K to 675 MPa at 873 K. When the temperature exceeds 1,073 K, the high diffusion rate promotes the crystallization at the grain boundary. In this scenario, the unstable subgrain boundary becomes the nucleation point of the crack, as well as a channel that allows crack growth. Consequently, the crack propagates rapidly, resulting in the instability of the material. Therefore, the strength of the TaNbHfZrTi HEA decreases sharply from 535 MPa at 1,073 K to 295 MPa when the temperature is 1,273 K. Furthermore, the $\sigma_y$ decreases to 92 MPa when the temperature reaches 1,473 K.

At high temperatures that are greater than the ductile-brittle transition temperature[256] (the exact value was not obtained, but surely it is above room temperature), the NbMoTaW HEA, which is another kind of the bcc-structured refractory alloy, exhibits brittle to ductile behavior[68]. The direction of cracking in the alloy is approximately 40° from that of the compression direction, while the fracture of the material was completed by shear. Such behavior indicates that the plasticity of the material was improved[68]. Based on the above discussion, refractory HEAs possess natural advantages that make them potential candidates for applications in ultra-high temperature environments (873 K to 1,873 K).

## 8. Fatigue, creep, and fracture properties of HEAs

For this section, some of the mechanical properties, *e.g.* fatigue[44-49, 257-259], creep[260-262], and fracture[50, 258, 263] of HEAs are briefly discussed. Lam *et al.*[49] improved the fatigue resistance of an FeCoNiCrMn HEA by overload-induced deformation twinning. Thurston *et al.*[48] studied the effect of temperature on the fatigue behavior of FeCoNiCrMn HEAs. With a decrease in the temperature from 293 to 198 K, the fatigue threshold stress intensity factor range, $\Delta K_{th}$, increases from 4.8 to 6.3 MPa·m$^{1/2}$, and the Paris exponent, *m,* increases from 3.5 to 4.5. The dominant fatigue mechanism changes from transgranular crack propagation at 293 K to intergranular failure at 198 K. The main factor that influences the fatigue-crack-growth behavior is the roughness-induced fatigue crack closure. Li *et al.*[47] summarized the fatigue behaviors of HEAs and compared them with some traditional alloys, as shown is Figs. 39(a)-(b). Figure 39(a) indicates that the range of fatigue strength and UTS of HEAs is located between those of Cu alloys and steels, while Fig. 39(b) indicates that the combination fatigue ratio and UTS of HEAs is between those of Cu alloys, Ti alloys, and steels.

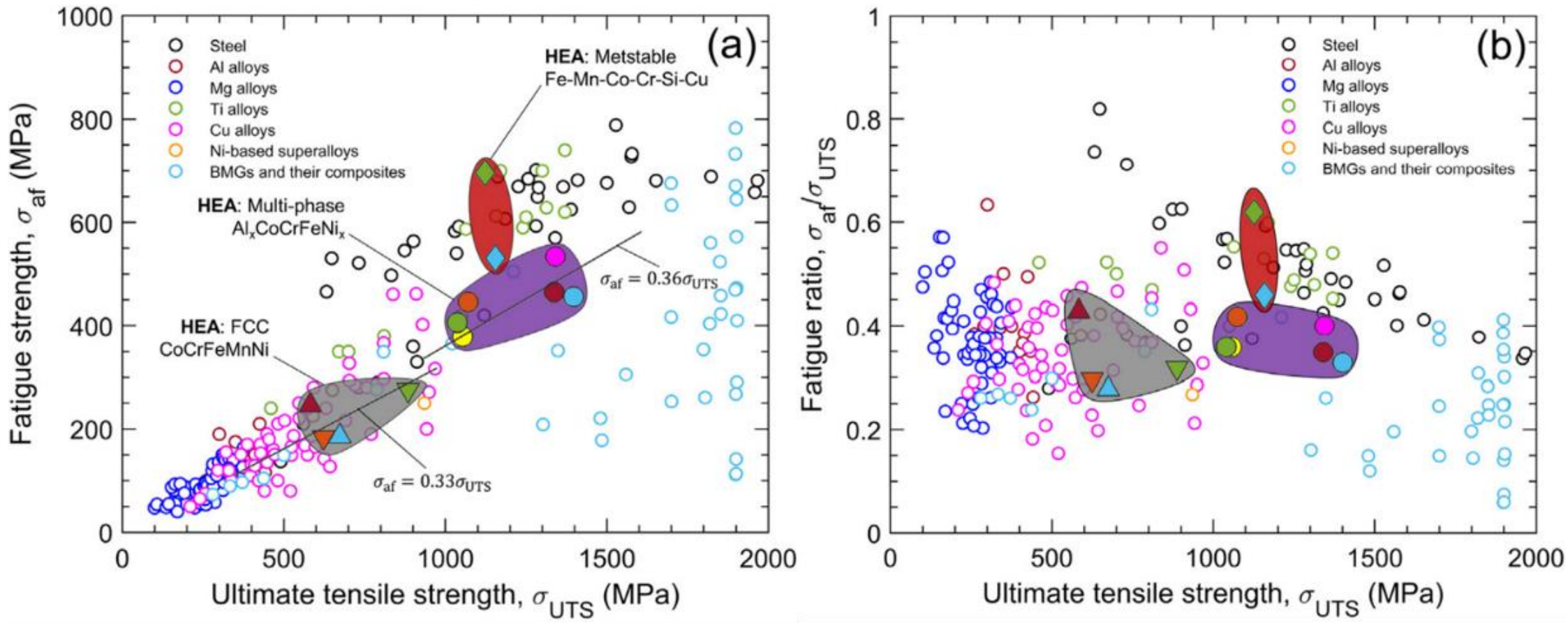

**Fig. 39.** (a) Fatigue strength-UTS and (b) fatigue ratio-UTS maps of HEAs and some traditional alloys. Figures from Li et al. [46, 47, 264-276].

To consider the crack-growth behavior, the linear-elastic fracture mechanics (LEFM)[277, 278] model was used. The fatigue-crack-growth-rate behavior of cracks was found to be longer than 1 mm[279] as shown in Fig. 40(a), where *da/dN* refers to the crack-growth rate, and $\Delta K$ refers to the applied stress intensity factor range. The *da/dN* - $\Delta K$ curve is bounded by the threshold stress intensity factor range ($\Delta K_{th}$) and the critical cyclic fracture toughness ($\Delta K_c$). The curve can be divided into three regions, where in region I, the crack growth rate is asymptotical and close to 0 when $\Delta K$ is approaching $\Delta K_{th}$. Furthermore, $\Delta K_{th}$ is determined by the fatigue limit ($\sigma_{af}$), which means that below this value, the crack growth will not occur, as expressed in the following equation[279, 280]:

$$\sigma_{af} = \frac{\Delta\sigma}{2} = \gamma \frac{\Delta K_{th}}{2\sqrt{\pi a}} \tag{8.1}$$

where $\gamma$ is a scaling factor, $\Delta\sigma$ is the applied stress range, and $a$ is the crack length. In region II, the crack-growth rate follows the Paris’ crack-growth law[281, 282]:

$$\frac{da}{dN} = A(\Delta K)^m = A\left(Y\Delta\sigma\sqrt{\pi a}\right)^m \tag{8.2}$$

where $A$ and $m$ are material constants, $Y$ is a geometry factor related to the loading and cracked body configuration. In region III, the $\Delta K$ has a limit of the critical cyclic fracture toughness, $\Delta K_c$.

The experimental crack-growth rate versus the stress intensity factor range for cracks longer than 1mm in HEAs and some conventional metals are presented in Fig. 40(b). The $\Delta K_{th}$ value of the metastable HEAs and two duplex $Al_xCrFeNi_y$ HEAs are higher than that of the fcc-structured CoCrFeMnNi HEA, and much greater than that of the bcc-structured HfNbTaTiZr HEA. This result indicates that the fatigue resistance is in the following order: metastable HEA > duplex HEA > fcc-structured HEA > bcc-structured HEA. When comparing HEAs with some conventional alloys, the results indicate that the fatigue resistance of fcc and bcc HEAs are higher than the Al alloy, Mg alloy, and bulk metallic glass, but lower than those of the steel, Ni-based superalloy, Cu, and Ti alloy. On the other hand, the values of the metastable and duplex HEAs are higher than that of the conventional metals, except the Cu and Ti alloys.

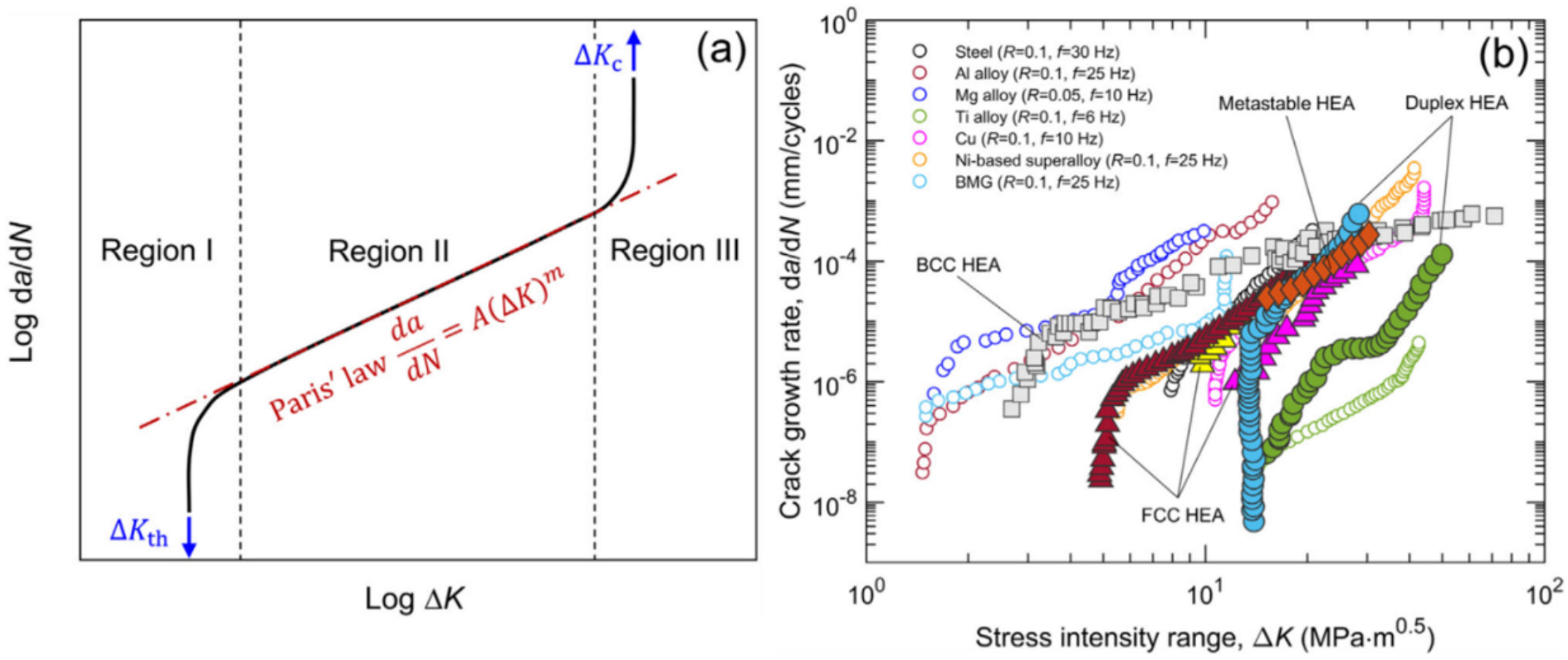

**Fig. 40.** (a) The schematic logarithm of crack-growth rate versus the stress intensity factor range for cracks longer than 1mm. (b) The experimental crack-growth rate versus the stress intensity factor range for cracks longer than 1mm of HEAs and some conventional metals. Figures from Li et al. [47, 48, 258, 283-292].

The fatigue mechanisms of HEAs with different crystal structures are displayed in Figs. 41(a)-(d). Figure 41(a) shows that the fatigue-cracks tend to initiate along slip lines or at the deformation twin boundaries for the fcc-structured HEAs[293-295]. As can be seen in Fig. 41(b), for the refractory bcc-structured HEAs, the cracks initiate at grain boundaries and then propagate inside the grains. Such behavior results in the low fatigue limit and high crack-growth rate of this kind of HEAs[296]. For the metastable HEAs [Fig. 41(c)], in which the fcc phase transforms into the hcp phase, the crack blunting occurs in the fcc phase. Subsequently, the crack branching and deflection takes place at the twin boundary in the hcp phase, resulting in the high fatigue resistance[265]. The fatigue mechanism of multi-phase HEAs can be observed in Fig. 41(d), where the cracks initiate at the persistent slip bands and phase boundaries. If there were some nano-precipitates in the HEA, the cracks will deflect at particles, triggering tortuous crack propagation, which is indicative of the increased fatigue resistance[267, 297, 298].

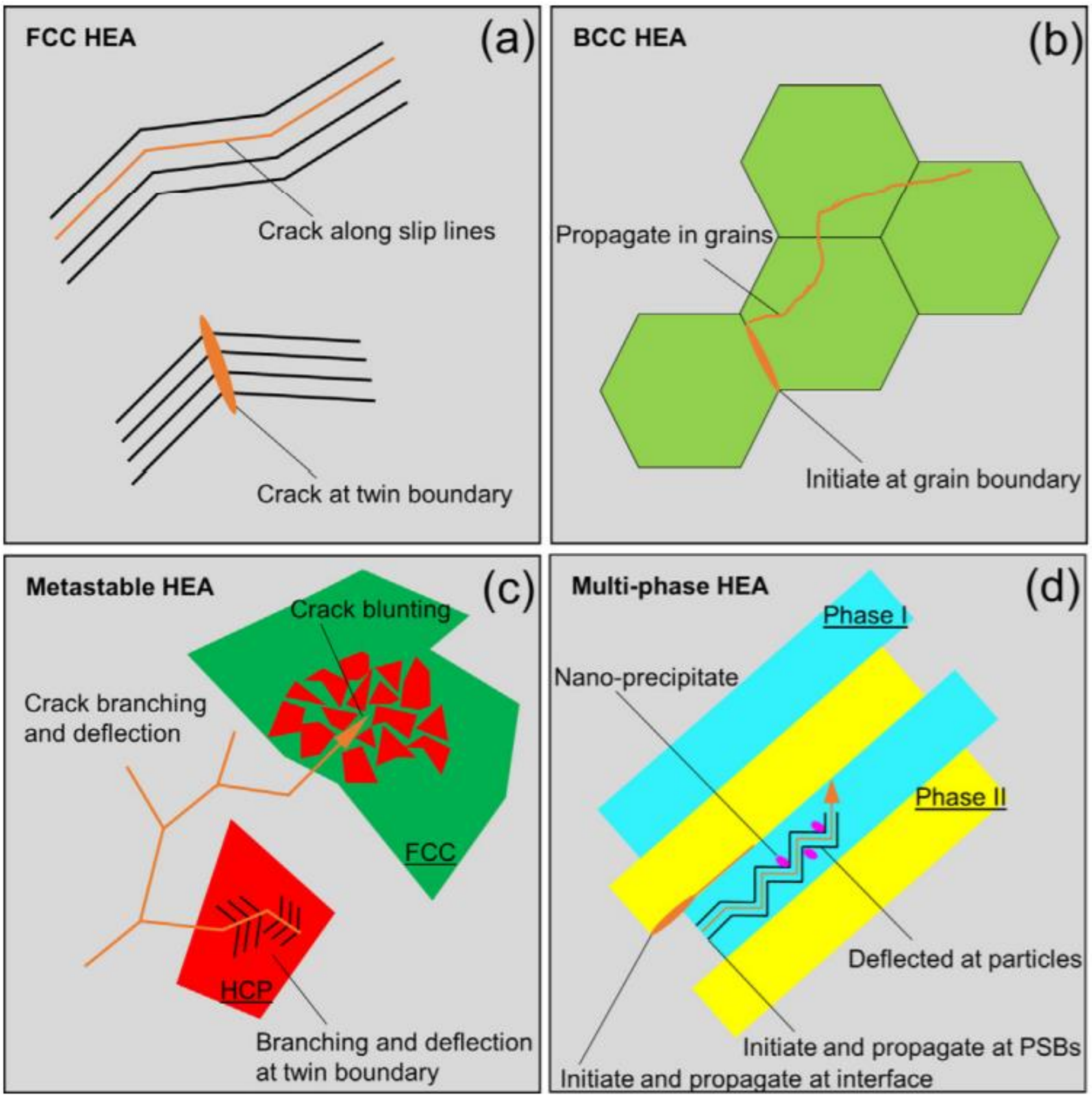

**Fig. 41.** The fatigue-mechanism schematic illustrations of (a) fcc-structured; (b) bcc-structured; (c) metastable; and (d) multi-phase HEAs. Figures from Li et al. [47].

Recently, Xie *et al.*[261] found that the NiCoCr MEA exhibited superior tensile creep behavior, as compared to an FeCoNiCrMn HEA and comparable creep behavior to some commercial steels [as shown in Figs. 42(a)-(b)], at the temperature of 973 K. It was determined that the solid-solution strengthening effect and low SFE were the primary factors in improving the creep properties. Finally, the plate-like Cr-rich $\sigma$ phases were observed at the grain boundaries during the creep process, suggesting that during the creep test, the decomposition phenomenon occurs in the NiCoCr alloy.

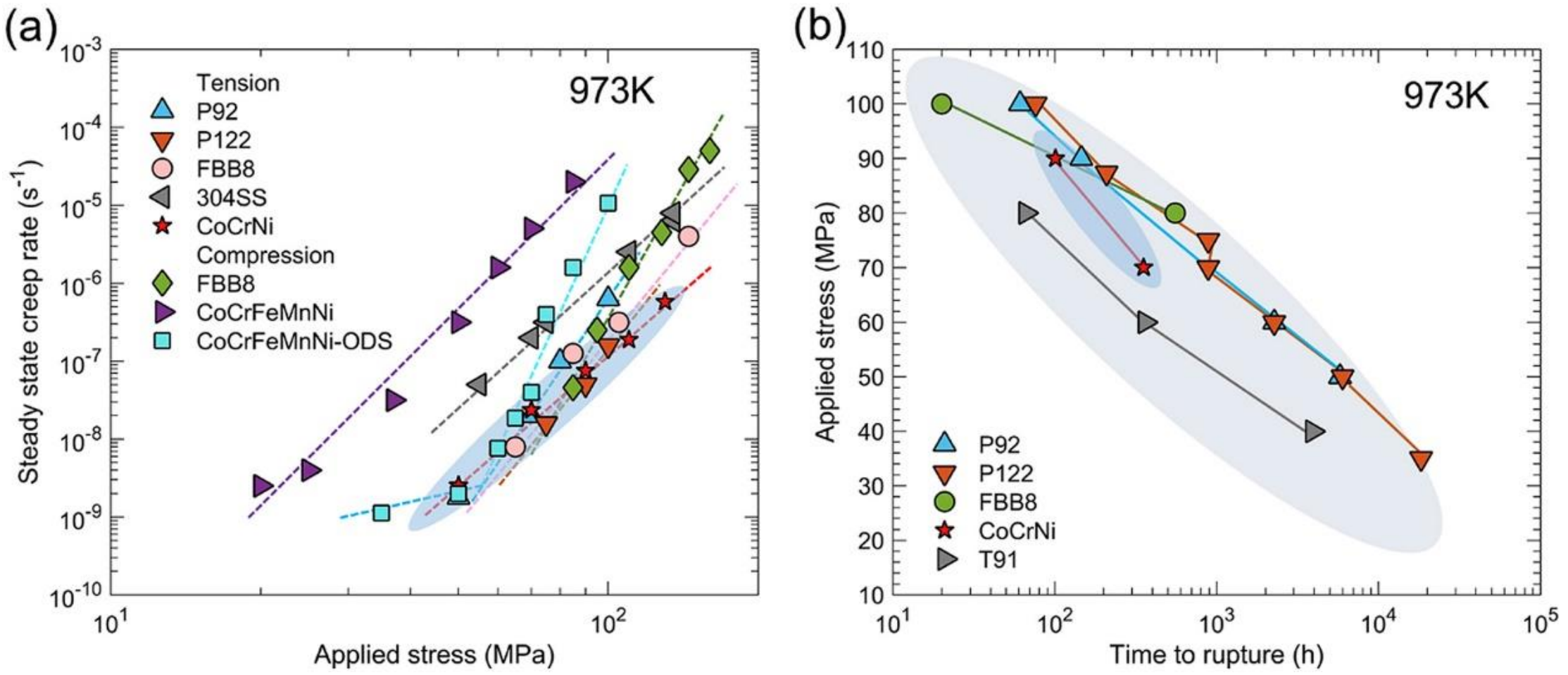

**Fig. 42.** The tensile-creep behavior of the NiCoCr MEA, FeCoNiCrMn HEA, and commercial steels at 973 K: (a) the relationship between the steady-state creep rates and applied stress (ranging from $10^1$ to $10^2$ MPa); and (b) the relationship between the applied stress and time to rupture (ranging from $10^1$ to $10^5$ h). Figures from Xie et al. [261, 299-304].

Fracture toughness is an important metric for measuring the fracture resistance of a material. For example, Gludovatz *et al.* studied the fracture toughness of FeCoNiCrMn HEA[23] and NiCoCr MEA[305]. For the FeCoNiCrMn HEAs, the average grain size was approximately 6 μm, while at room temperature, the crack-initiation fracture toughness $K_{JIC}$ was 217 MPa·m$^{1/2}$. Moreover, the crack-growth fracture toughness, $K$, was greater than 300 MPa·m$^{1/2}$ after a crack propagation of about 2 mm. When lowering the temperature to 200 K and 77 K, the $K_{JIC}$ values were 221 and 219 MPa·m$^{1/2}$, respectively. For the NiCoCr MEA with grain sizes of 5 ~ 50 μm, the $K_{JIC}$ value was 208 MPa·m$^{1/2}$, and the $K$ value increased to 290 MPa·m$^{1/2}$ when the crack was extended to 2 mm. When decreasing the temperature to 77 K, the $K_{JIC}$ and $K$ values increased by 31% and 48%, respectively. The high fracture toughness results from the microvoid coalescence, the strain hardening, and deformation nanotwinning. Nanotwinning is the predominant factor for the enhanced fracture toughness at low temperature of both the FeCoNiCrMn and NiCoCr alloys[23, 305].

Li *et al.*[263] summarized the fracture resistance of some reported HEAs and compared these values with those of some conventional materials. The results are plotted in Fig. 43. The fcc HEA has better fracture resistance than the fcc + bcc mixture and bcc HEAs. HEAs have superior fracture resistance when compared with foams, polymers and elastomers, glasses, technical ceramics, non-technical ceramics, as well as some traditional metals and alloys (such as Pd alloys, Mg alloys, Zn alloys, Al alloys, Cu alloys, and Ni alloys). While the HEAs have comparable fracture resistance to stainless steels, low alloy steels, cryogenic steels, Ni-based superalloys, bulk metallic glasses, W alloys, carbon steels, Ti alloys, and cast irons, which indicates that HEAs have great potential to be used as structural materials.

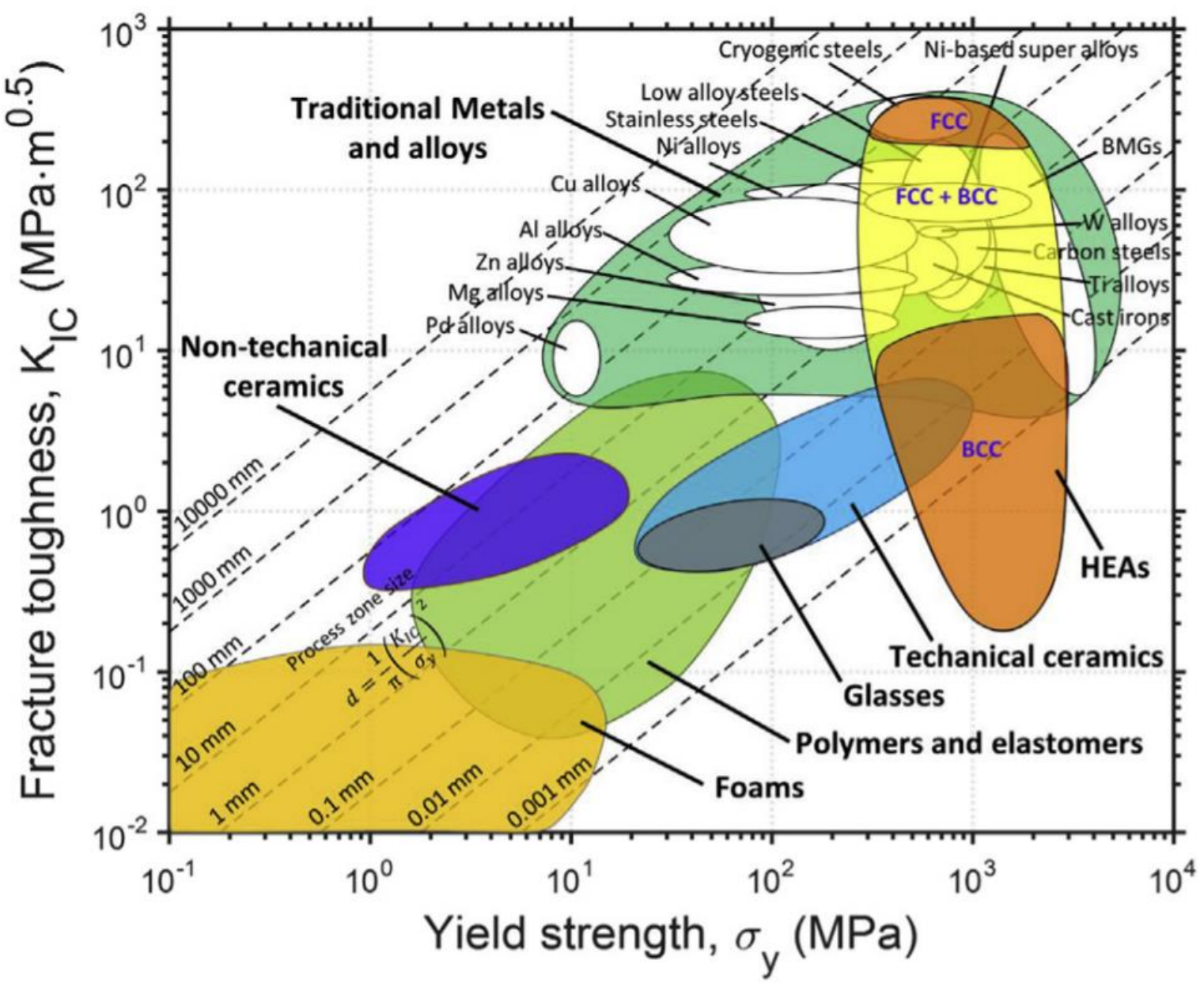

**Fig. 43.** The fracture toughness versus yield strength maps of HEAs and some conventional materials. Figure from Li et al. [263].

# 9. Future prospects

## 9.1. Lightweight HEAs

At present, structural materials tend to be lightweight, and the typical high density of HEAs limits their applications in certain fields, such as the aerospace industry. In order to broaden the application range of HEAs, more and more emphasis has been placed on the production of lightweight HEAs[162, 306-312]. Hence, the new challenge is to develop lightweight HEAs with excellent mechanical properties.

Feng *et al.*[306] assessed some existing solid-solution phase-formation rules to design new lightweight HEAs. The results show that for lightweight HEAs, the identified criteria are: $\phi_c \geq 7$, $\delta_r \leqslant 4.7$ %, and $\Delta H_{mix} \geqslant -16.25$ kJ·mol$^{-1}$, where $\phi_c$ is the critical single-parameter value needed to separate the single phase from the multi-phase, $\delta_r$ is the atomic-radius difference, and $\Delta H_{mix}$ is the enthalpy of mixing. The Al-containing HEAs typically undergo strong interatomic interactions among the constituent elements. Hence the empirical rules are not affected the enthalpy change for the solid-solution phase, leading to a severe underestimation of the free energies for the solid-solution phases. Moreover, the Al-containing HEAs typically have densities greater than 5.5 g·cm$^{-3}$, which are still higher than that of the Ti alloys, *e.g.* 4.42 g·cm$^{-3}$ of the Ti-6Al-4V alloy[313]. Thus, future research should pay attention to increasing the amount of these lightweight elements in the matrix, such as Al, Ti, and Mg.

## 9.2. HEA films and coatings

Nowadays, the high cost of constituent elements such as Ni, Co, Cr, Nb, and W, have limited the industrial applications of HEAs. Hence, using the HEA films and coatings on low-cost metal substrates are widely studied[314-321]. Moreover, the mechanical properties of HEA films and coatings are also of great importance for its serviceability. The widely-used technologies implemented to fabricate HEA films and coatings include sputter deposition[322-324], electrochemical deposition[325, 326], laser cladding[327-329], plasma-transferred-arc cladding[330], and spraying[331-333].

As an example, Figs. 44 display the typical cross-sectional microstructures of FeCoNiCrCu and $FeCoNiCrCuAl_{2.5}$ HEA films prepared by sputter deposition[334]. From the TEM results in Figs. 44(a)-(b), we can see that the FeCoNiCrCu HEA film exhibits an fcc structure. From the TEM results in Figs. 44(c)-(d), we can find that the $FeCoNiCrCuAl_{2.5}$ HEA film consists of a bcc structure, indicating that the structure of the HEA film is the same as that of the bulk HEA materials, whereas the SAD patterns help determine the phase. Once the HEA films and coatings are produced, they have attracted extensive attention. However, since the preparation processes of HEA films and coatings are not well developed, their applications are currently limited. However, it is expected that an improvement in the preparation methods of HEA films and coatings will further enhance their properties. For example, the properties of HEA films and coatings can also be improved by adding trace elements[335, 336] or implementing specific heat treatments[337, 338]. Furthermore, it is anticipated that such processes will also reduce their production cost. Thus, it is expected that successful implementation of these methods will make large-scale applications of HEA films and coatings possible.

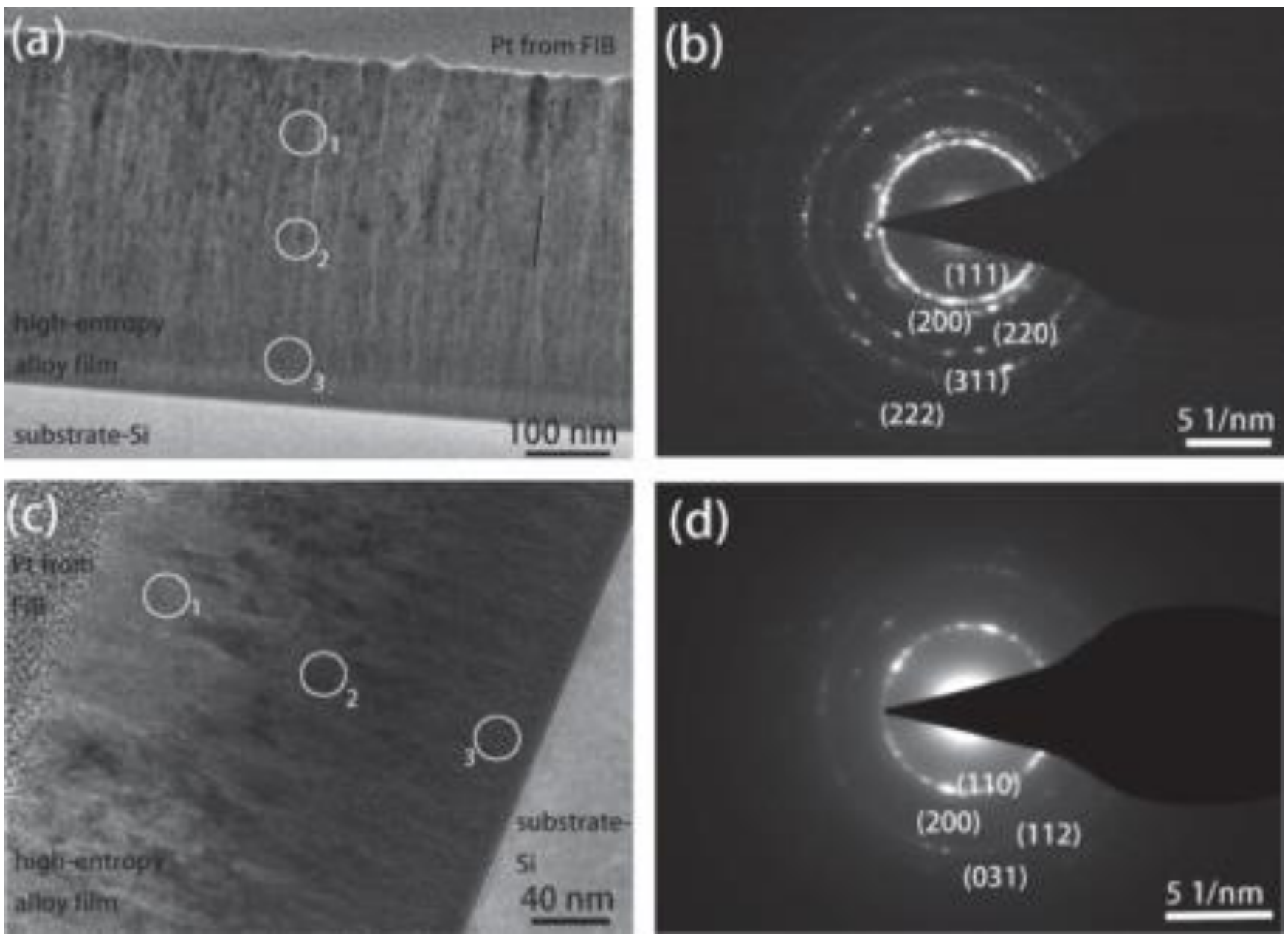

**Fig. 44.** (a) Bright-field TEM image and corresponding (b) SAD pattern of the FeCoNiCrCu HEA film; (c) Corresponding bright-field TEM image; and (d) SAD pattern of the $FeCoNiCrCuAl_{2.5}$ HEA film. Figures from Wu et al. [334].

### 9.3. Additive manufacturing

More recently, additive manufacturing (AM)[339, 340] methods have been used to fabricate the materials via 3D models and layering techniques. The precise control of the shape and the rapid solidification rate make this method a possible avenue for the practical applications of HEAs. This technology consists of three main techniques, namely direct laser deposition (DLD), selective laser melting (SLM), and selective electron beam melting (SEBM). Joseph *et al.*[341] compared the microstructures and mechanical properties of $FeCoNiCrAl_x$ HEAs fabricated by DLD and arc melting. Figure 45(a) presents the EBSD maps, and Fig. 45(b) shows the high-magnification SEM pictures of as-cast $FeCoNiCrAl_x$ HEAs fabricated by DLD and arc melting. The results indicate that the $FeCoNiCrAl_{0.3}$ and $FeCoNiCrAl_{0.85}$ alloys exhibit a <001> fiber texture, and the $FeCoNiCrAl_{0.6}$ has a coarse dendritic-grain structure. The particle size is larger in the DLD samples, but the morphology is the same as arc melting, which indicates that DLD is a successful technique for the fabrication of HEAs.

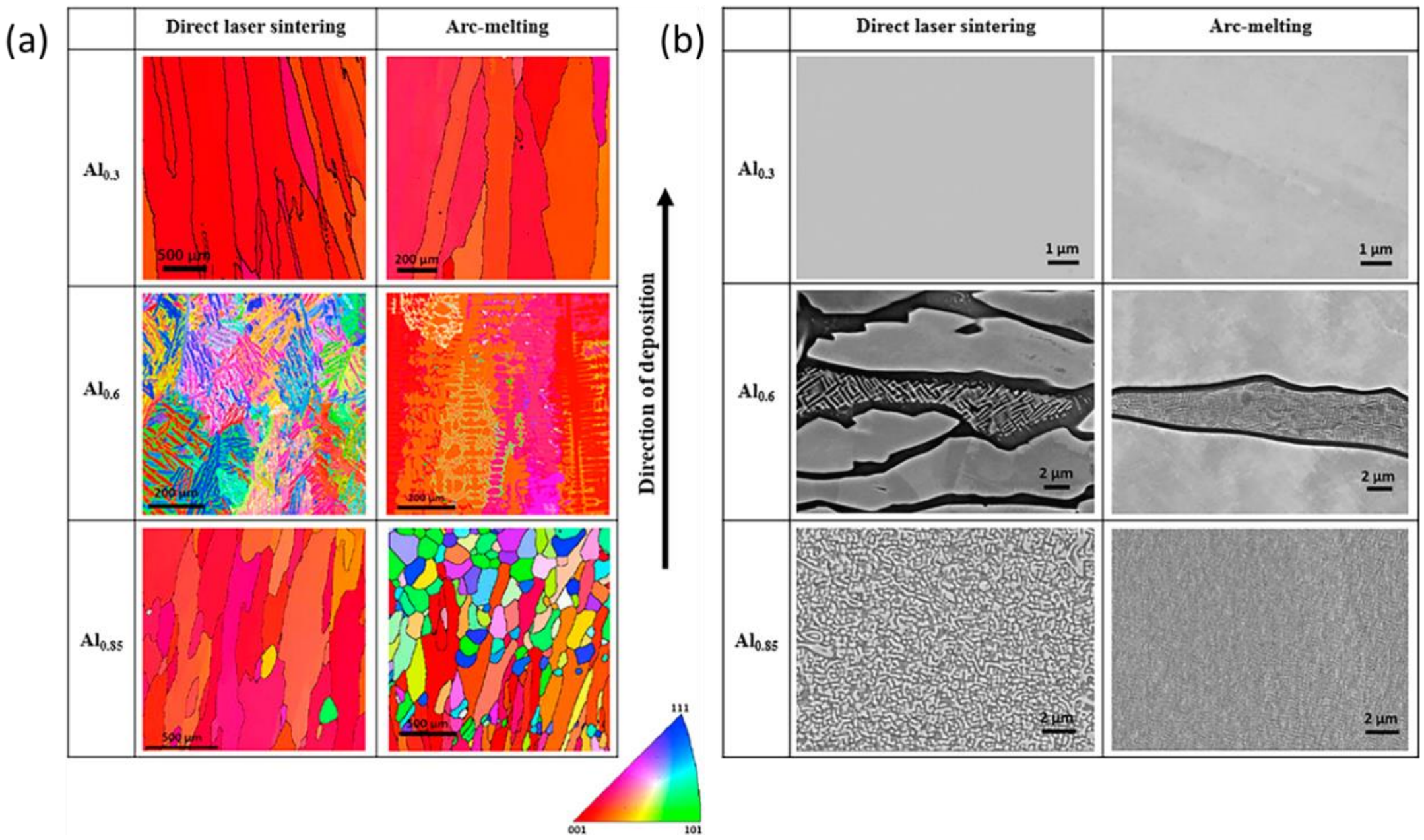

**Fig. 45.** (a) EBSD maps and (b) high-magnification SEM pictures of $FeCoNiCrAl_x$ alloys fabricated by DLD and arc melting, respectively. Figures from Joseph et al. [341].

Brif *et al.*[342] used the SLM method to prepare the FeCoNiCr HEA. Both the XRD and EDX results show that the material is single-phased and solid-solutionized. The FeCoNiCr HEA prepared by AM with a laser spot size of 50 μm has a $\sigma_y$ of 402 MPa, ultimate tensile strength of 480 MPa, and fracture elongation of 8%, while the values for the FeCoNiCr HEA prepared by additive manufacturing with a laser spot size of 25 μm are 600 MPa, 745 MPa, and 32%, respectively. The combination of the tensile properties for these two additively-manufactured HEAs is superior to that of the FeCoNiCr HEA that is prepared by an arc melter (of which the $\sigma_y$, ultimate tensile strength, and fracture elongation are 188 MPa, 457 MPa, and 50%, respectively), indicating that using SLM for manufacturing HEAs is possible.

Afterwards, Kenel *et al.*[343] applied the 3D ink-extrusion printing method to develop FeCoNiCr HEA micro-lattices. Figure 46(a) shows the tensile properties of single filaments sintered in flowing $H_2$ at 1,573 K for 1 h and then tension tested at 293 K and 130 K, respectively. The inserted SEM pictures feature the fracture morphology (scale bar = 1 μm) and the cross-section (scale bar = 25 μm) of the filaments. Figure 46(b) displays the compressive behavior of 0/90° cross-ply micro-lattices that were sintered at 1,573 K 1 h from 25% (extruded with a 200 μm nozzle) to 63% (250 μm nozzle) relative density, and tested at 293 K and 77 K. The inserted picture shows the sample tested at 77 K and compressed to 50 %, which suggests that the deformation during compression is uniform. Additionally, the results indicate that the mechanical properties of the 3D ink-extrusion-printed HEAs are excellent at low and room temperatures, which offer some insight into the additive manufacturing of HEAs.

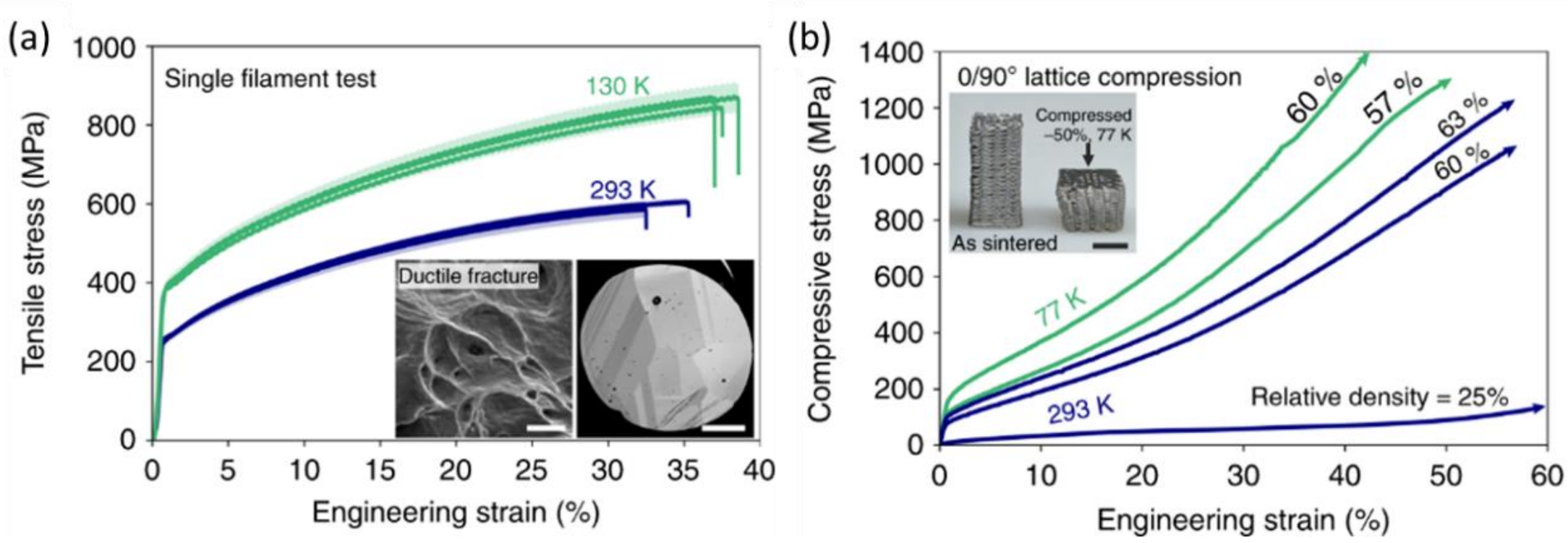

**Fig. 46.** (a) EBSD maps and (b) High-magnification SEM pictures of $FeCoNiCrAl_x$ alloys fabricated by DLD and arc melting, respectively. The inset SEM images in (a) are the fracture morphology (scale bar = 1 μm) and the cross-section (scale bar = 25 μm) of the filaments. The inset image in (b) shows the as-sintered sample tested at 77 K and compressed to 50 %. Figures from Kenel et al. [343].

Recently, Peng *et al.*[344] combined the direct ink writing and thermal sintering methods and successfully synthesized the 3D-architectured FeCoNiCrMn HEAs. Previous experiments revealed that the 3D-architectured FeCoNiCrMn HEAs have the superior energy-absorption property. For example, it was determined that for a density of 6 $g \cdot cm^{-3}$, the energy absorption per unit volume (upon densification) for the alloy was 333 $MJ \cdot m^{-3}$. As a comparison, values for the Ti-6Al-4V-architected material and stainless-steel foam are only 116 $MJ \cdot m^{-3}$ and 2 $MJ \cdot m^{-3}$, respectively. The deformation mode of the 3D-architectured FeCoNiCrMn HEAs is bend-dominated, while the microstructure is homogenous. Thus, the strain hardening during deformation is pronounced, indicating that 3D-printing architected materials can shed some lights on exploring new material fabrication methods.

## 9.4. HT method

Producing HEAs with advantageous combinations of high strength and excellent ductility has always been a major goal in terms of the mechanical properties of this alloy system[172, 174, 345-349]. Hence, many new techniques have been used to design HEAs with these desirable properties. Recently, the HT method has been used to develop HEAs with the aforementioned properties[350-357]. This method[60] is used to design materials by performing many experiments or calculations at once and is low-cost and time-efficient when compared with the traditional experimental alloy-design strategy. Based on the combination of quantum-mechanical-thermodynamic methods and techniques involving the database construction and intelligent data mining, the HT method gives a new research direction in the materials-science field[60, 358]. The aim of the HT material design is to find new materials by developing the current supercomputer architecture, and then generating, managing, and analyzing the database.

For example, Shi *et al.*[359] successfully synthesized nanocrystalline $Al_x(CoCrFeNi)_{100-x}$ (where x = 4.5 to 40) by an HT sputter deposition method. With the increase of the Al content, the crystal structures of HEAs changes from fcc to bcc phases. At the same time, the general and localized corrosion resistance decrease, which results from the weak compactness and protection of the oxide film with higher Al concentration. Li *et al.*[360] successfully used the laser processing as an HT method to study the relationships among the microstructure, processing, and properties of $Al_xCoCrFeNi$ (x = 0.51 ~ 1.25) HEAs. Their results indicated that HT laser processing could be one possible method to reduce the time and cost for preparing HEAs with a wide range of composition concentration.

Troparevsky *et al.*[351] predicted the formation of single-phase HEAs using first-principles HT-DFT calculations, which is based on the conception of enthalpy (see Fig. 47). The value of the enthalpy of formation ($\Delta H_f$) for Fe, Co, Ni, Cr, and Mn binaries is relatively small, less than 100 meV·atom$^{-1}$, favoring the single-phase formation. On the other hand, the V, Ti, and Mo binaries have negative $\Delta H_f$. Therefore, they tend to form ordered intermetallics. Finally, the Cu binaries have positive $\Delta H_f$, leading to the instability of the alloys.

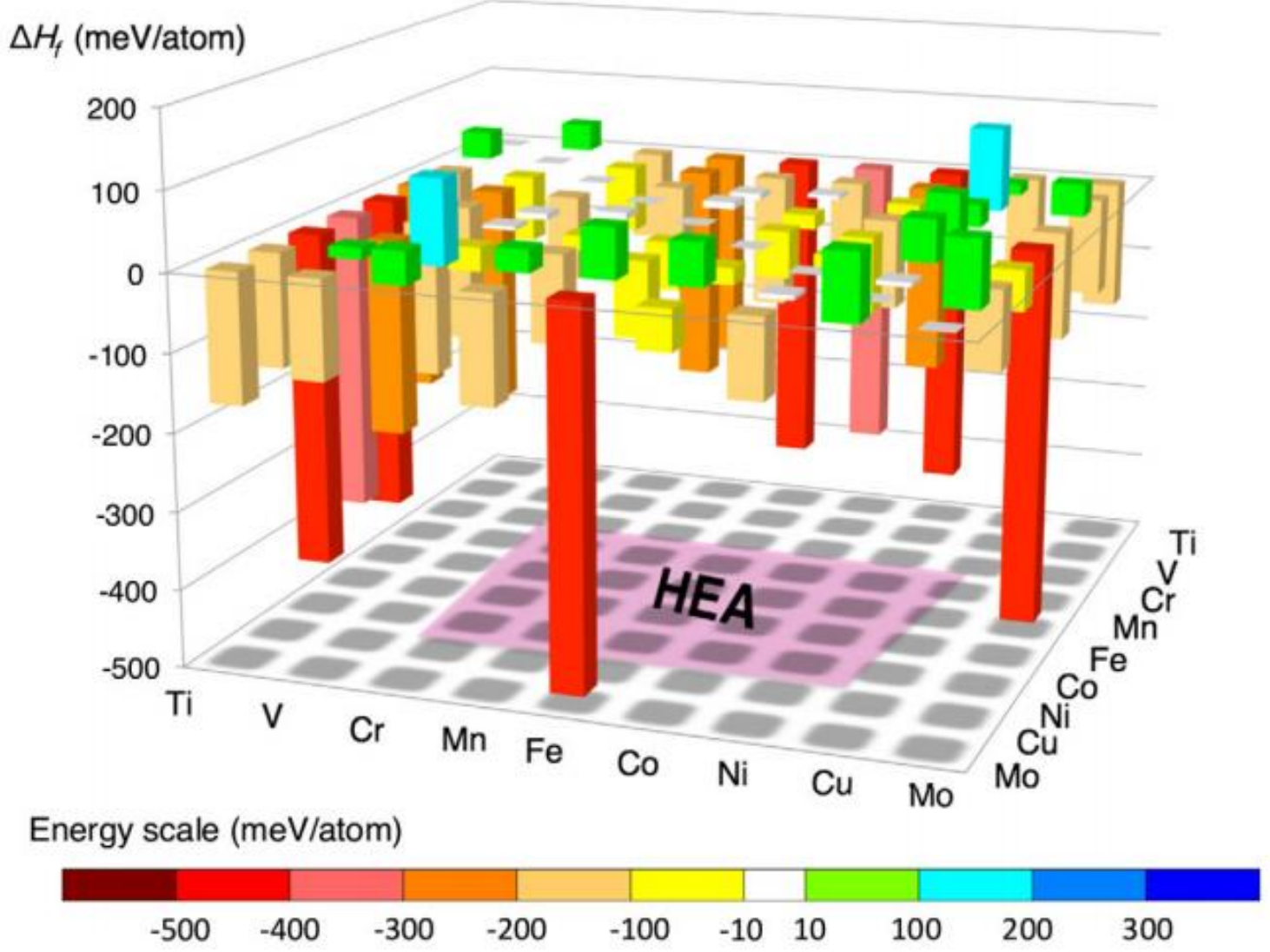

**Fig. 47.** Enthalpies needed to form the binary compounds based on the experimental alloys. Figure from Troparevsky et al. [351].

The HT-modeling method is also used to characterize solid-solution strengthening effects[352], search for new HEAs[350], and predict the multiphase evolution in Al-containing lightweight HEAs[361]. Miracle *et al.*[362] used the HT theory to explore new structural materials which can advance the design of new structural HEAs. This particular application involves a new testing procedures to evaluate these materials, which is illustrated in Fig. 48. Stage 0 represents the evaluation stage, in which the microstructures are determined by the HT calculations of phase diagrams. Stage 1 consists of changing the composition gradients and measuring the properties that are not affected by the microstructure. Finally, Stage 2 corresponds to the evaluation of properties that are sensitive to both microstructures and compositions, and is based on the change of microstructures and composition gradients.

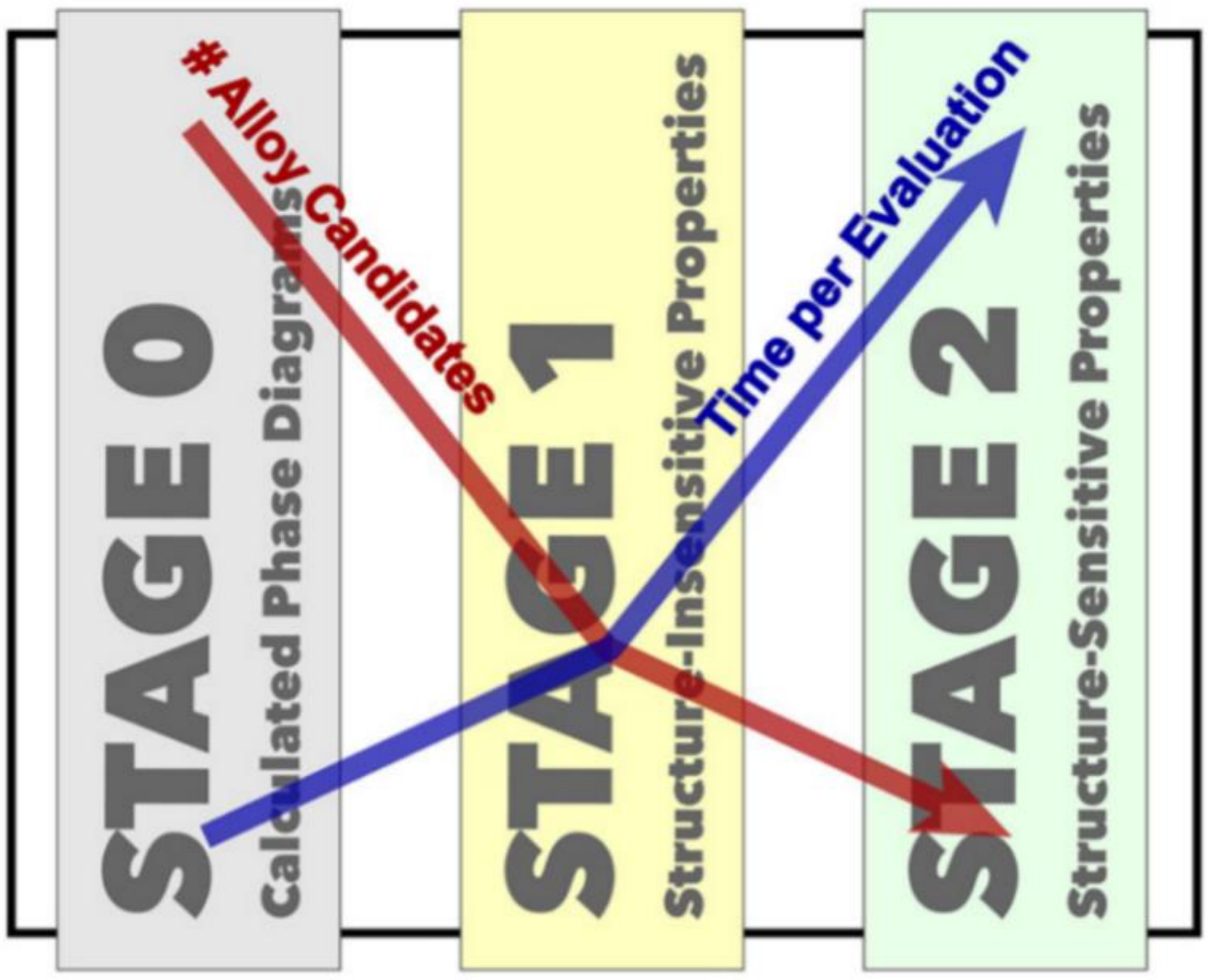

**Fig. 48.** Illustrations of the HT method to search for new structural materials. Figure from Miracle et al. [362].

Xu *et al.*[354] investigated the *in-situ* HT synthesis of HEAs by TEM, for which the results can be seen in Figs. 49(a)-(e). Figures 49(a)-(c) show the melting process of the Al element in the $FeCoNiCrCuAl_x$ (x = 0.75 and 1.05) HEA. Figure 49(d) presents the EDS result of the circled region in Fig. 46(b), which presents that the composition is close to $FeCoNiCrCuAl_{0.75}$. Figure 49(e) exhibits the high-resolution TEM (HRTEM) image of the joint, which confirms that the alloy is fcc-structured. The inset figure in Fig. 49(e) is the corresponding fast Fourier transform (FFT) pattern under the [011] zone, confirming the fcc-structure of the synthesized $FeCoNiCrCuAl_{0.75}$ HEA. Based on the results it can be said that the FeCoNiCrCuAl0.75 HEA with a single fcc structure and the $FeCoNiCrCuAl_{1.05}$ HEA with a mixed fcc + bcc structure were successfully obtained. Thus, it can be said that their method exhibits an effective and new way to synthesize HEAs.

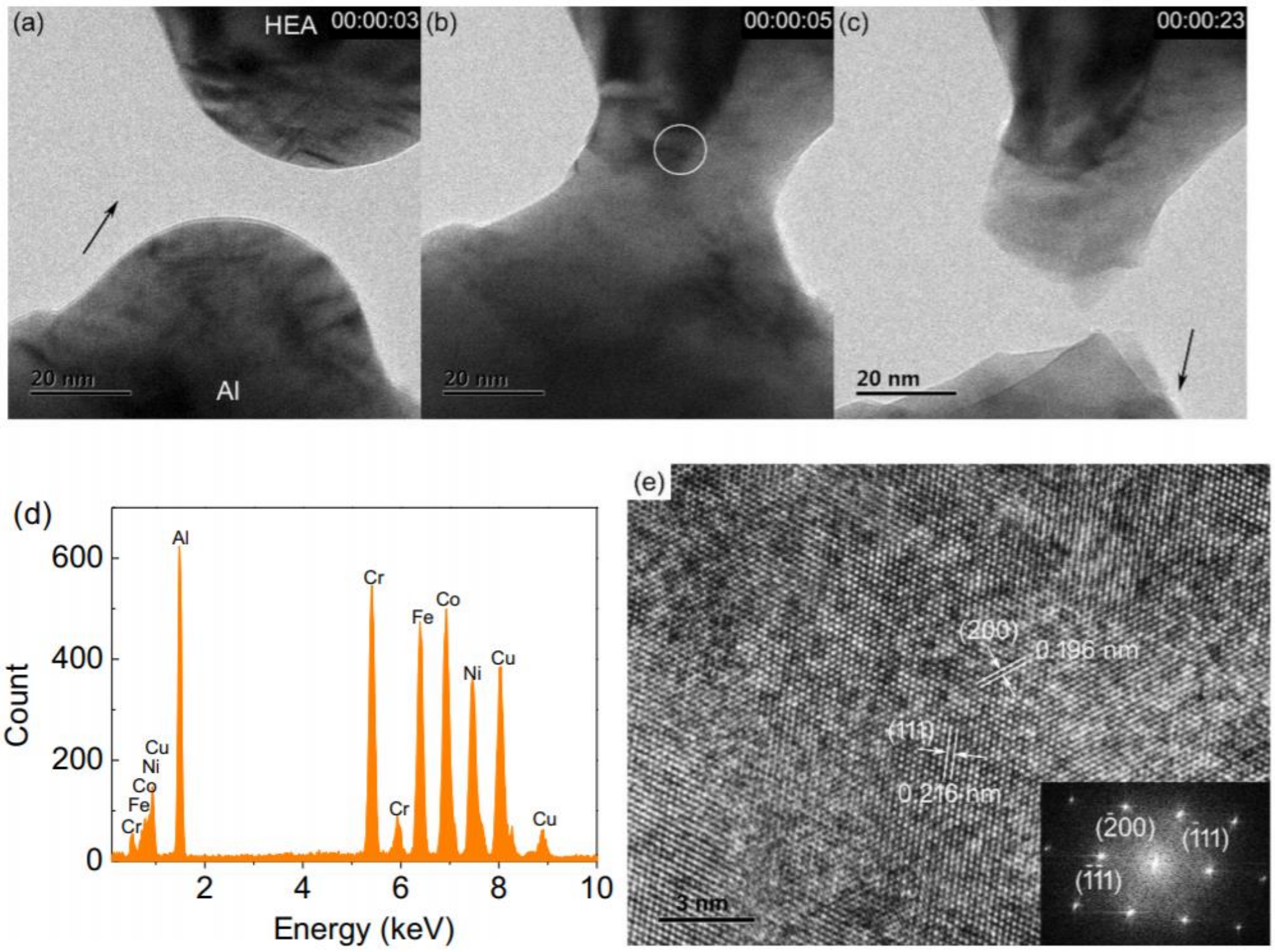

**Fig. 49.** (a)-(c) Dynamic-melting process of Al in the $FeCoNiCrCuAl_{0.75}$ HEA. (d) EDS result of the region of the circle in (b). (e) HRTEM image of the joint, the inserted picture is the corresponding FFT pattern under the [011] zone. Figures from Xu et al. [354].

Scientists also used the HT method in screening experiments. For instance, Li *et al.*[355] produced the $FeCoNiCrAl_x$ (x = 0.15 ~ 1.32) HEAs by the HT-laser deposition. The preparation process can be seen in Fig. 50. The alloy library (based on the FeCoNiCr substrate) consists of a 14 mm × 14 mm grid, which contains 25 patches that are 2 mm × 2 mm in size for each. Furthermore, each patch contains different amounts of Al powders. For the experiments, lasers were used to melt and bind the powders and the substrate together. During the deposition process, the laser power and travel speed were 150 W and 12.7 $mm \cdot s^{-1}$ (30 in/min), respectively. To make the compositional distribution more homogeneous, a laser power of 200 W was applied to remelt the patches twice in a direction that was perpendicular to the deposition. The role of powder nozzles was to add more Al powders to make subsequent layers. Then they successfully characterized the microstructures and properties of the alloys by SEM, EBSD, HAADF-STEM, and nanoindentation. While SEM, EBSD, and HAADF-STEM show that the microstructures for $FeCoNiCrAl_x$ (x = 0.15 ~ 0.37) HEAs consist of a single fcc phase, while for $FeCoNiCrAl_x$ (x = 0.41 ~ 1.32) HEAs the microstructures contain multiple fcc + bcc/B2 phases. The nanoindentation results indicate that the hardness of the $FeCoNiCrAl_x$ (x = 0.15 ~ 1.32) HEAs ranges from 4 to 9 GPa. These findings suggest that this method reduced the time to synthesize many compositions, and can be extended to other complex alloys. Furthermore, the huge database that can be obtained by HT suggesting that more attention should be given to the combination of the HT method with other experimental techniques, such as the synchrotron X-ray and neutron diffraction. It is anticipated that these methods will generate a large amount of data, showing the excellent potential for HT laser-deposition devices in the fields of HEA research.

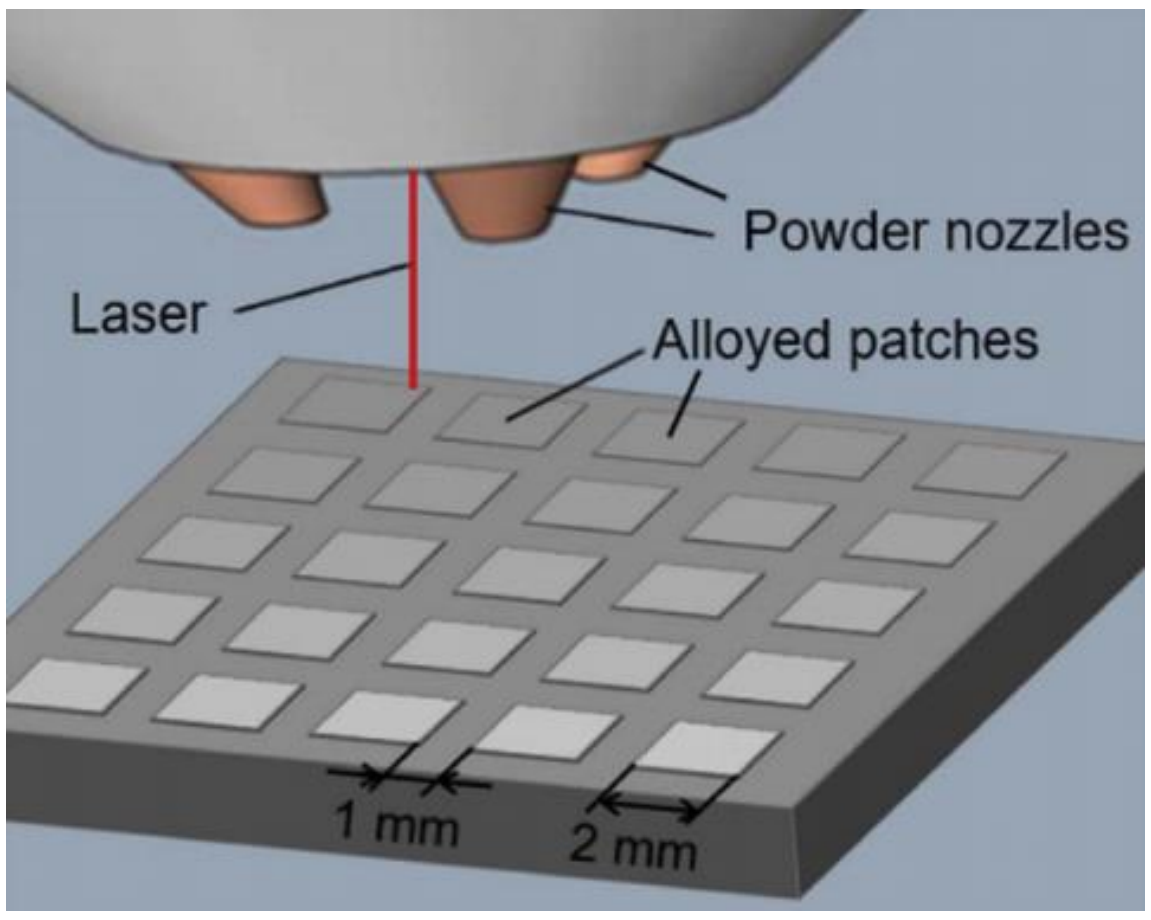

**Fig. 50.** The schematic diagram of the HT-laser-deposition method. Figure from Li et al. [355].

## 9.5. ML method

ML is an interdisciplinary subject that combines computer science, statistics, mathematics, and engineering knowledge to form an important branch of artificial intelligence[61, 363-365]. This field encompasses how computer programs can predict the performance of a material that can be improved, based on experimental data that is accumulated over time[366-368]. In other words, ML is a method that uses existing data to automatically optimize and improve a given physical model under the guidance of different algorithms such that it can judge and predict new outcomes.

In materials-computing simulations, large-scale, high-dimensional data sets are often produced[369]. ML can provide an extensible method to identify the pattern of the big data and extract the laws and trends from the data. Such a methodology, in theory, can provide a strategy to reverse design materials. At present, ML has been applied to the research and design of HEAs with excellent mechanical properties, such as the hardness[370]. Figure 51 provides a schematic of the design loop for HEAs using the ML method. In the iteration loop I, the first stage consists of a training dataset with known properties. From the data set, a feature pool and the composition are obtained. Next, the ML-surrogate model will be applied to the search space, where the ML-surrogate model is based on the function of $y_i = f(c_i)$, where $c_i$ represents the composition. Then a virtual space of nearly two million HEAs[370] will be obtained. A utility function is then used to select the next experiment. Subsequently, the preparation and measurement of the candidate will be added to the database. After the preparation and measurement of recommended candidates, the new obtained data are added into the training dataset to further improve the surrogate model. The iteration loop II is nearly the same as the iteration loop I, except that the iteration loop I does not contain the feature pool, and the ML-surrogate model is based on the function of $f(c_i, p_i)$ in the loop II, where $p_i$ represents the preselected physical features.

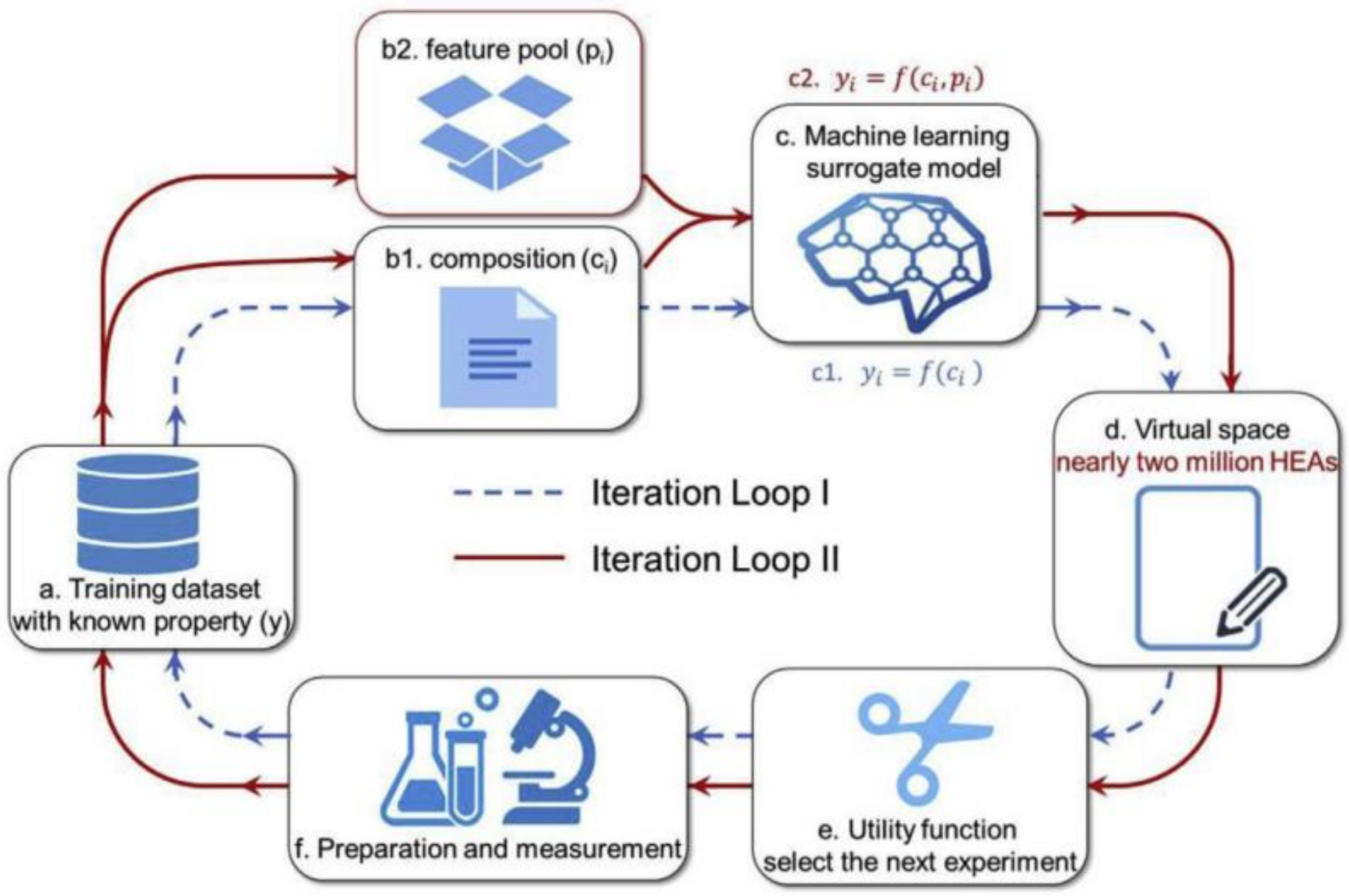

**Fig. 51.** Schematic sketch of the design loop for HEAs using the ML method. Figure from Wen et al. [370].

After the above-mentioned ML-iteration loop is completed, the alloys with the highest hardness were selected, *e.g.* $Fe_7Co_{22}Ni_5Cr_{33}Al_{43}$ and $Fe_5Co_{20}Ni_5Cr_{18}Cu_5Al_{47}$ HEAs. The two alloys both have a disordered bcc phase and an ordered B2 phase, as shown in the XRD patterns of Fig. 52(a). Figure 52(b) presents the dual-phase structure, where the darker region is the ordered phase. For the elemental distribution, as can be seen in Fig. 52(c), the ordered phase is composed of Al and Ni, confirming the feasibility of applying the ML method. Hence in the future ML design, the strategy can be extended to optimize other properties, such as designing lightweight HEAs and preparing HEA coatings.

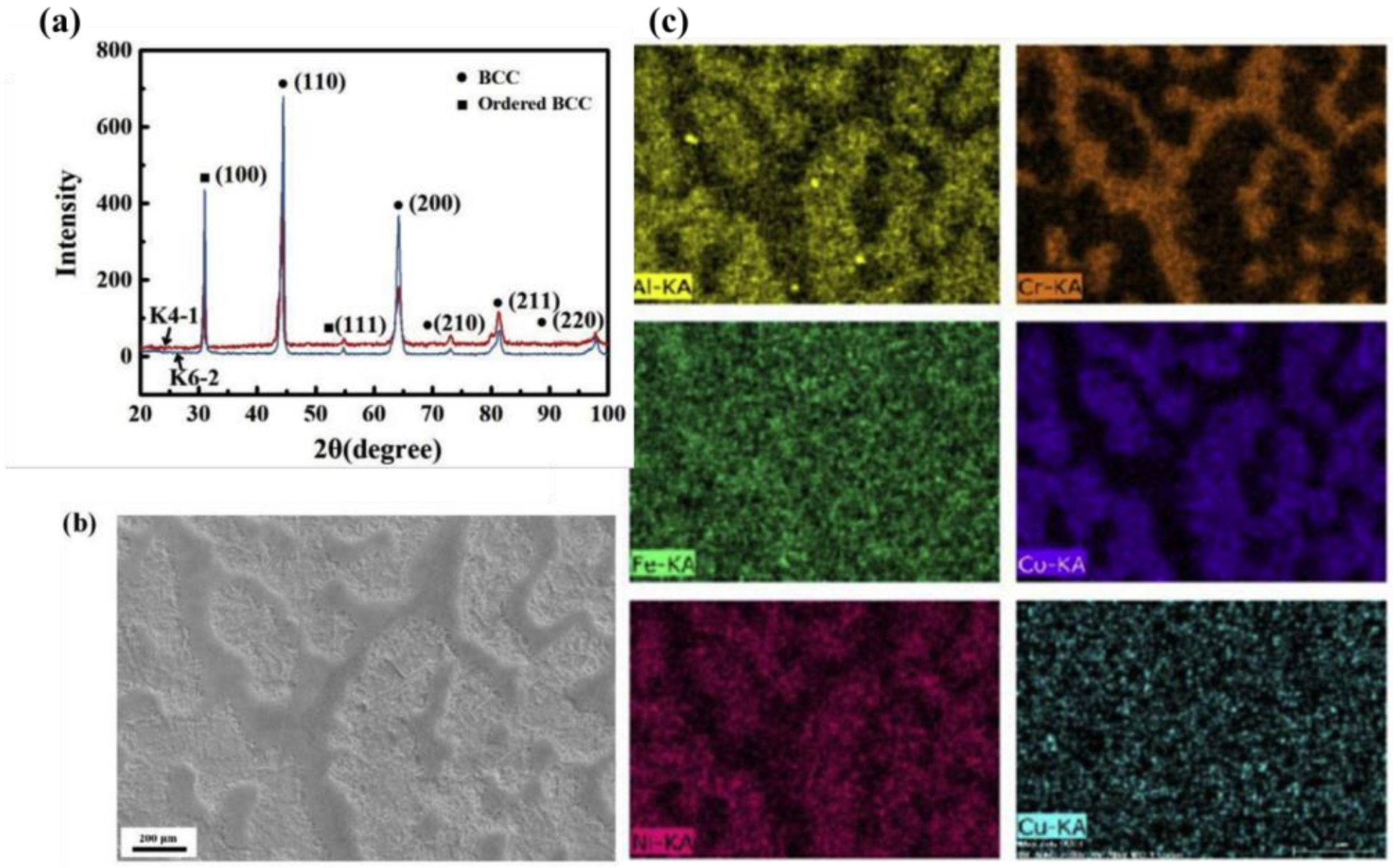

**Fig. 52.** (a) XRD patterns of the $Fe_7Co_{22}Ni_5Cr_{33}Al_{43}$ and $Fe_5Co_{20}Ni_5Cr_{18}Cu_5Al_{47}$ alloys; (b) SEM images; and (c) EDS maps of the $Fe_7Co_{22}Ni_5Cr_{33}Al_{43}$ alloy. Figures from Wen et al. [370].

The ML method can also be used to predict the phase[371], select the phase[372], and elucidate the deformation mechanism[373]. This technique can also be applied to predict the elasticity[374], phase stability, and chemical ordering that are affected by local lattice relaxations [375], thus predicting the compositions and hardness[376] of HEAs[373, 377]. For example, Kim *et al.*[374] used the ML method to find the most reliable and efficient models to predict the elastic constants of the $FeCoNiCrAl_{0.3}$ HEA. Here, they used the gradient-boosted trees (GB-Trees)[378] algorithm to build the model. In particular, the model was used to find out the effect of the bulk and shear moduli on different material features. The ML model is benchmarked according to the model used in Ref.[379]. It is worth noting that compared to the Gradient Boosting Machine Local Polynomial Regression (GBM-Locfit) benchmark model[378, 380], the training database of the GB-Tree model is different, which contains neither the random alloy data nor the HEA data. In order to compare the two models better, based on the size of the training database, the learning curve is applied to determine the correlation of the learning behavior and the model. Figures 53(a) and (b) show the learning curves produced via the bulk and shear moduli prediction models, respectively. Note that the y-axis represents the mean-squared error. In the training data, the red curve is the average 5-fold cross-validated score of the model, while the red-shaded area indicates that it is higher than or below a standard deviation. The green curve and shaded area represent the values of the test database, respectively. For the 1,940 training samples, the average cross-validated mean-square-error values of the shear bulk models on the test database are 0.048 and 0.032, respectively. It should be noted that these values are lower than those calculated by the GBM-Locfit models, *e.g.* 0.058 and 0.064, respectively. This trend suggests that under the condition of a given number of samples, the GB-Trees model improves the prediction accuracy.

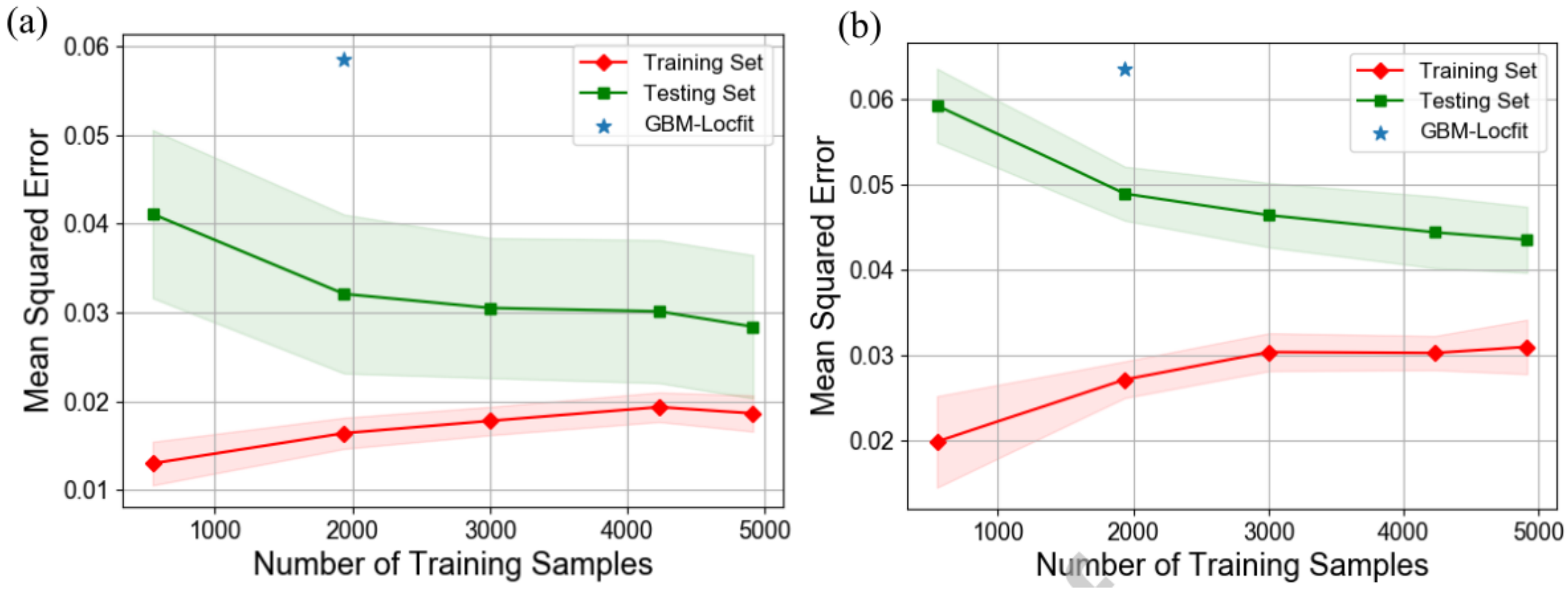

**Fig. 53.** The ML curves of the relationships between the mean-squared error and number of training samples in the database for (a) shear modulus and (b) bulk modulus, respectively. Figures from Kim et al. [374].

Li *et al.*[381] used a support vector machine (SVM) model[382] to determine 267 fcc and 369 bcc HEAs based on 16 chosen elements. First, they built a database of the 322-reported binary, ternary, quaternary, and quinary as-cast HEAs[62]. Then they chose the parameters of the atomic-size difference ($r_d$), mixing enthalpy ($\Delta H$), VEC, configurational entropy ($S_C$), and melting temperature ($T_m$) as features to build a SVM model. Figure 54(a) shows the results of a forward-search process, which indicates that the balanced accuracy of the cross-verification (CV) for the five parameters is 90.69 %, while the accuracy of training can reach 96.55 %. Figure 54(b) reports the relationship between the *C* and *G* (where *C* is the penalty parameter for the model, and *G* is the kernel coefficient), and when $C = 2^8$ and $G = 2^3$, which are the highest CV that can be obtained. The relationships between the accuracies and the number of features are shown in Fig. 54(c). At the initial stage, the accuracies of the training, CV, and testing all increase and until they saturate. Figure 54(d) shows the relationships between the accuracies and the number of features obtained by the synthetic minority over-sampling technique (SMOTE) method[383]. The technique is used to exclude the unbalanced samples, and the results indicate that due to the overfitting problem of the SMOTE method[384], the original SVM model is preferred. After the machine training and testing, the accuracy of the CV is more than 90 %. Based on the above discussion, it is clear that this SVM model can provide guidance for future applications of ML in the alloy design. Therefore, with the further development of the theory and method, ML will have more in-depth and extensive applications in the field of HEAs.

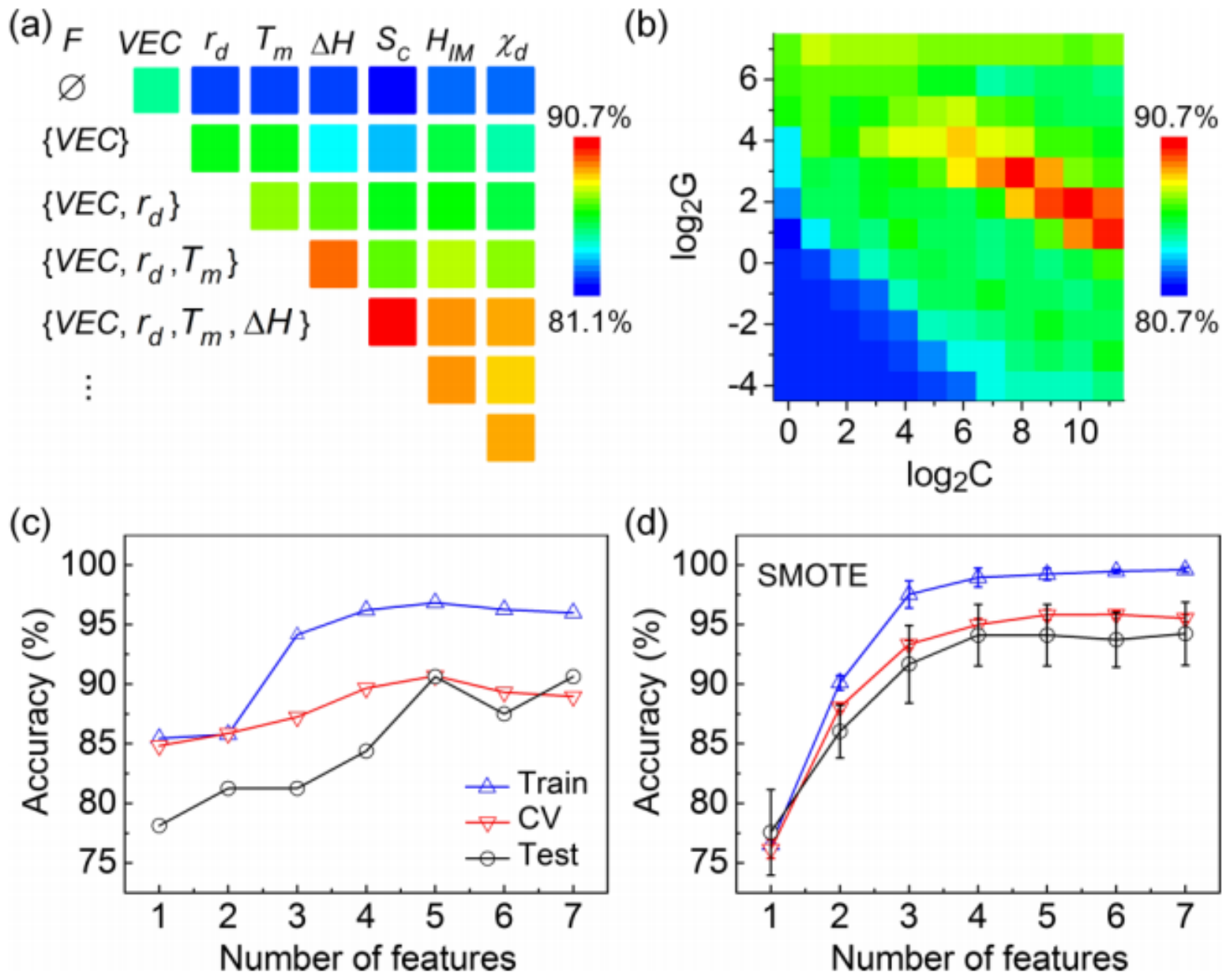

**Fig. 54.** (a) A forward-search process. (b) The relationship between the penalty parameter for the model (*C*) and the kernel coefficient (*G*). The relationships between the accuracies and the number of features obtained by (c) the SVM model and (d) the SMOTE model, respectively. Figuers from Li et al. [381].

## 10. Summary

As a new kind of advanced alloys, HEAs exhibit desirable properties for industry, such as high hardness, wear, fatigue, and corrosion resistance. As such, HEAs are potential candidates for use in different applications, such as cutting tools, molds, and welding materials. For this chapter, the major characteristics of HEAs were detailed and discussed, *e.g.* the formed single phases in HEAs and the super solid-solution features. Furthermore, an overview was provided on the microstructures, elastic properties, hardness, compressive behaviors, and tensile behaviors of HEAs. However, up to now, many experiments have shown that HEAs do not possess a good combination of strength and ductility. Hence, discovering a golden rule regarding an effective strengthening method for HEAs requires further study. Some strengthening strategies used in HEAs, such as strain hardening, grain-boundary strengthening, solid-solution strengthening, particle strengthening, and TWIP strengthening were also reviewed. The unusual dominant presence of edge dislocations in deformed bcc refractory HEAs needs to be further studied. Moreover, it is difficult to design HEAs with special properties and large sizes. Furthermore, based on the discussion of mechanical properties of HEAs at both low and elevated temperatures, as well as the fracture, fatigue, and creep behaviors described here, HEAs can also be potentially used in some extreme environmental conditions, *e.g.* cryogenic conditions. Therefore, all of these critical issues pose many questions with the need for further research.

It is important to note that suitable design strategies and processing technologies can help develop HEAs with excellent mechanical properties. Hence, there is potential for these materials to be used in many fields. For example, HEAs can be used as lightweight materials, films, coatings, and additive-manufactured 3D materials. To better understand the mechanical properties of HEAs, some new methods, such as HT and ML techniques, can also be applied to this field. Such methods can facilitate the discovery and design HEAs with desirable properties, and also to predict the phases and microstructures of HEAs before fabrication. However, it should be noted that continued efforts are needed to make these methods more accessible for practical applications of HEAs. Moreover, because of their great potential and broad application prospects, it is believed that HEAs will be important players in future industrial endeavors. It is also anticipated that the increase in the potential applications of HEAs will greatly bring great economic benefits to society.